\documentclass[12pt]{article}
\pdfoutput=1

\usepackage{array} 
\usepackage{amssymb}
\usepackage{graphics,graphpap}
\usepackage{graphicx}
\usepackage{color}
\usepackage{graphicx}
\usepackage{dcolumn}
\usepackage{epsfig}
\usepackage{epstopdf}
\usepackage{bbm}
\usepackage{amsmath}
\usepackage{amsfonts}
\usepackage{textcomp}
\usepackage{subcaption}
\usepackage{setspace}
\usepackage{slashed}
\usepackage{multirow}
\usepackage{footnote}
\makesavenoteenv{table}
\usepackage{url}
\usepackage{slashed}
\usepackage{mathtools}
\newcommand*\diff{\mathop{}\!\mathrm{d}}
\DeclareMathOperator\arctanh{Arctanh}

\DeclareMathOperator\Li{Li}
\usepackage{tikz}
\definecolor{dunkelgruen}{rgb}{0,0.7,0}
\definecolor{dunkelblau}{rgb}{0,0,0.7}

\usepackage{xcolor}
\definecolor{red}{rgb}{1,0,0}
\definecolor{purple}{rgb}{0.5,0,0.5}
\definecolor{blue}{rgb}{0,0,1}

\newcommand{\beq}{\begin{eqnarray}}
\newcommand{\eeq}{\end{eqnarray}}

\newcommand{\bmp}{\noindent\begin{minipage}{16cm}}
\newcommand{\emp}{\end{minipage}\vskip 7mm} 

\def\drawbox#1#2{\hrule height#2pt
        \hbox{\vrule width#2pt height#1pt \kern#1pt
              \vrule width#2pt}
              \hrule height#2pt}

\def\Asym#1#2{\vcenter{\vbox{\drawbox{#1}{#2}
              \kern-#2pt 
              \drawbox{#1}{#2}}}}

\def\simge{\mathrel{%
   \rlap{\raise 0.511ex \hbox{$>$}}{\lower 0.511ex \hbox{$\sim$}}}}

\def\simle{\mathrel{
   \rlap{\raise 0.511ex \hbox{$<$}}{\lower 0.511ex \hbox{$\sim$}}}}

\def\s#1{\setbox0=\hbox{$#1$}%
\rlap{\ifdim\wd0>.7em\kern.22\wd0\else\kern.1\wd0\fi /}#1}

\newcommand{\R}[1]{\text{\bf#1}}

\usepackage{cite}

\begin{document}

\begin{titlepage}
\title{\vspace*{-2.0cm}
\hfill {\small MPP-2015-282}\\[20mm]
\vspace*{-1.5cm}
\bf\Large
Conversions of Bound Muons: Lepton Flavour Violation from Doubly Charged Scalars\\[5mm]\ \vspace{-1cm}}

\author{
Tanja Geib\thanks{email: \tt tgeib@mpp.mpg.de}~~~and~~Alexander Merle\thanks{email: \tt amerle@mpp.mpg.de}
\\ \\
{\normalsize \it Max-Planck-Institut f\"ur Physik (Werner-Heisenberg-Institut),}\\
{\normalsize \it F\"ohringer Ring 6, 80805 M\"unchen, Germany}\\
}
\date{\today}
\maketitle
\thispagestyle{empty}

\begin{abstract}
\noindent
We present the first detailed computation of the conversion of a bound muon into an electron mediated by a doubly charged $SU(2)$ singlet scalar. Although such particles are not too exotic, up to now their contribution to $\mu$-$e$ conversion is unknown. We close this gap by presenting a detailed calculation, which will allow the reader not only to fully comprehend the discussion  but also to generalise our results to similar cases if needed. We furthermore compare the predictions, for both the general case and for an example model featuring a neutrino mass at 2-loop level, to current experimental data and future sensitivities. We show that, depending on the explicit values of the couplings as well as on the actual future limits on the branching ratio, $\mu$-$e$ conversion may potentially yield a lower limit on the doubly charged singlet scalar mass which is stronger than what could be obtained by colliders. Our results considerably strengthen the case for low-energy lepton flavour violation searches being a very valuable addition to collider experiments.
\end{abstract}

\end{titlepage}

\section{\label{sec:Intro}Introduction}

The Standard Model (SM) of elementary particle physics is the most well-tested description of Nature we know. While some parts are amazingly precise, such as the quantitative explanation of the anomalous magnetic moment of the electron being accurate to ten digits~\cite{Gorringe:2015cma}, other sectors of the SM still seem mysterious and/or incomplete. For example, the SM suffers from internal inconsistencies such as the hierarchy problem~\cite{Barbieri:1987fn} or the strong $CP$ problem~\cite{Peccei:1977hh}, it does not feature any good candidate to explain the Dark Matter in the Universe~\cite{Bertone:2004pz}, and it also fails to explain neutrino masses and mixings~\cite{Mohapatra:2005wg}. More generally, the last point illustrates that the \emph{flavour structure} of the SM is not well understood, i.e., how the three generations of fermions combine to mass eigenstates. In particular in the lepton sector, we know from the observation of neutrino oscillations~\cite{Fukuda:1998mi,Ahmad:2002jz,Araki:2004mb,Michael:2006rx,An:2012eh,Ahn:2012nd,Abe:2011sj,Abe:2011fz} that lepton flavour is not conserved, e.g.\ in processes like $\overline{\nu}_\mu \to \overline{\nu}_e$. Yet, in the charged lepton sector, we have not observed any flavour changing reaction -- even though all fundamental conservation laws such as energy, momentum, and angular momentum would not forbid lepton flavour violating (LFV) decays like $\mu \to e\gamma$ or $\tau \to \mu \gamma$. On the contrary, experimental limits on the branching ratios of these processes are extremely strong, e.g.: ${\rm BR}(\mu \to e\gamma)<5.7\cdot 10^{-13}$@90\%~C.L.~\cite{Adam:2013mnn}, ${\rm BR}(\tau \to e\gamma)<3.3\cdot 10^{-8}$@90\%~C.L.~\cite{Aubert:2009ag}, and ${\rm BR}(\tau \to \mu \gamma)<4.4\cdot 10^{-8}$@90\%~C.L.~\cite{Aubert:2009ag}.

However, there is no fundamental reason for lepton flavour to be conserved. While in the SM it is \emph{accidentally} conserved at tree level~\cite{ChengLi}, already when augmenting the SM by massive neutrinos, LFV decays such as $\mu \to e\gamma$ are generated at 1-loop level (albeit strongly suppressed by the Glashow-Iliopoulos-Maiani (GIM) mechanism~\cite{Glashow:1970gm}, such that even in the most optimistic case the corresponding branching ratio for $\mu \to e\gamma$ would be not more than a daunting $10^{-45}$~\cite{Cheng:1980tp,Cheng:1980qt}). More generally, due to no reason being present for lepton flavour to be conserved, any type of physics beyond the SM has a strong tendency to create LFV reactions~\cite{Blum:2007he}. Accordingly, once we experimentally observe any type of LFV process, it would be an unambiguous and groundbreaking signal for physics beyond the SM -- which up to now is only verified in the lab by neutrino oscillations.

Thus, the experimental hunt for LFV reactions is regarded to be a high-priority matter in experimental advances alternative to high-energy colliders. While experiments like MEG~\cite{Adam:2013mnn} ($\mu \to e\gamma$), BaBar~\cite{Adam:2013mnn} ($\tau \to e \gamma$, $\tau \to \mu \gamma$), SINDRUM~\cite{Bellgardt:1987du} ($\mu \to 3e$), or Belle~\cite{Hayasaka:2010np} ($\tau \to 3e$, $\tau \to 3\mu$, $\tau^- \to \mu^- e^+ e^-$, $\tau^- \to e^-\mu^+ \mu^-$) obtain their best limits from ``clean'' decays with initial and final states only containing elementary particles, in the near future the most dramatic experimental advances are to be expected for the conversion of muons bound on atomic nuclei to electrons ($\mu$-$e$ conversion), with sensitivities quoted in experimental proposals improving current limits by up to seven orders of magnitude~\cite{Raidal:2008jk}.\footnote{Note that, although $\mu$-$e$ conversion does intrinsically contain nuclear physics uncertainties which make it more difficult to interpret experimental limits, it is nevertheless clear that this process will yield a limit by far better than what we could possibly expect from experiments on $\mu \to e\gamma$.} It is this process we focus on in this paper.

While $\mu$-$e$ conversion was proposed more than fifty years ago~\cite{Weinberg:1959zz,Marciano:1977cj}, it is surprising that it has not even been computed explicitly for some relatively generic settings. The rate for $\mu$-$e$ conversion has been calculated for channels like light or heavy Majorana neutrino exchange~\cite{Dinh:2012bp}, $Z'$-exchange~\cite{Bernabeu:1993ta}, some specific extended scalar sectors~\cite{Crivellin:2014cta}, or several supersymmetric settings~\cite{Frank:2000sn,Arganda:2007jw,Zhang:2013jva}, however, the generic example of this decay being mediated by a doubly charged $SU(2)$ singlet scalar has only been briefly estimated~\cite{Raidal:1997hq}. In this paper, we will close this gap by presenting the first detailed computation of that very process. Up to now, not much technical information is available in the literature, which is why we chose to present the computation in great detail and illustrate all important steps and subtleties involved. Our results are fully general and hold for \emph{any} doubly charged singlet scalar $S^{++}$ coupling to pairs of right-handed charged leptons by $\mathcal{L}_{\rm LFV} = f_{ab} S^{++} \overline{(l_{Ra})^c} l_{Rb} + h.c.$ (such a coupling cannot be forbidden in practice). We will in passing also investigate the validity of the approximation applied in Ref.~\cite{Raidal:1997hq} revealing that, while we generally confirm the results obtained there, the estimate based on effective field theory (EFT) turns out to be not as accurate as anticipated. Furthermore, even if the doubly charged scalar was, say, a component of a Higgs triplet field, the principal computation would not change very significantly, so that our results could even be extended to this case. Thus, also to maximise the applicability of our results and the interest to a wide readership, we have decided to present our computation in a fairly detailed manner, to ease the comparison with similar frameworks.

However, the purpose of our work is two-fold. On top of a very general computation, we will also present an application of our results to one particular example model. This model, first presented in Ref.~\cite{King:2014uha}, features a doubly charged singlet scalar field $S^{++}$ which, in addition to the coupling to right-handed charged leptons $l_{Ra}$ and $l_{Rb}$ with strengths $f_{ab}$, also features an effective coupling of strength $\xi$ to a pair of $W$-bosons:
\begin{equation}
 \mathcal{L}_{S^{++}}  = \mathcal{L}_{\rm SM} - \frac{g^2 \,v^4\,\xi}{4\,\Lambda^3} S^{++} W^-_\mu {W^-}^\mu + f_{ab} S^{++} \overline{(l_{Ra})^c} l_{Rb} + h.c. - V',
 \label{eq:Lagrangian}
\end{equation}
where $V' = M_S^2 \, S^{++} S^{--} + \lambda_S (S^{++} S^{--})^2 + \lambda_{HS} (H^\dagger H) (S^{++} S^{--})$ and $v=246~$GeV. This model is in some sense the simplest setting one could possibly write down to generate a light neutrino mass, because it contains only one single particle with certain couplings in addition to the SM.\footnote{Alternatively, one could view the setting as a whole class of models which are at low energies described by the effective theory defined by Eq.~\eqref{eq:Lagrangian}.}  Light neutrinos then receive a mass at 2-loop level, by a diagram containing $S^{++}$ as crucial ingredient~\cite{King:2014uha}. This implies that both the couplings ($f_{ab}$ \& $\xi$) as well as the mass ($M_S$) of the doubly charged scalar are constrained by phenomenology. While in Ref.~\cite{King:2014uha} all neutrino observables and nearly all low-energy LFV observables, as well as neutrinoless double beta decay and collider limits, have been taken into account, the crucial process of $\mu$-$e$ conversion had not been investigated so far. This is another gap we will close with this paper on the technicalities of the process, which complements Ref.~\cite{Geib:2015tvt} that focuses in particular on the complementarity between high- and low-energy bounds.\footnote{We are furthermore preparing a study of the lepton number violating conversion $\mu^-$ to a $e^+$~\cite{Geib:2016atx}, which comprises an experimental alternative to neutrinoless double beta decay.}

This paper is structured as follows. We first discuss the long-range (i.e., photonic) contributions to $\mu$-$e$ conversion in great detail in Sec.~\ref{sec:mue_minus-photonic}, which serves as a first approximation to the true result. We then include the short-range (non-photonic) contributions in Sec.~\ref{sec:mue_minus-non-photonic}, which will only slightly modify the branching ratios. We conclude in Sec.~\ref{sec:conc}. Finally, technical details are summarised in Appendices~\ref{app:FeynmanRules} (Feynman rules) and~\ref{app:Passarino} (details on the scalar three-point function).

\section{\label{sec:mue_minus-photonic}Long-range (photonic) contributions}

The goal of this section is to derive the particle physics part of the branching ratio for coherent $\mu$-$e$ conversion in a muonic atom, for the moment focusing on the \emph{long-range contributions} only, i.e., those diagrams which basically attach a diagram for $\mu \to e\gamma$ to a nucleus. As we will see, this already comes very close to our final result because the photonic contributions turn out to dominate the non-photonic short-range contributions by far. This is very convenient, because for the case of long-range contributions being dominant, the total amplitude factorises into a particle physics and a nuclear physics part. Thus, the nuclear physics factor (which quantifies all nuclear physics contributions) can be computed separately and it can easily be updated once improved computations become available -- as done for neutrinoless double beta decay.

\subsection{\label{sec:mue_minus-photonic_general}The physics of $\boldsymbol{\mu}$--$\boldsymbol{e}$ conversion}

Taking into account gauge invariance\footnote{Note that, due to the (Abelian) Ward identity, it holds that $f_3=g_3=0$ for the photonic case. This is an additional cross check for our computation and was confirmed when determining the form factors.}, the most general form for the photonic matrix element (i.e., for the $\mu^-$--$e^-$--$\gamma$ vertex) can be written as~\cite{Kuno:1999jp,Kitano:2002mt,Lavoura:2003xp,ChengLi,Patel:2015tea}:
\begin{eqnarray}
 i\mathcal{M} &=& -i\,e\,A^*_\nu (q)\,\overline{u}_e (p_e)\,\Big[\Big(f_{\rm E0}(q^2)+\gamma_5\,f_{\rm M0}(q^2)\Big)\Big(\gamma^\nu-\frac{\slashed{q} q^\nu}{q^2}\Big)\ +\Big(f_{\rm M1}(q^2)+\gamma_5\,f_{\rm E1}(q^2)\Big)\frac{i\,\sigma^{\nu\rho}\,q_\rho}{m_\mu} \nonumber\\
 && +2\frac{q^\nu}{m_\mu}\,f_3(q^2)+2\frac{q^\nu}{m_\mu}\,\gamma_5\,g_3(q^2)\Big]\,u_\mu(p_\mu)\,,
  \label{eq:MatrixElPhotonic}
\end{eqnarray}
where $q=p_\mu-p_e$ is the photon momentum and $\sigma^{\nu\rho}\equiv \frac{i}{2} [\gamma^\nu,\,\gamma^\rho]$.\footnote{In order to prevent any confusion, we do not use the letter ``$\mu$'' as Lorentz index, but instead we only use it to refer to the muon.} The functions $f$ are \emph{form factors} that in general depend on the momentum transfer. They are the quantities which ultimately encode the loop structures involved in the diagrams. Note that the amplitude as reported in Eq.~\eqref{eq:MatrixElPhotonic} is the same for both $\mu \to e\gamma$ and $\mu$-$e$ conversion. However, both processes nevertheless yield qualitatively different information. The reason is that $\mu \to e\gamma$ is strongly simplified by on-shell relations being applicable only for external photons, in particular $q^2 = 0$ (the photon is massless) and $\epsilon_\nu q^\nu = 0$ (the photon is transversal). On the contrary, in $\mu$-$e$ conversion, the off-shell part of the amplitude strongly contributes, which is reflected in the resulting bounds on the effective model used as an example here being very different for both processes~\cite{Geib:2015tvt}.

The decisive observable is the branching ratio of $\mu$-$e$ conversion with respect to ordinary muon capture, which is simple if the long-range contributions dominate~\cite{Kuno:1999jp}:
\begin{equation}
 {\rm BR}(\mu^- N \to e^- N)|_\text{long-range} = \frac{8 \alpha^5 m_\mu Z_{\rm eff}^4 Z F_p^2}{\Gamma_{\rm capt}} \,\Xi_{\rm particle}^2,
 \label{eq:mu-e_BR_long}
\end{equation}
where $\alpha$ is the fine structure constant and $\Gamma_{\rm capt}$ is the rate for ordinary muon capture (with emission of a $\nu_\mu$) on the nucleus under consideration, which is quasi identical to the total rate. Furthermore, the effective atomic charge $Z_{\rm eff}^4 = \frac{\pi Z}{\alpha^3 m_\mu^3} \cdot 4\pi \int\limits_{r=0}^\infty dr\ r^2 |\Phi_{1s,\mu}(r)|^2 \rho_p(r)$ [with $\Phi_{1s,\mu}(r)$ being the $1s$ wave function of the muon bound to a nucleus of atomic number $Z$] and the nuclear matrix element (NME) $F_p = 4\pi \int\limits_{r=0}^\infty dr\ \frac{r}{m_\mu} \sin (r m_\mu) \rho_p(r)$ can both be calculated easily if the proton charge density $\rho_p(r)$ inside the nucleus is known.

Let us discuss the physics of $\mu$-$e$ conversion before entering the actual computation. In Eq.~\eqref{eq:mu-e_BR_long}, all the particle physics is contained in the factor $\Xi_{\rm particle}^2$, which is our main quantity of interest. It is explicitly given by~\cite{Kuno:1999jp}:
\begin{equation}
 \Xi_{\rm particle}^2 = | f_{\rm E0}(-m_\mu^2) + f_{\rm M1}(-m_\mu^2) |^2 + | f_{\rm E1}(-m_\mu^2) + f_{\rm M0}(-m_\mu^2) |^2.
 \label{eq:Xi_particle}
\end{equation}
Thus, in our computation, we ``only'' need to extract the form factors $f_{\rm E0, E1, M0, M1}$ from the amplitude and to evaluate them at a 4-momentum transfer of $q^2 = -m_\mu^2$. Once we achieve that, we can immediately use Eq.~\eqref{eq:mu-e_BR_long} to obtain the branching ratio for $\mu$-$e$ conversion.

However, there are several other aspects to the process which have to be discussed before we can start our computation. While the basic principle behind $\mu$-$e$ conversion, the capture of a bound muon with subsequent emission of a fast electron, is easy to grasp, several subtleties make this process comparatively difficult to compute in practice. Further (technical) details on this discussion can be found e.g.~in~\cite{Berestetsky:1982aq,Kitano:2002mt,Chiang:1993xz,Shanker:1979ap}.

First, let us have a look at the initial state muon. It is not free but in the $1s$ bound state of a muonic atom. Also the final state electron is not free, as it does feel the influence of the electric field of the remainder of the atom present in the final state. Thus, to take into account all resulting effects, it is easiest to perform the computation in real space and to use the solutions of the Dirac equation in a Coulomb potential instead of the spinors corresponding to free particles: $\overline{u}_e(p_e)\to \overline{\psi}_e(p_e,r)$ and $u_\mu(p_\mu) \to \psi_\mu (p_\mu,r)$.

Second, a simplification arises from the muon mass being the dominant energy scale compared to the binding energies $E_b$ involved or to the electron mass: $m_\mu \gg m_e > E_b \approx 13.6~\mathrm{eV}\cdot \frac{m_\mu}{m_e} Z$. Thus, we can set the electron mass to zero, $m_e \approx 0$, and we can treat the muon non-relativistically. This furthermore implies that the kinematics of the process are in effect very similar to those of a $t$-channel diagram, with both the initial state muon and the initial (and final) state nucleus being nearly at rest; we can thus approximate $q^2 \simeq -m^2_\mu$.

Third, given the nature of the process, it is unavoidable to consider some atomic and nuclear physics aspects. Fortunately a standard formalism exists to take them into account. For example, the photon couples to electric charges (no matter if it is on- or off-shell), which means that the corresponding part of the matrix element must be proportional to the proton charge density $\rho^{(P)}(r)$ in the nucleus: $\langle N|\overline{q}\,\gamma_\nu\,q|N\rangle \propto Z e \rho^{(P)}(r)\,\delta_{\nu 0}$. Thus, the full amplitude for the process must have the following structure:
 \begin{equation}
  \mathcal{M}\propto \int \diff^3 r\,\overline{\psi^e_{jlm}}(p_e,r)\,\Gamma^\nu\,\psi^\mu_{j_\mu l_\mu m_\mu} (p_\mu,r) Z e \rho^{(P)}(r)\,\delta_{\nu 0},
  \label{eq:mu-e-con_full}
 \end{equation}
where $\Gamma^\nu$ includes the form factors and Lorentz structure displayed explicitly between the two spinors in Eq.~\eqref{eq:MatrixElPhotonic}. Given that the nucleus is taken to be non-relativistic, its 4-current density consists of only the 0-component to a good approximation, which is why effectively only $\Gamma^0$ contributes to the amplitude.\footnote{Note that at this point we have in fact broken Lorentz invariance, because we have chosen a particular system -- namely the rest frame of the nucleus. However, for a non-relativistic bound system this makes perfect sense because all relevant quantities can be expressed easily and, after all, we can compute a Loretz-invariant amplitude in any frame.} This implies further simplifications: the pre-factor $q^\nu=p_\mu^\nu - p_e^\nu$ in front of the form factors $f_3$ and $g_3$ reduces to $q^0 \simeq m_\mu - m_\mu =0$ for the case of a non-relativistic muon in the initial state dictating the electron energy in the final state. Thus, even for non-vanishing $f_3$ and $g_3$, they would not contribute to the conversion process.

Finally, we need to discuss the forms of the muon and electron wave functions. They depend on the details of the atomic physics configuration. We follow the standard approach taken in textbooks~\cite{Berestetsky:1982aq}, and write the fermion spinor in terms of ``upper'' and ``lower'' radial components $f$ and $g$. Since we work in the Dirac representation, only the upper component survives in the non-relativistic limit (i.e.\ for the muon). Encoding the angular part in spherical harmonic spinors $\Omega_{jlm}$, we can thus describe the physics of both the muon and the electron by wave functions of the following form:
 \begin{equation}
  \psi_{jlm} =
  \begin{pmatrix}
  f(r)\,\Omega_{jlm}\\
  (-1)^{1/2(1+l-l')}\,g(r)\,\Omega_{jl'm}
  \end{pmatrix},
  \label{eq:spinor}
 \end{equation}
with total angular momentum $j$, orbital angular momenta $l$ and $l'=2j-l$, and spin projection $m$. In the 1$s$ state, the muon has quantum numbers $(j, l, l',m) = (1/2, 0, 1, \pm 1/2)$. Thus, angular momentum conservation dictates quantum numbers of $(1/2, 0, 1, \pm 1/2)$ or $(1/2, 1, 0, \pm 1/2)$ for the final electron. Depending on the configuration, different parts of the amplitude in Eq.~\eqref{eq:MatrixElPhotonic} will contribute (e.g., only structures featuring $\gamma_5$ survive for $l=1$). Exploiting that the initial state muon is nearly at rest, while the final state electron is highly relativistic, we can furthermore set $g_\mu^{l=0} \simeq 0$ as well as $f_e^{l=1}=-g_e^{l=0}$ and $g_e^{l=1}=f_e^{l=0}$. Finally, because the two final states with $l=0$ and $l=1$ are distinguishable, we have to sum over probabilities rather than amplitudes; hence the form in Eq.~\eqref{eq:Xi_particle}.

\subsection{\label{sec:mue_minus-matching}Determination of the Form Factors}

\begin{figure}[th!]
  \vspace{-2cm}
  \begin{minipage}[c]{14.6cm}
    \begin{subfigure}[c]{7.3cm}
    \includegraphics[width=7.3cm]{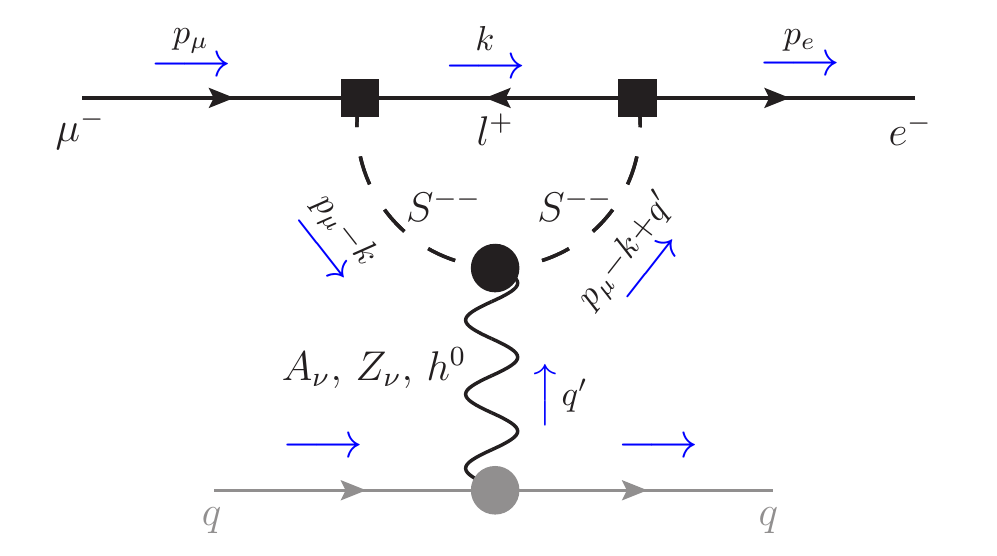}
    \caption{Diagram I}
    \label{fig:Diagram_I}
    \end{subfigure}
   \begin{subfigure}[c]{7.3cm}
    \includegraphics[width=7.3cm]{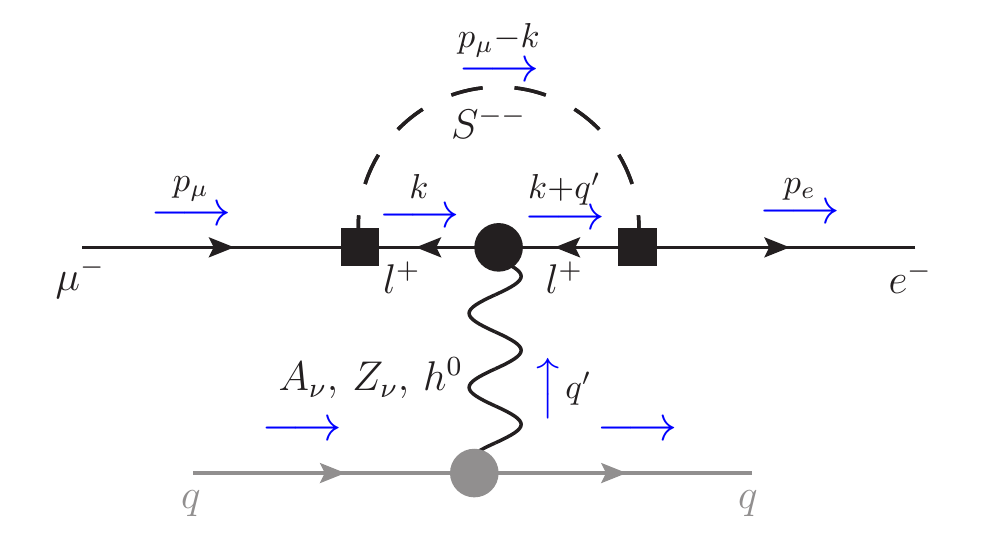}
    \caption{Diagram II}
    \label{fig:Diagram_II}
     \end{subfigure}  
  \end{minipage}
  \hspace{1\textwidth}
    \begin{minipage}[c]{14.6cm}
    \begin{subfigure}[c]{7.3cm}
    \includegraphics[width=7.3cm]{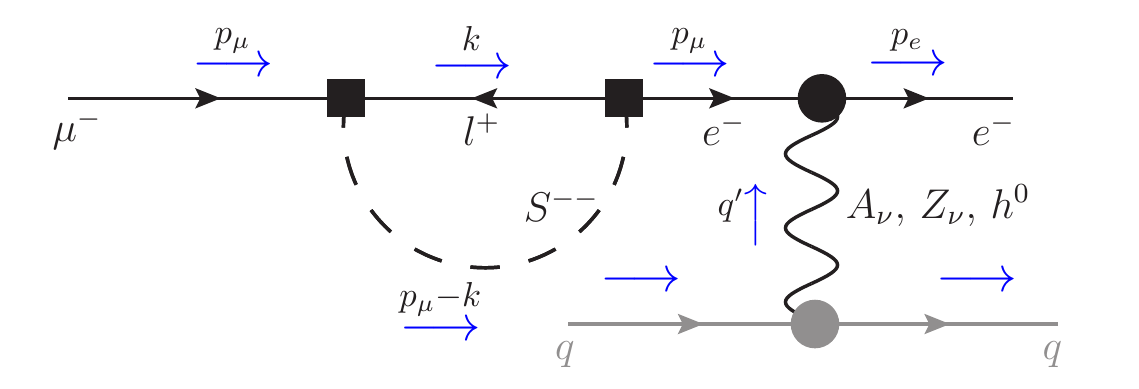}
    \caption{Diagram III}
    \label{fig:Diagram_III}
    \end{subfigure}
   \begin{subfigure}[c]{7.3cm}
    \includegraphics[width=7.3cm]{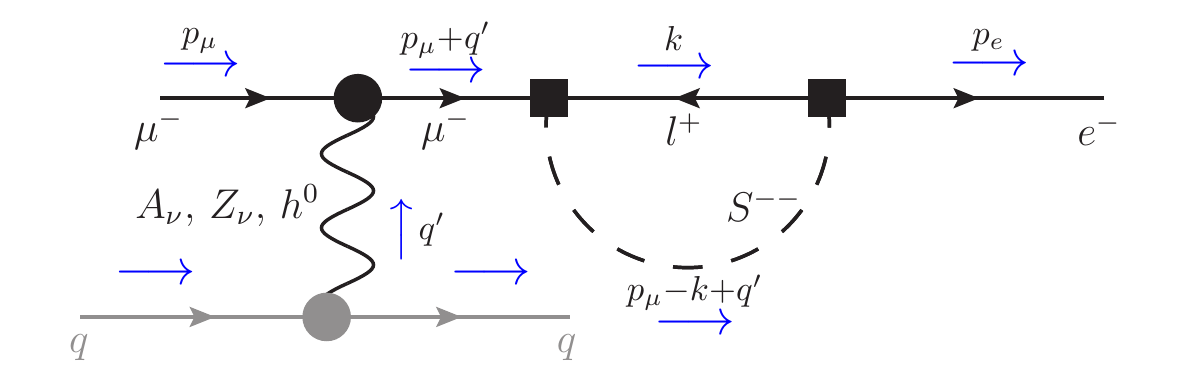}
    \caption{Diagram IV}
    \label{fig:Diagram_IV}
     \end{subfigure}  
  \end{minipage}
   \hspace{1\textwidth}
   \begin{minipage}[c]{14.6cm}
    \begin{subfigure}[c]{7.3cm}
    \includegraphics[width=7.3cm]{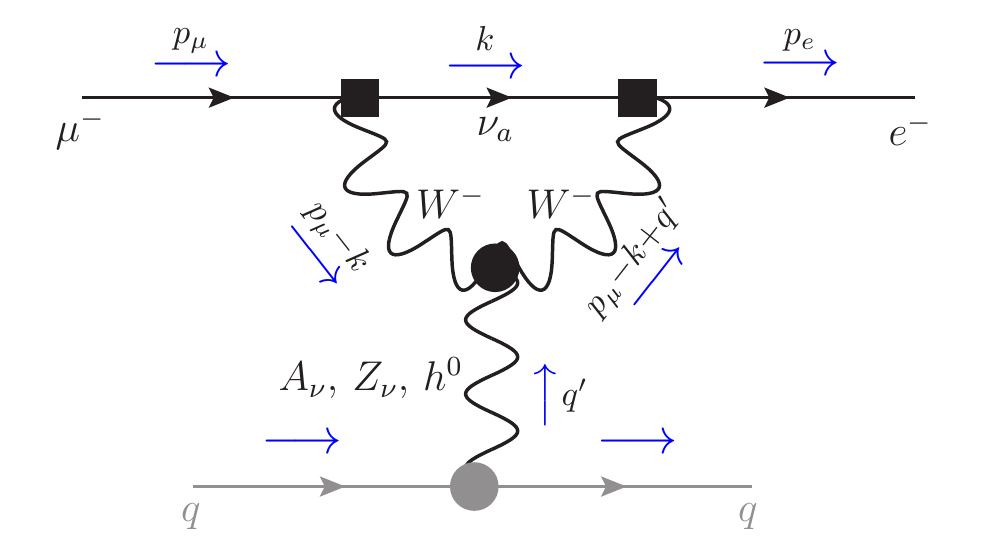}
    \caption{Diagram V}
    \label{fig:W_Diagram_I}
    \end{subfigure}
   \begin{subfigure}[c]{7.3cm}
    \includegraphics[width=7.3cm]{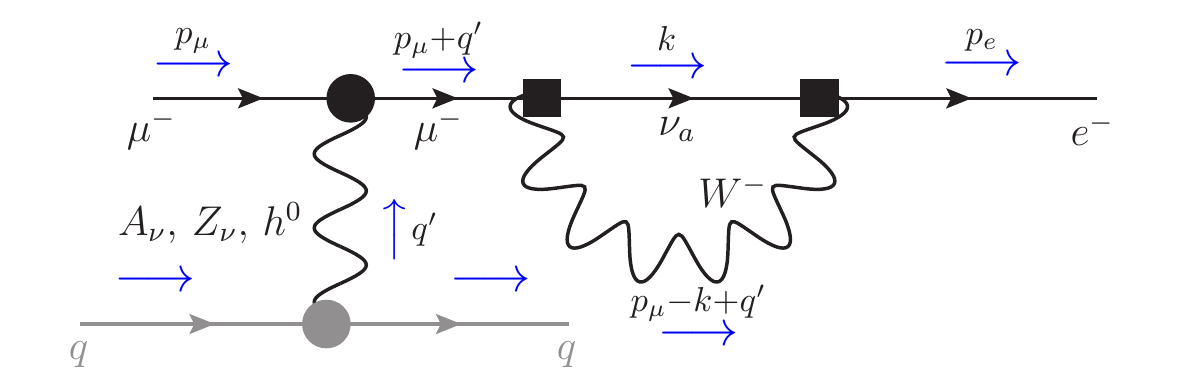}
    \caption{Diagram VI}
    \label{fig:W_Diagram_II}
     \end{subfigure}  
  \end{minipage}
  \hspace{1\textwidth}
    \begin{minipage}[c]{14.6cm}
    \begin{subfigure}[c]{7.3cm}
    \includegraphics[width=7.3cm]{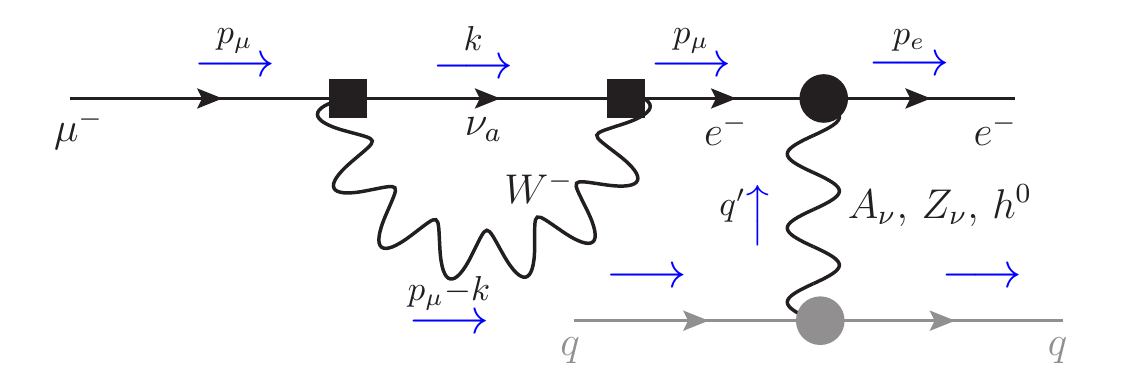}
    \caption{Diagram VII}
    \label{fig:W_Diagram_III}
    \end{subfigure}
     \begin{subfigure}[c]{7.3cm}
    \includegraphics[width=7.3cm]{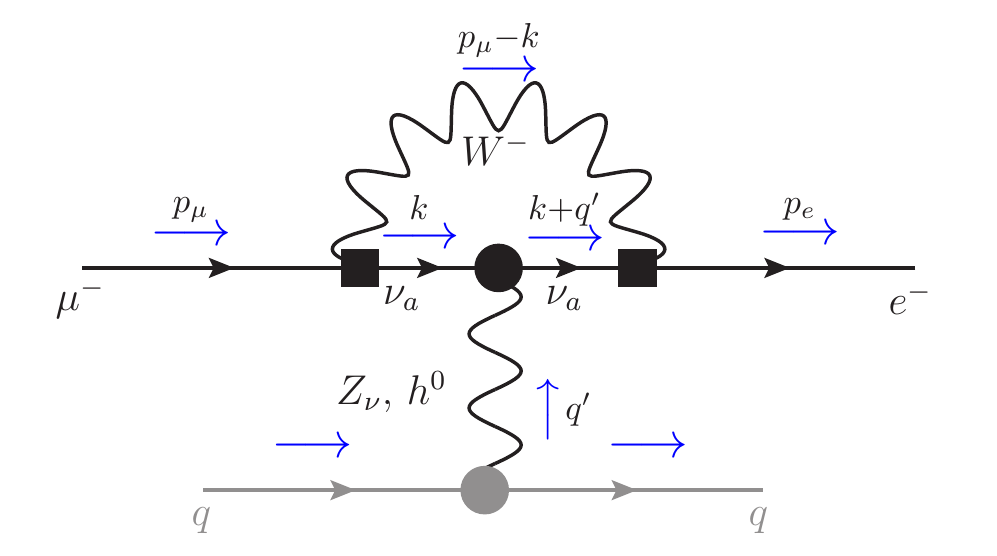}
    \caption{Diagram VIII}
    \label{fig:W_Diagram_VIII}
     \end{subfigure}  
    \end{minipage}
    \hspace{1\textwidth}
    \begin{minipage}[c]{14.6cm}
    \begin{subfigure}[c]{7.3cm}
    \includegraphics[width=7.3cm]{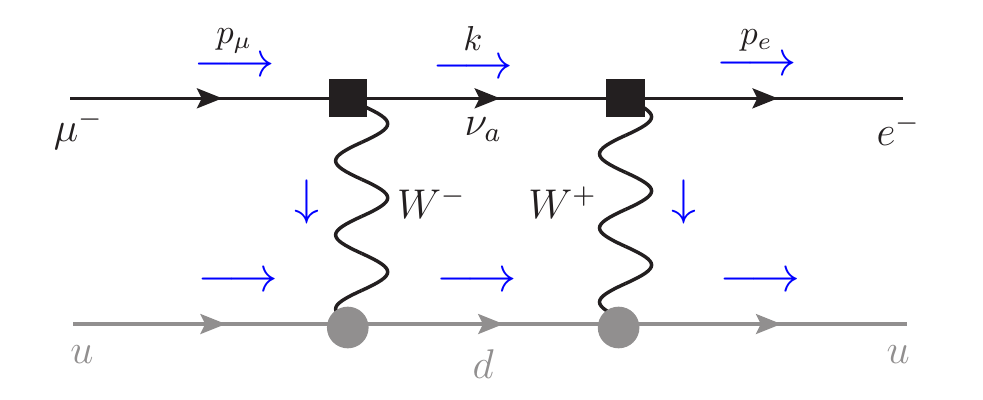}
    \caption{Diagram IX}
    \label{fig:Box_I}
    \end{subfigure}
     \begin{subfigure}[c]{7.3cm}
    \includegraphics[width=7.3cm]{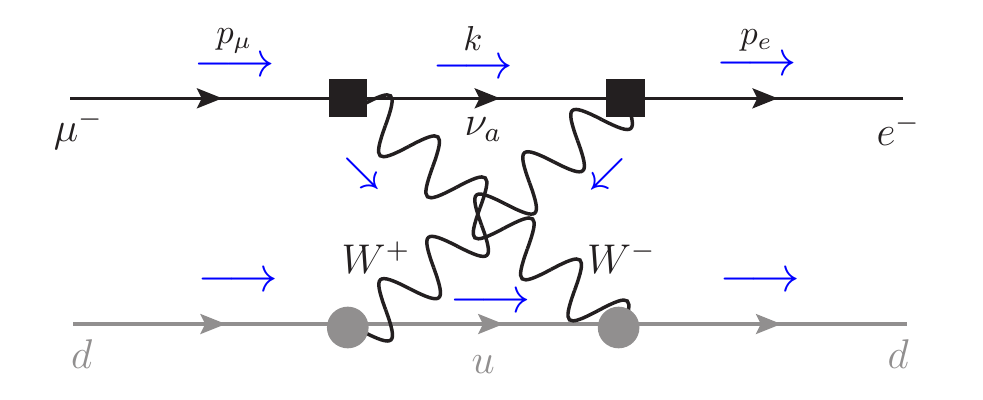}
    \caption{Diagram X}
    \label{fig:Box_II}
     \end{subfigure}  
    \end{minipage}

    \caption{One-loop contributions to $\mu$-$e$ conversion.}
    \label{fig:Diagrams_mu_e_conversion}
  \end{figure}

In our example model, or more generally in any setting featuring a doubly charged scalar coupling to right-handed charged leptons as in Eq.~\eqref{eq:Lagrangian}, $\mu$-$e$ conversion is realised at one-loop level only. The decisive diagrams are those in which the initial state muon turns into a virtual anti-lepton/$S^{--}$ combination, which then turns into an electron. A photon can couple to either of these particles, thus implying four different diagrams (see Fig.~\ref{fig:Diagrams_mu_e_conversion}, Diagrams~I to~IV).\footnote{In Figs.~\ref{fig:Diagram_I} to~\ref{fig:Box_II}, the grayish parts indicate that the quarks are bound within the nucleus. We will solely need the black part of each diagram to determine the form factors, so that we are displaying the hadronic part only for the sake of illustration.} In principle one could also have a loop containing a $W$-boson and a neutrino, with three possibilities to couple a photon to (see Fig.~\ref{fig:Diagrams_mu_e_conversion}, Diagrams~V to~VII). The latter three diagrams are, however, strongly suppressed by the GIM mechanism~\cite{Glashow:1970gm}.

Furthermore, one could in either of these diagrams trade the photon for a $Z$-boson, which yields another seven diagrams. In addition, a $Z$-boson could also couple to the neutrino line (which the photon could not), see Diagram~VIII in Fig.~\ref{fig:Diagrams_mu_e_conversion}. One could also replace all $Z$-boson lines by Higgs bosons, thus producing another eight diagrams. Note that also for Z-bosons and Higgs bosons mediating the process, Diagrams~V to~VIII are GIM-suppressed in contrast to Diagrams~I to~IV. In addition, all these diagrams with heavy exchange particles contribute to the short-range part of the amplitude, cf.\ Sec.~\ref{sec:mue_minus-non-photonic}, which is by far subdominant. Finally, there could also be box-diagrams with two $W$-bosons each, see Diagrams~IX and~X in Fig.~\ref{fig:Diagrams_mu_e_conversion}. These could mediate the process but are GIM-suppressed, too~\cite{Alonso:2012ji}. Thus, starting with the long-range/photonic part, the only relevant diagrams are~I to~IV as displayed in Fig.~\ref{fig:Diagrams_mu_e_conversion}. We will compute these in the following.

Beginning with momentum assignments, we have chosen the photon momentum to be incoming, i.e., we use $q'=p_e-p_\mu=-q$ in order to adapt a notation consistent with our tool of choice, \texttt{Package-X}~\cite{Patel:2015tea}, to reliably compute the loop-integrals. We furthermore use the approximation of a massless electron, which only introduces an error at the sub-$\%$ level. We also use the fact that the electron is on-shell and the muon is approximately on-shell (as it is only bound non-relativistically): $p^2_e = m^2_e\approx 0$, $p^2_\mu\simeq m^2_\mu$, and $q'^2 \simeq -m^2_\mu$.

In order to obtain the decisive matrix elements, we make use of the Feynman rules given in Figs.~\ref{fig:FeynRule1} to \ref{fig:FeynRule4}, see Appendix \ref{app:FeynmanRules}. Let us now go through all contributions in detail. From Diagram~I in Fig.~\ref{fig:Diagram_I}, we obtain the matrix element:
\begin{eqnarray}
 i\mathcal{M}_{\rm I} &=& -4Q_S\,e\,f^*_{ea}\,f_{a\mu}\,A_\nu(q')\,\\
 && \overline{u}_e (p_e)\, \int \frac{\diff^d k}{(2\pi)^d}\,\frac{P_L \slashed{k} (2p_\mu-2k+q')^\nu}{[k^2-m^2_a+i\epsilon][(p_\mu-k+q')^2-M^2_S+i\epsilon][(p_\mu-k)^2-M^2_S+i\epsilon]}\,u_\mu(p_\mu)\,, \nonumber
\end{eqnarray}
where $d=4-2\varepsilon$ is the dimension of the integral, and we have written the matrix element in terms of the charge $Q_S=-2$.\footnote{This seemingly too formal notation serves to display the cancellation of divergences more clearly.} We use \texttt{Package-X}~\cite{Patel:2015tea}, where the most general form of the matrix element given in Eq.~\eqref{eq:MatrixElPhotonic} is put in the form of:
\begin{eqnarray}
 i\mathcal{M}&=&i\,e\,A_\nu (q')\,\overline{u}_e (p_e)\,\Big[\Big(\gamma^\nu-\frac{\slashed{q}'q'^\nu}{q'^2}\Big)F_1(q'^2)+\frac{i\,\sigma^{\nu\rho}\,q'_\rho}{m_\mu}\,F_2(q'^2)+2\frac{q'^\nu}{m_\mu}\,F_3(q'^2)\nonumber\\
 &&+\Big(\gamma^\nu-\frac{\slashed{q}'q'^\nu}{q'^2}\Big)\,\gamma_5\,G_1(q'^2)+\frac{i\,\sigma^{\nu\rho}\,q'_\rho}{m_\mu}\,\gamma_5\,G_2(q'^2)+2\frac{q'^\nu}{m_\mu}\,\gamma_5\,G_3(q'^2)\Big]\,u_\mu(p_\mu)\,,
\end{eqnarray}
to compute the form factors $F_1,\,F_2,\,F_3,\,G_1,\,G_2,\,\text{and}\,G_3$.
The form factors obtained from the \texttt{Package-X} computation are related to the ones from Eq.~\eqref{eq:MatrixElPhotonic} by:
\begin{equation}
 \begin{split}
 f_{E0}(q^2)&= -F_1(q'^2)\,,\\
 f_{M0}(q^2)&= G_1(q'^2)\,,\\
 f_{E1}(q^2)&= G_2(q'^2)\,,\\
 f_{M1}(q^2)&=  F_2(q'^2)\,,\\
 f_3(q^2)&= -F_3(q'^2)\,,\\
 g_3(q^2)&= -G_3(q'^2)\,.
 \label{}
 \end{split}
\end{equation}
Before calculating the factor $\Xi_{\rm particle}^2$ from the form factors, we will first check our computation by taking a closer look at the UV divergences. Since there is no tree level 3-point vertex connecting muon, electron, and photon, and thus no counterterm in the Lagrangian, the combination of Diagrams~I~--~IV in Fig.~\ref{fig:Diagrams_mu_e_conversion} must be finite. We thus need to extract the divergent part from each matrix element, which for Diagram~I is given by:
\begin{equation}
 i\mathcal{M}^{\text{div}}_{\rm I}=\frac{i}{(4\pi)^2}\,\frac{2}{\varepsilon}\,Q_S\,e\,f^*_{ea}\,f_{a\mu}\,A_\nu(q')\,\overline{u}_e (p_e)\,P_L\,\gamma^\nu\,u_\mu(p_\mu)\,.
 \label{eq:div_I}
\end{equation}
The matrix element for the second diagram given in Fig.~\ref{fig:Diagram_II} yields:
\begin{eqnarray}
 i\mathcal{M}_{\rm II}&=&-4Q_{l^+}\,e\,f^*_{ea}\,f_{a\mu}\,A_\nu(q')\,\\
 && \overline{u}_e (p_e)\,\int \frac{\diff^d k}{(2\pi)^d}\,\frac{P_L (\slashed{k}+\slashed{q}'+m_a)\,\gamma^\nu\,(\slashed{k}+m_a)P_R}{[k^2-m^2_a+i\epsilon][(p_\mu-k)^2-M^2_S+i\epsilon][(k+q')^2-m^2_a+i\epsilon]}\,u_\mu(p_\mu)\,,\nonumber
\end{eqnarray}
where $Q_{l^+}=1$, and it adds
\begin{equation}
 i\mathcal{M}^{\text{div}}_{\rm II}=-\frac{i}{(4\pi)^2}\,\frac{1}{\varepsilon}\,Q_{l^+}\,e\,f^*_{ea}\,f_{a\mu}\,A_\nu(q')\,\overline{u}_e (p_e)\,P_L\,\gamma^\rho\,\gamma^\nu\,\gamma_\rho\,P_R\,u_\mu(p_\mu)\,
 \label{eq:div_II}
\end{equation}
to the divergent part.

From Fig.~\ref{fig:Diagram_III}, we extract:
\begin{eqnarray}
 i\mathcal{M}_{\rm III}&=&-4Q_{e^-}\,e\,f^*_{ea}\,f_{a\mu}\,A_\nu(q')\,\\
 && \overline{u}_e (p_e)\,\int \frac{\diff^d k}{(2\pi)^d}\,\frac{\gamma^\nu\,\slashed{p}_\mu\,P_L \,\slashed{k}}{[p_\mu^2+i\epsilon][(p_\mu-k)^2-M^2_S+i\epsilon][k^2-m^2_a+i\epsilon]}\,u_\mu(p_\mu)\,,\nonumber
\end{eqnarray}
with $Q_{e^-}=-1$, and obtain:
\begin{equation}
 i\mathcal{M}^{\text{div}}_{\rm III}=-\frac{i}{(4\pi)^2}\,\frac{2}{\varepsilon}\,\frac{Q_{e^-}}{m^2_\mu}\,e\,f^*_{ea}\,f_{a\mu}\,A_\nu(q')\,\overline{u}_e (p_e)\,\gamma^\nu\,\slashed{p}_\mu\,P_L\,\slashed{p}_\mu\,u_\mu(p_\mu)\,.
 \label{eq:div_III}
\end{equation}

Finally, the matrix element of Fig.~\ref{fig:Diagram_IV} leads to:
\begin{eqnarray}
 i\mathcal{M}_{\rm IV}&=&-4Q_{\mu^-}\,e\,f^*_{ea}\,f_{a\mu}\,A_\nu(q')\,\\
 && \overline{u}_e (p_e)\,\int \frac{\diff^d k}{(2\pi)^d}\,\frac{P_L \,\slashed{k}\,(\slashed{p}_\mu+\slashed{q}'+m_\mu)\,\gamma^\nu}{[(p_\mu+q')^2-m^2_\mu+i\epsilon][(p_\mu-k+q')^2-M^2_S+i\epsilon][k^2-m^2_a+i\epsilon]}\,u_\mu(p_\mu)\,,\nonumber
\end{eqnarray}
with $Q_{\mu^-}=-1$ and a divergent contribution of
\begin{equation}
 i\mathcal{M}^{\text{div}}_{\rm IV}=\frac{i}{(4\pi)^2}\,\frac{2}{\varepsilon}\,\frac{Q_{\mu^-}}{m^2_\mu}\,e\,f^*_{ea}\,f_{a\mu}\,A_\nu(q')\,\overline{u}_e (p_e)\,P_L \,(\slashed{p}_\mu+\slashed{q}')\,(\slashed{p}_\mu+\slashed{q}'+m_\mu)\,\gamma^\nu\,u_\mu(p_\mu)\,.
 \label{eq:div_IV}
\end{equation}
In $d=4$ dimensions, the Lorentz structures simplify due to the relations $\gamma^\rho\,\gamma^\nu\,\gamma_\rho=-2\gamma^\nu$ and $\slashed{p}\,\slashed{p}=p^2$ and upon employing the approximate on-shell conditions. As a consequence, the divergent part of the $\mu$-$e$ conversion amplitude takes the form:
\begin{equation}
  i\mathcal{M}^{\text{div}}=\frac{i}{(4\pi)^2}\,\frac{1}{\varepsilon}\,e\,f^*_{ea}\,f_{a\mu}\,A_\nu(q')\,\overline{u}_e (p_e)\,[(2Q_S+2Q_{l^+}-Q_{e^-}-Q_{\mu^-})P_L\,\gamma^\nu]\,u_\mu(p_\mu)\,,
\end{equation}
which indeed vanishes as soon as we enter the charges explicitly, as to be expected. 

Checking with \texttt{Package-X} confirms that all form factors are finite. It also shows that, under the assumption of both muon and electron being approximately on-shell in combination with kinematic relations following a vanishingly small momentum of the nucleus, both $F_3$ and $G_3$ vanish exactly, which confirms the general structure in Eq.~(\ref{eq:MatrixElPhotonic}) for the photonic case and agrees with the considerations of the previous section, where the same arguments led to $q^0=-q'^0\to 0$ and thereby to the disappearance of these structures from the branching ratio.

We have also extracted the finite parts of the form factors, which are the actual physics contributions. They take the following forms:
\begin{eqnarray}
 &&F_1(-m_\mu^2)=\,G_1(-m_\mu^2)=\nonumber\\
 &&=-\frac{1}{128\,\pi^2 m_\mu^4}\,\sum_{a=e,\,\mu,\,\tau} f^*_{ea}\,f_{a\mu}\,\Big[2\,m_\mu^2\big(-5m_a^2+6m_\mu^2+5M_S^2\big)-2\, S_a\,m_\mu^2\big(m_a^2+3m_\mu^2-M_S^2\big)\nonumber\\
 && \ln \Big[\frac{2m_a^2}{2m_a^2+m_\mu^2(1+S_a)}\Big]+4\,S_S\,m_\mu^2\big(m_a^2+m_\mu^2-M_S^2\big)\,\ln \Big[\frac{2M_S^2}{2M_S^2+m_\mu^2(1+S_S)}\Big]+\Big(3m_a^2\big(2m_a^2-m_\mu^2\nonumber\\
 && -4M_S^2\big)+5m_\mu^4-7m_\mu^2\,M_S^2+6M_S^4\Big) \ln \Big[\frac{m_a^2}{M_S^2}\Big]+2\,T_a\big(-6m_a^2+m_\mu^2+6M_S^2\big)\ln \Big[\frac{2m_a\, M_S}{m_a^2-m_\mu^2+M_S^2-T_a}\Big]\nonumber\\
 &&  +2\,m_\mu^2\Big[\Big(m_a^4+8m_a^2\, m_\mu^2+M_S^4-2M_S^2\big(m_a^2+2m_\mu^2\big)\Big)\,C_0\big[0,\,-m_\mu^2,\,m_\mu^2;\,m_a,\,M_S,\,m_a\big]\nonumber\\
 &&   +2\Big(m_a^4-2M_S^2\big(m_a^2-2m_\mu^2\big)+M_S^4\Big)\, C_0\big[0,\,-m_\mu^2,\,m_\mu^2;\,M_S,\,m_a,\,M_S\big]\Big]\Big] \,,
 \label{eq:FFexpl_1}
\end{eqnarray}
as well as:
\begin{eqnarray}
 && F_2(-m_\mu^2)=-G_2(-m_\mu^2)=\nonumber\\
 && =-\frac{1}{128\,\pi^2 m_\mu^4}\,\sum_{a=e,\,\mu,\,\tau} f^*_{ea}\,f_{a\mu}\,\Big[2\,m_\mu^2\big(-m_a^2+6m_\mu^2+M_S^2\big)+2\, S_a\,m_\mu^2\big(3m_a^2+m_\mu^2-3M_S^2\big)\nonumber\\
 && \ln \Big[\frac{2m_a^2}{2m_a^2+m_\mu^2(1+S_a)}\Big]+4\,S_S\,m_\mu^2\big(-3m_a^2+m_\mu^2+3M_S^2\big)\,\ln \Big[\frac{2M_S^2}{2M_S^2+m_\mu^2(1+S_S)}\Big]\nonumber\\
 && +\Big(m_a^2\big(-2m_a^2-7m_\mu^2+4M_S^2\big)+m_\mu^4+5m_\mu^2\,M_S^2-2M_S^4\Big) \ln \Big[\frac{m_a^2}{M_S^2}\Big]\nonumber\\
 && +2\,T_a\big(2m_a^2-3m_\mu^2-2M_S^2\big)\,\ln \Big[\frac{2m_a\, M_S}{m_a^2-m_\mu^2+M_S^2-T_a}\Big] \nonumber\\
 && +2\,m_\mu^2 \Big[\Big(-3m_a^4-3M_S^4+2M_S^2\big(3m_a^2+2m_\mu^2\big)\Big)\,C_0\big[0,\,-m_\mu^2,\,m_\mu^2;\,m_a,\,M_S,\,m_a\big]\nonumber\\
 && +2\Big(-3m_a^4 +2m_a^2\big(3M_S^2+2m_\mu^2\big)-3M_S^4\Big)\,C_0\big[0,\,-m_\mu^2,\,m_\mu^2;\,M_S,\,m_a,\,M_S\big]\Big]\Big] \,.
 \label{eq:FFexpl_2}
\end{eqnarray}
Here, we have used the following abbreviations:
\begin{eqnarray}
 && S_i=\sqrt{1+4m_i^2/m_\mu^2}\,, \ \ \ \ S_S=\sqrt{1+4 M_S^2/m_\mu^2}\,,\qquad \text{and}\qquad\\
 && T_a=\sqrt{(m_a-m_\mu-M_S)(m_a+m_\mu-M_S)(m_a-m_\mu+M_S)(m_a+m_\mu+M_S)}.\nonumber
\end{eqnarray}
Moreover, the scalar three-point function in four dimensions is given by~\cite{Patel:2015tea}:
\begin{eqnarray}
 && C_0\big[p^2_1,\,p^2_2,\,Q^2;m_2,\,m_1,\,m_0\big] =-\int_0^1 \diff x \int_0^{1-x} \diff y \, \Big[p_1^2\, x^2+p_2^2\, y^2+\big(p_1^2+p_2^2-Q^2\big)xy\nonumber\\
 && +\big(m_1^2-m_0^2-p_1^2\big)x +\big(m_2^2-m_0^2-p_2^2\big)y+m_0^2-i\epsilon\Big]^{-1}\,, \label{eq:C_0}
\end{eqnarray}
which corresponds to the assignment given in Fig.~\ref{fig:C0_Notation} in Appendix \ref{app:Passarino} and which makes use of $Q\equiv p_1-p_2$. 

The scalar three-point function in Eq.~\eqref{eq:C_0} agrees with the original one from Passarino and Veltman~\cite{Passarino:1978jh,'tHooft:1978xw,Bardin:1999ak} upon rearranging the mass terms and considering the change of metric,\footnote{In order to compare the scalar three-point function from Passarino and Veltman with the one given in Eq.~\eqref{eq:C_0}, one needs to switch the Minkowski metric from $(-1,1,1,1)$ to $(1,-1,-1,-1)$. One also needs to shift the outer Feynman parameter $x= 1-x'$, such that $\int_0^1 \diff x \int_0^x \diff y \to \int_0^1 \diff x' \int_0^{1-x'} \diff y$.} such that
\begin{equation*}
 C_0[p^2_1,\,p^2_2,\,Q^2;m_2,\,m_1,\,m_0]=-C_0^{\text{Passarino-Veltman}}[-p^2_1,\,-p^2_2,\,-Q^2;m_1,\,m_0,\,m_2]\,.
\end{equation*}
Inserting the form factors listed in Eqs.~\eqref{eq:FFexpl_1} and~\eqref{eq:FFexpl_2} into Eq.~\eqref{eq:Xi_particle}, we eventually obtain:
\begin{eqnarray}
 && \Xi_{\rm particle}^2 =\,\Big|f_{E0}(-m^2_\mu)+f_{M1}(-m^2_\mu)\Big|^2+\Big|f_{M0}(-m^2_\mu)+f_{E1}(-m^2_\mu)\Big|^2\label{eq:Xi_full}\\
 && =\,\Big|-F_{1}(-m^2_\mu)+F_{2}(-m^2_\mu)\Big|^2+\Big|G_{1}(-m^2_\mu)+G_{2}(-m^2_\mu)\Big|^2 = 2\,\Big|F_{1}(-m^2_\mu)-F_{2}(-m^2_\mu)\Big|^2 \nonumber\\
 && =\,\frac{1}{512\,\pi^4\,m^8_\mu}\,\Bigg|\sum_{a=e,\,\mu,\,\tau}f^*_{ea}\,f_{a\mu}\,\Bigg(2\,m^2_\mu\big(m_a^2-M^2_S\big) +4\,S_S\,m^2_\mu \big(M^2_S-m^2_a\big)\ln \Big[\frac{2M_S^2}{2M_S^2+m_\mu^2(1+S_S)}\Big]\nonumber\\
 && +2\,S_a\,m^2_\mu\big(m^2_a+m^2_\mu-M^2_S\big)\,\ln \Big[\frac{2m_a^2}{2m_a^2+m_\mu^2(1+S_a)}\Big] \nonumber\\
 && -\big(2m^4_a+m^4_\mu-3m^2_\mu\,M^2_S+2M^4_S+m^2_a\,m^2_\mu-4m^2_a\,M^2_S\big) \ln \Big[\frac{m_a^2}{M_S^2}\Big]\nonumber\\
 && +2\,T_a\big(2m_a^2-m_\mu^2-2M_S^2\big)\,\ln \Big[\frac{2m_a\, M_S}{m_a^2-m_\mu^2+M_S^2-T_a}\Big]+2\,m_\mu^2\big(m_a^2-M^2_S\big)\Big[\big(-m^2_a-2m^2_\mu+M^2_S\big)\nonumber\\
 && C_0\big[0,\,-m_\mu^2,\,m_\mu^2;\,m_a,\,M_S,\,m_a\big]+2\big(-m^2_a+m^2_\mu+M^2_S\big)\,C_0\big[0,\,-m_\mu^2,\,m_\mu^2;\,M_S,\,m_a,\,M_S\big]\Big]\Bigg)\Bigg|^2\,.\nonumber
\end{eqnarray}
We can greatly simplify this expression by exploiting the mass hierarchy $M_S \gg m_{e,\mu,\tau}$. Hence, each term in Eq.~\eqref{eq:Xi_full} is expanded around $M_S \to \infty$ up to $\mathcal{O}(1/M^2_S)$, which has to be done in a careful manner.\footnote{While the expansion of the first few terms does not make a problem, the Passarino-Veltman functions require a cautious treatment. To this end, we rewrite the Passarino-Veltman functions in terms of dilogarithms. Instead of the {\tt Mathematica} function {\tt PolyLog[2,x]}, \texttt{Package-X}~\cite{Patel:2015tea} uses its own function {\tt DiLog[x,A]}. The latter has a branch cut discontinuity in the complex $x$ plane running from $1$ to $\infty$. For real $x \leq 1$ or complex $x$ the {\tt DiLog[x,A]} is equivalent to {\tt PolyLog[2,x]}. However, for real $x \textgreater 1$, the side of the branch cut which {\tt DiLog[x,A]} evaluates is given by the prescription $\lim_{\epsilon \to 0} \Li_2[x+i A \epsilon]$. Thus, the sign of $A$ fixes where {\tt DiLog} evaluates. To expand the {\tt DiLog} functions in the limit $M_S \to \infty$, we need to insert numerical values for $A$. Since the $A$'s all consist of combinations of $m_a,\,m_\mu$, and $M_S$, we fix the scalar mass within $A$ to an arbitrary value (considering $M_S \gg m_a$), and expand the remaining function.} That way, we observe delicate cancellations at the orders $M^4_S$, $M^2_S$, and $M^0_S$, such that the remaining expression takes the form:
\begin{eqnarray}
 && \Xi_{\rm particle}^2=\,\frac{1}{288\,\pi^4\,m^2_\mu\,M^4_S}\, \Bigg|\sum_{a=e,\,\mu,\,\tau}f^*_{ea}\,f_{a\mu}\,\Big(4m^2_a\,m_\mu-m^3_\mu \label{eq:Xi_simpl}\\
 && +2\big(-2m^2_a+m^2_\mu\big)\sqrt{4m^2_a+m^2_\mu}\,\arctanh \Big[\frac{m_\mu}{\sqrt{4m^2_a+m^2_\mu}}\Big]+m^3_\mu\,\ln\Big[\frac{m^2_a}{M^2_S}\Big]\Big)\Bigg|^2\,,\nonumber
\end{eqnarray}
at leading order. Including the next-to-leading contribution would change our result by roughly $4\%$/at the per mille level for the $\tau$ contribution/the $\mu$ or $e$ contributions being dominant, as we have checked numerically. Note that the cancellations mentioned may not materialise numerically when employing the full expression in Eq.~\eqref{eq:Xi_full} in case large numbers are not treated with sufficient accuracy in a numerical computation.

\begin{figure}[th]
\hspace{-1.7cm}
\begin{tabular}{lcr}
\includegraphics[width=6.1cm]{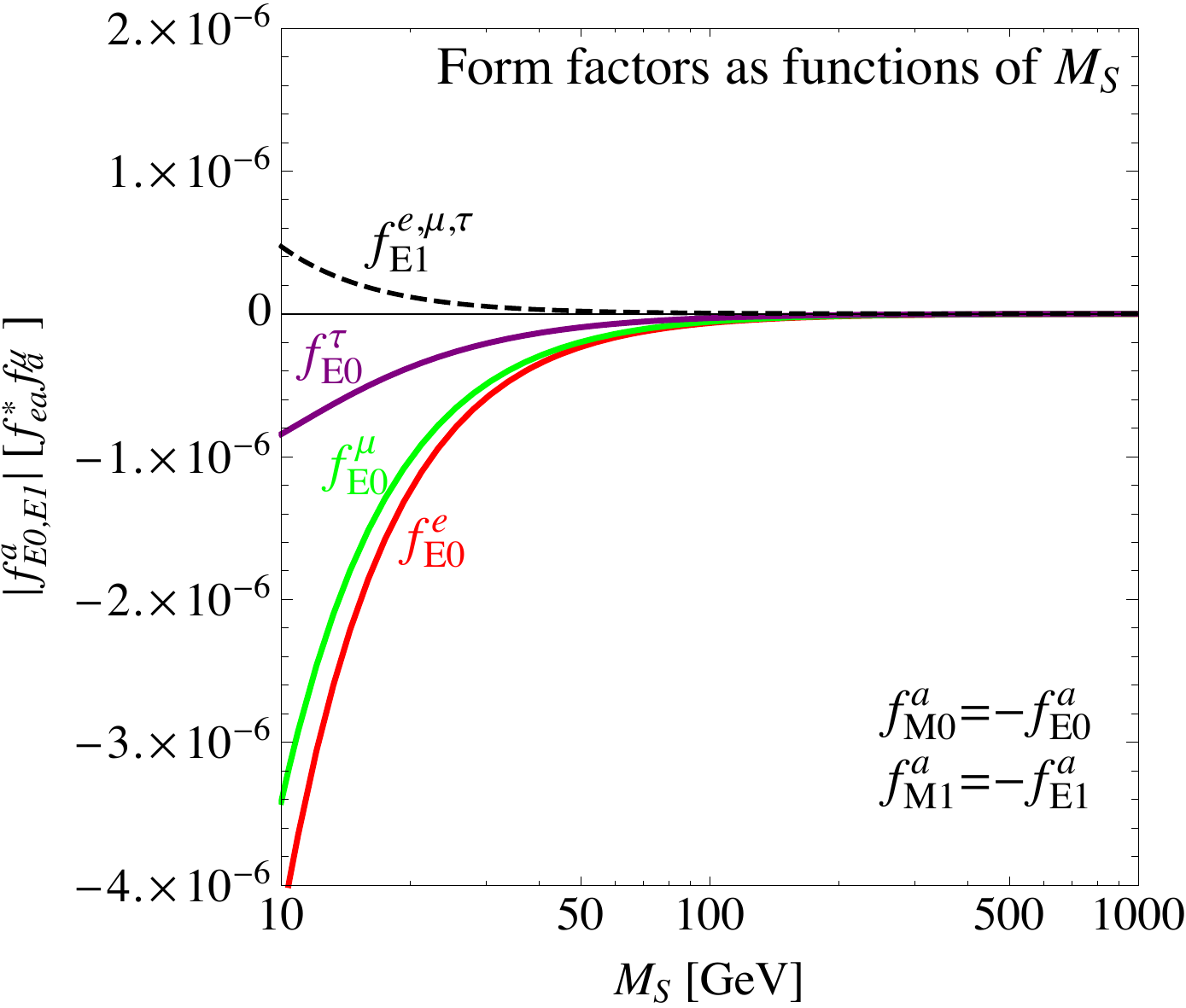} & \includegraphics[width=5.9cm]{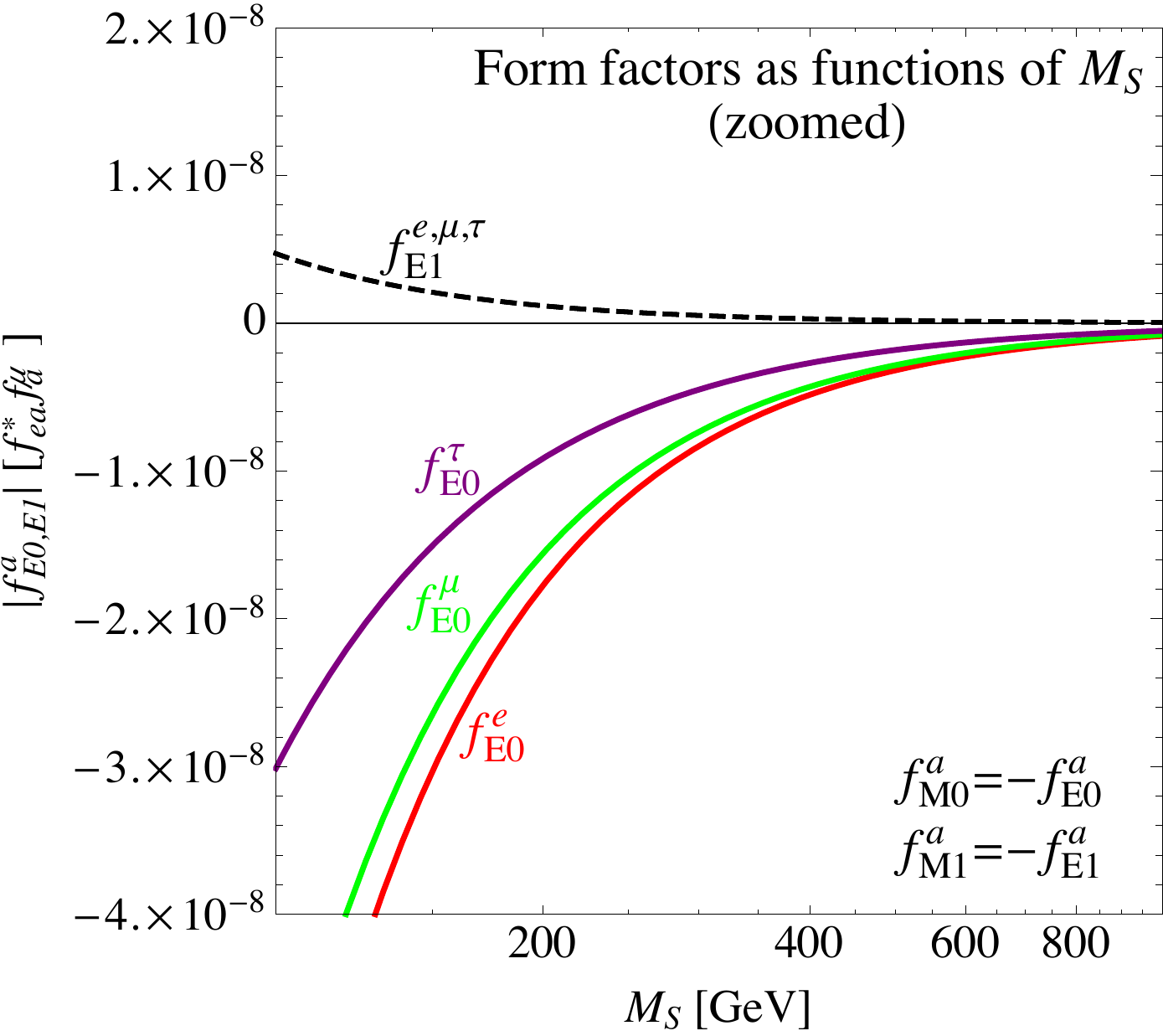} & \includegraphics[width=5.3cm]{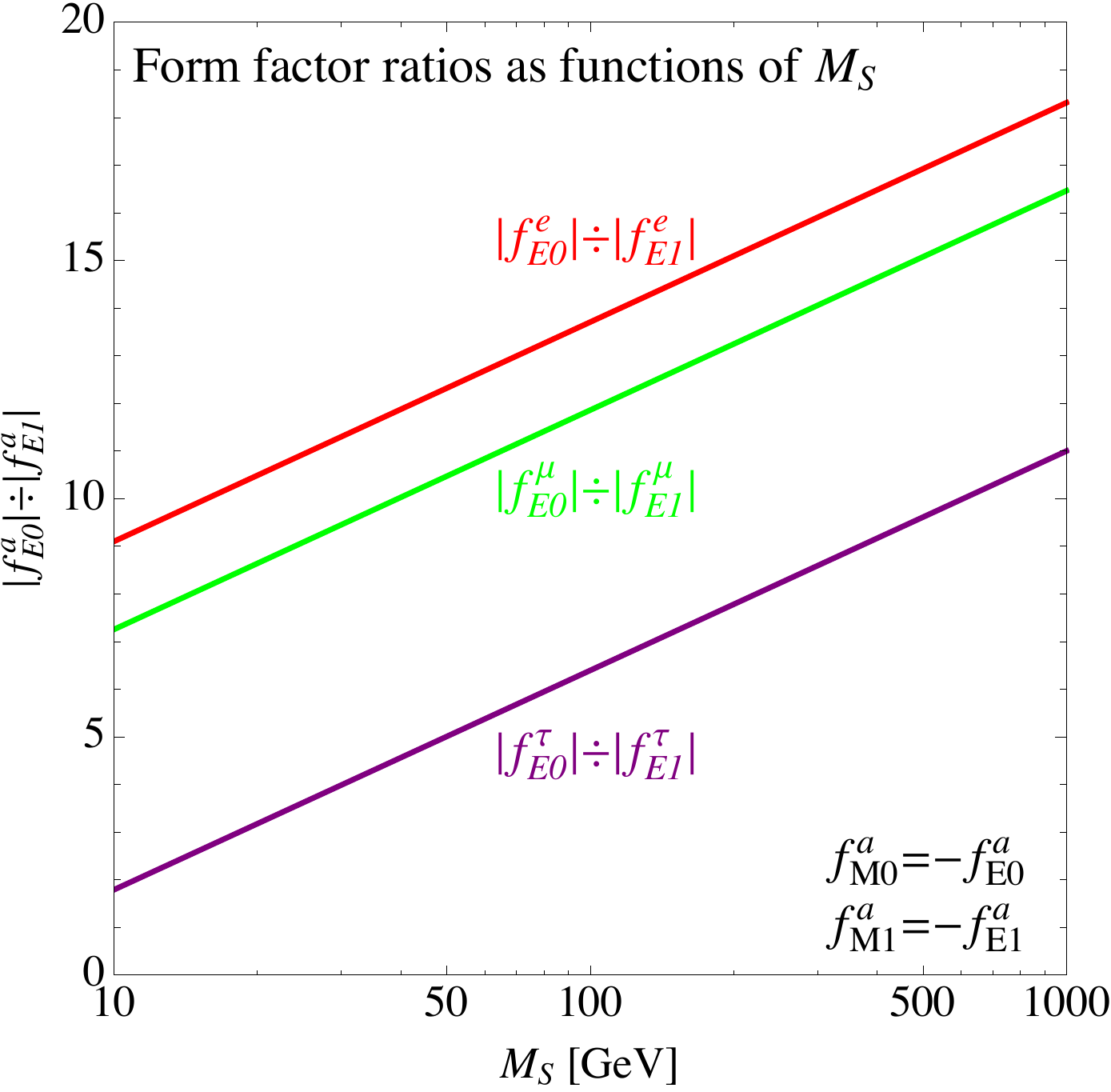}
\end{tabular}
\caption{\label{fig:form-factors}Form factors and ratios of form factors as functions of $M_S$.}
\end{figure}

Let us take a moment to compare our results to the previous ones obtained in Ref.~\cite{Raidal:1997hq}, based on an estimate using EFT. We should in fact recover the results obtained there in the limit of a sufficiently heavy scalar. To perform this consistency check, it is first of all useful to look at the form factors themselves, which are displayed in the left and middle panels of Fig.~\ref{fig:form-factors} (in a zoomed version in the latter case), in units of $f_{ea}^* f_{a\mu}$. As can be seen, the magnitudes of the form factors $f_{E0}^a$ ($=-f_{M0}^a$) are in all cases $a = e, \mu, \tau$ bigger for smaller scalar masses, however, they later on decrease from $\mathcal{O}(10^{-8})$ -- $\mathcal{O}(10^{-7})$ for $M_S \sim 100$~GeV to $\mathcal{O}(10^{-10})$ -- $\mathcal{O}(10^{-9})$ for $M_S \sim 1000$~GeV. The form factors $f_{E1}^a=-f_{M1}^a$, in turn, do not depend on the charged lepton masses and they decrease from about $\mathcal{O}(10^{-9})$ for $M_S \sim 100$~GeV to $\mathcal{O}(10^{-11})$ for $M_S \sim 1000$~GeV. That already implies that the approximation for the numerical values of the form factors used in Ref.~\cite{Raidal:1997hq} for the case of doubly charged scalars is only accurate to about $10\%$. This can also be seen from the right panel of Fig.~\ref{fig:form-factors}, displaying the ratio between the form factors $f_{E0}^a$ and $f_{E1}^a$, and it implies a percent accuracy of the photonic decay rate when computed with $f_{E1}^a$ and $f_{M1}^a$ being neglected. Note that, however, as we will see in Sec.~\ref{sec:mue_minus-non-photonic}, short-range contributions lead to a modification of the same size.

For completeness, let us display the explicit versions of the purely photonic form factors in the limit of a very large $M_S$:
\begin{eqnarray}
 && \frac{f_{E0}^a}{f_{ea}^* f_{a\mu}} = - \frac{f_{M0}^a}{f_{ea}^* f_{a\mu}} = \frac{2 m_a^2 + m_\mu^2 \log\left(\frac{m_a}{M_S}\right)}{12 \pi^2 M_S^2} + \frac{\sqrt{m_\mu^2 + 4 m_a^2}(m_\mu^2 - 2 m_a^2)}{12 \pi^2 m_\mu M_S^2} \arctanh \left( \frac{m_\mu}{\sqrt{m_\mu^2 + 4 m_a^2}} \right),\nonumber\\
 && \frac{f_{E1}^a}{f_{ea}^* f_{a\mu}} = - \frac{f_{M1}^a}{f_{ea}^* f_{a\mu}} = \frac{m_\mu^2}{24 \pi^2 M_S^2} ,
 \label{eq:form-factors}
\end{eqnarray}
evaluated at $q^2 = - m_\mu^2$. While our formulae for the form factors are basically identical to those obtained in Ref.~\cite{Raidal:1997hq}, note that this reference seems to contain a relative sign difference between $f_{E0}$ and $f_{M0}$ compared to our results, which can alter the resulting numerical predictions. Given that we have automatised our computation to a high degree and that we have explicitly performed several decisive cross-checks, such as showing that the divergent parts of the loop amplitudes contained in Eqs.~\eqref{eq:div_I}, \eqref{eq:div_II}, \eqref{eq:div_III}, and~\eqref{eq:div_IV} do indeed cancel, we are confident that all our relative signs should be correct.\\

The expression displayed in Eq.~\eqref{eq:Xi_simpl} is our final result for the photonic contribution of the doubly charged scalar to $\mu$-$e$ conversion. In combination with Eq.~\eqref{eq:mu-e_BR_long}, it can be used to compute the corresponding branching ratio for any choice of Yukawa couplings $f_{ab}$ and scalar mass $M_S$, as long as the nuclear physics quantities entering the equations are known. However, these quantities suffer from uncertainties which we currently cannot resolve. Thus, when aiming at a bound on the squared particle physics amplitude displayed in Eq.~\eqref{eq:Xi_simpl}, it is easiest to absorb all uncertainties into the experimental bounds, meaning that an experimental upper bound on the branching ratio translates into a range of upper bounds on $\Xi_{\rm particle}^2$. This one can do as long as the nuclear physics and particle physics parts factorise, as is the case in Eq.~\eqref{eq:mu-e_BR_long}.

\subsection{\label{sec:mue_minus-nuclear}Nuclear physics, experimental aspects, and resulting bounds}

The main nuclear physics quantities entering the branching ratio in Eq.~\eqref{eq:mu-e_BR_long} are $Z$, $Z_{\rm eff}$, and $F_p$. Out of those, the atomic number $Z$ can be trivially looked up, however, the computation of the effective atomic charge $Z_{\rm eff}$ and of the nuclear matrix element $F_p$ require knowledge of the proton charge density $\rho_p (r)$, with $r$ being the distance to the centre of the nucleus. A good reference summarising the nuclear physics aspects is Ref.~\cite{Kitano:2002mt}: based on the classic Refs.~\cite{DeJager:1987qc,Fricke:1995zz}, they assign different simplified nuclear models (such as harmonic oscillator models as well as different Fermi- and Gaussian-type models) to the different nuclei. In order to use values which are as updated as possible, we have however instead relied on the online database called \emph{The Nuclear Charge Density Archive}~\cite{NuclearData}, whose data are to the greatest extent identical to those used in the previous references, but they nevertheless contain some updates or smaller corrections. We would like to stress that, from a nuclear physics point of view, the process of $\mu$-$e$ conversion would certainly deserve more attention. Although some example computations of NMEs exist~\cite{Gonzalez:2013rea,Faessler:2005hx,Faessler:2004ea,Faessler:2004jt,Kosmas:1997eca}, they still seem not as advanced and/or up to date as the comparatively involved computations of NMEs for neutrinoless double beta decay (see, e.g., Refs.~\cite{Faessler:2012ku,Barea:2015kwa,Hyvarinen:2015bda,Engel:2015wha,Simkovic:2013qiy,Barea:2013bz,Suhonen:2012ii,Suhonen:2012zzc}), and in particular they do not cover all relevant cases. On the other hand, the process of $\mu$-$e$ conversion was recognised by parts of the nuclear physics community also in recent years~\cite{Gonzalez:2013rea}, so that hopefully, at some point, it will be clear how safe the bounds obtained truly are.

\begin{figure}[t]
\centering
 \includegraphics[width=8cm]{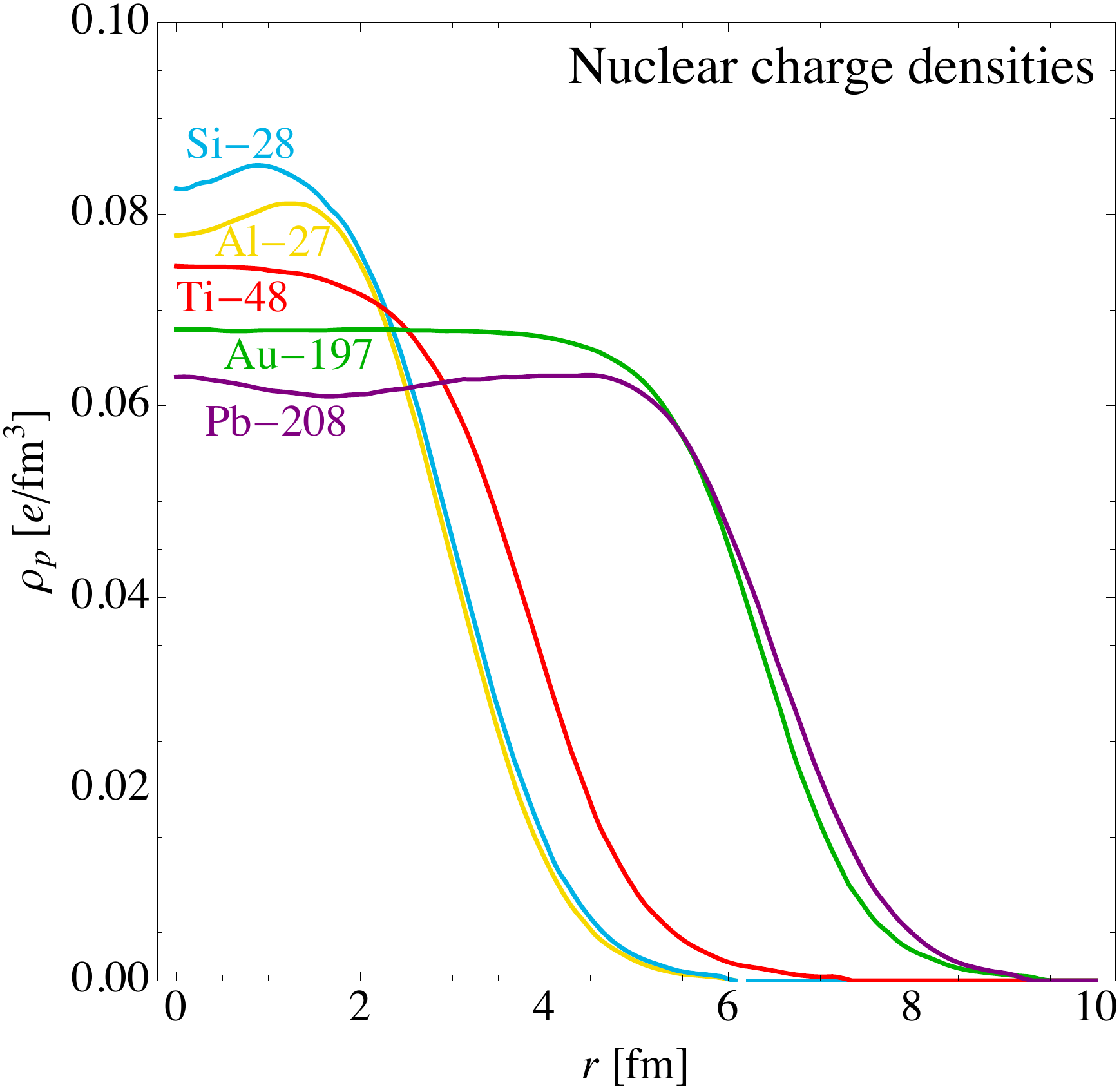}
 \caption{\label{fig:ChargeDensities}Electric charge densities of the isotopes under consideration. The normalisations are chosen such that $\int \diff^3 r\ \rho_p(r) \simeq Z$ for each isotope.}
\end{figure}

The relevant nuclear charge densities are displayed in Fig.~\ref{fig:ChargeDensities} for the isotopes under consideration. The corresponding effective atomic charges and NMEs are displayed in Tab.~\ref{tab:ZZeffFp}. Note that, as long as the particle physics and nuclear physics parts factorise, cf.\ Eq.~\eqref{eq:mu-e_BR_long}, all nuclear physics dependence can be absorbed into the experimental bounds. Hence, we can conveniently compare bounds from different experiments which constrain the same particle physics amplitude.

\begin{table}[t]
\centering
\begin{tabular}{|c||c|c|c|c|}\hline
Isotope & $Z$& $Z_{\rm eff}$ & $F_p$ & $\Gamma_{\rm capt}[10^{6}/{\rm s}]$\\\hline\hline
Al-27 & $13$ & $22.79$ & $0.633$ & $0.7054$\\\hline
Si-28 & $14$ & $24.37$ & $0.621$ & $0.8712$\\\hline
Ti-48 & $22$ & $35.85$ & $0.504$ & $2.59$\\\hline
Au-197 & $79$ & $75.86$ & $0.180$ & $13.07$\\\hline
Pb-208 & $82$ & $75.44$ & $0.151$ & $13.45$\\\hline
\end{tabular}
\caption{\label{tab:ZZeffFp}Atomic numbers $Z$, effective atomic charges $Z_{\rm eff}$ according to Eq.~(127) of Ref.~\cite{Kuno:1999jp}, and NMEs $F_p$ according to Eq.~(129) of Ref.~\cite{Kuno:1999jp} for the isotopes under consideration. We also quote the rates for ordinary muon capture, cf.\ Tab.~8 in Ref.~\cite{Kitano:2002mt} (note the typo ``Pb-207'' in that reference).}
\end{table}

\begin{figure}[ht]
\hspace{-1cm}
\begin{tabular}{lr}
\includegraphics[width=0.545\textwidth]{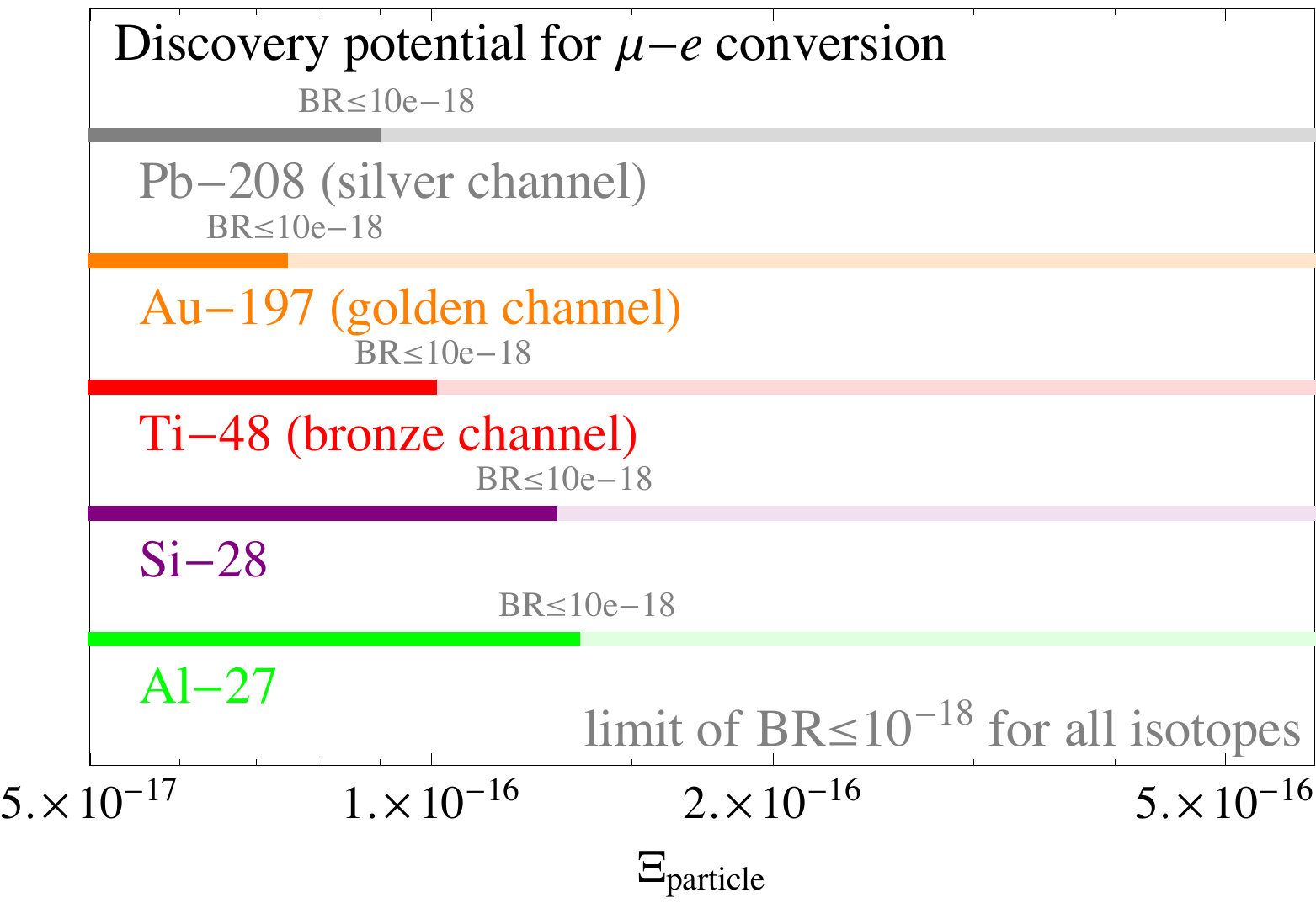} & \includegraphics[width=0.55\textwidth]{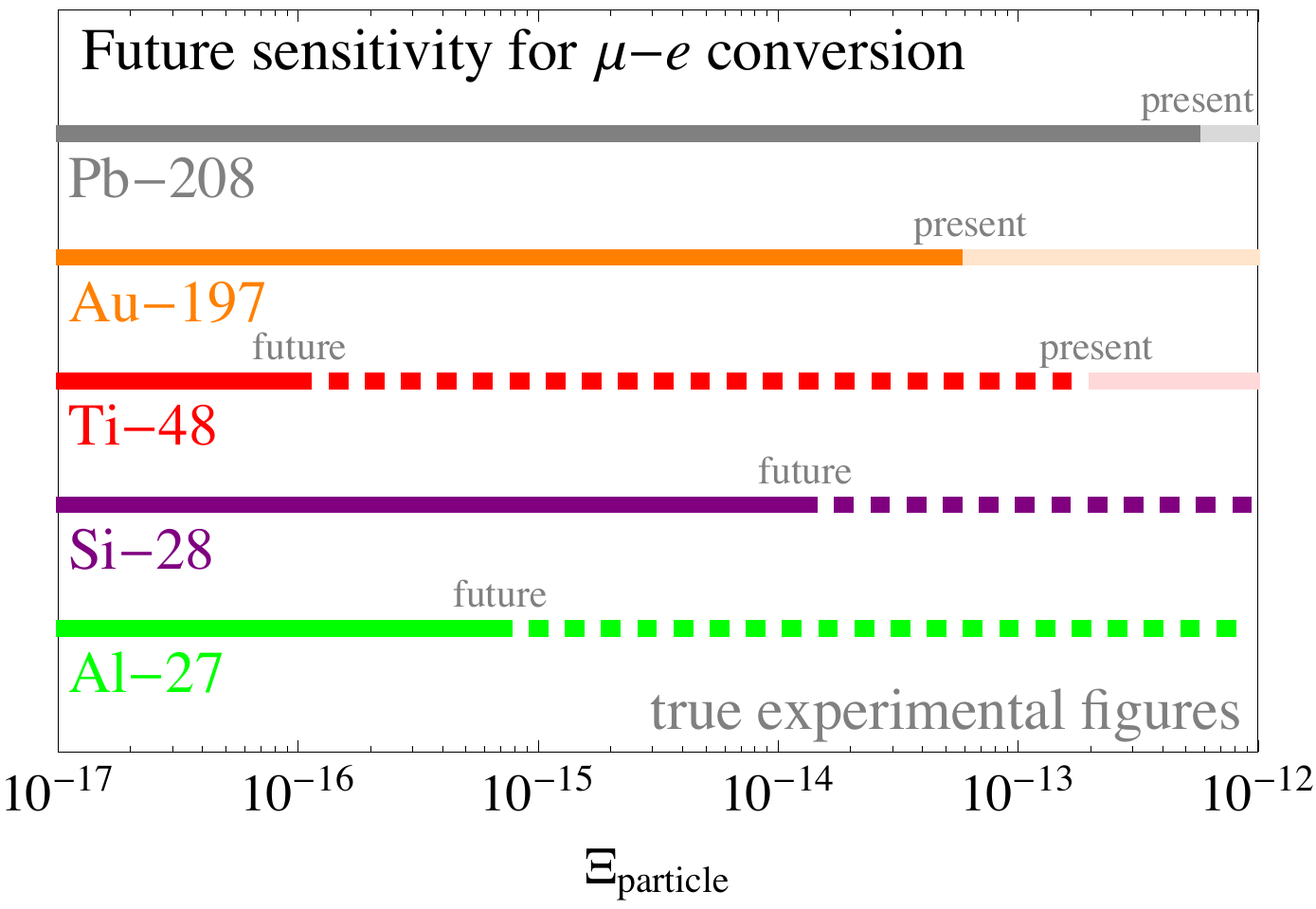}
\end{tabular}
\caption{\label{fig:me-con_potential}Discovery potential and future sensitivities/current limits on $\Xi_{\rm particle}$ for different isotopes under consideration for $\mu$-$e$ conversion.}
\end{figure}

The relevant nuclei we have taken into consideration are those for which either existing limits can be found or which are planned to be used in future experiments. The best existing limits were all obtained by the SINDRUM~II experiment: ${\rm BR}(\mu^- {\rm Ti} \to e^- {\rm Ti}) < 4.3\cdot 10^{-12}$@90\%~C.L. on ${}^{48}{\rm Ti}$~\cite{Dohmen:1993mp}, ${\rm BR}(\mu^- {\rm Au} \to e^- {\rm Au}) < 7\cdot 10^{-13}$@90\%~C.L. on ${}^{197}{\rm Au}$~\cite{Bertl:2006up}, and ${\rm BR}(\mu^- {\rm Pb} \to e^- {\rm Pb}) < 4.6\cdot 10^{-11}$@90\%~C.L. on ${}^{208}{\rm Pb}$~\cite{Honecker:1996zf}. Projections for future sensitivities are announced by DeeMe~\cite{Aoki:2010zz} for ${}^{28}{\rm Si}$, ${\rm BR}(\mu^- {\rm Si} \to e^- {\rm Si}) < 1\cdot 10^{-14}$, by COMET~\cite{COMET} for ${}^{27}{\rm Al}$, ${\rm BR}(\mu^- {\rm Al} \to e^- {\rm Al}) < 2.6\cdot 10^{-17}$,\footnote{Note that a slightly worse sensitivity of ${\rm BR}(\mu^- {\rm Al} \to e^- {\rm Al}) < 6\cdot 10^{-17}$ is announced by Mu2e~\cite{Kutschke:2011ux}.} and by PRISM/PRIME~\cite{Barlow:2011zza} for ${}^{48}{\rm Ti}$, ${\rm BR}(\mu^- {\rm Ti} \to e^- {\rm Ti}) < 1\cdot 10^{-18}$. However, due to the nuclear physics increasing or decreasing the rate for certain nuclei, it is not a priori clear whether the nuclei used in actual experiments have the greatest discovery potential. In order to disentangle these tendencies, we have depicted in Fig.~\ref{fig:me-con_potential} both the general discovery potential (i.e., the possible limit on the parameter $\Xi_{\rm particle}$) for a given limit on the branching ratio versus the actual future sensitivities and past limits. The left panel exhibits how far down a limit on $\Xi_{\rm particle}$ could go for a hypothetical bound of $1\cdot 10^{-18}$ on the branching ratio assumed for \emph{all} isotopes (which is identical to the quoted future sensitivity by PRISM/PRIME for ${}^{27}{\rm Ti}$). As one can see, the best isotope for $\mu$-$e$ conversion and thus the (quite literally) golden channel would be the transition on ${}^{197}{\rm Au}$, followed by ${}^{208}{\rm Pb}$ and ${}^{48}{\rm Ti}$. Glancing at the right panel, the true best future sensitivity is in fact expected to be reached for ${}^{48}{\rm Ti}$ by PRISM/PRIME. These simple considerations imply that, if it was possible to build a future experiment with ${\rm BR}(\mu^- {\rm Au} \to e^- {\rm Au}) < 1\cdot 10^{-18}$ instead of ${\rm BR}(\mu^- {\rm Ti} \to e^- {\rm Ti})$, we might even be able to boost our limit on $\Xi_{\rm particle}$ even further than currently planned.

To get a first impression of the limits one can obtain from this process, we ignore relative phases for the time being, i.e., we take $f^*_{ab}=f_{ab}$. To get a feeling for how strong the constraints could get, we choose the following scenarios: as limiting cases we take a rather optimistic scenario with comparatively large couplings, $f_{ab}=10^{-2}$ ($\forall a,b=e,\,\mu,\,\tau$), and a rather pessimistic scenario with small couplings, $f_{ab}=10^{-4}$. As we will see, these scenarios indeed comprise ``envelopes'' of the more concrete scenarios, although of course they comprise no strict boundaries. E.g., even more ``optimistic'' scenarios could be consistent with data if the scalar mass $M_S$ was chosen to be sufficiently large.

On the other hand, in Ref.~\cite{King:2014uha}, three categories of valid benchmark points were introduced. They have been found by numerically scanning the parameter space for 2-loop mass generation of light neutrinos using the Lagrangian given in our Eq.~\eqref{eq:Lagrangian}:
\begin{itemize}

\item \textcolor{red}{\bf red} points: $f_{ee}\simeq 0$ and $f_{e\tau}\simeq 0$,

\item \textcolor{purple}{\bf purple} points: $f_{ee}\simeq 0$ and $f_{e\mu}\simeq -\frac{f^*_{\mu\tau}}{f^*_{\mu\mu}}\,f_{e\tau}$,

\item \textcolor{blue}{\bf blue} points: $f_{e\mu}\simeq -\frac{f^*_{\mu\tau}}{f^*_{\mu\mu}}\,f_{e\tau}$.

\end{itemize}
These categories of points were chosen such that they reproduce all relevant low-energy phenomenology, i.e., all neutrino oscillation parameters as well as all LFV/LNV bounds, with $\mu$-$e$ conversion being the only exception. Note that the consistency of these benchmark categories partially arises from correlations, like $f_{e\mu}\simeq -\frac{f^*_{\mu\tau}}{f^*_{\mu\mu}}\,f_{e\tau}$ for the purple points, which lead to cancellations in the rate for $\mu \to e \gamma$. However, these cancellations do not appear anymore in $\mu$-$e$ conversion, as we will illustrate in the following. In order to not only show a few isolated points as found in Ref.~\cite{King:2014uha}, we will for illustrative purposes present idealised scenarios which roughly correspond to the three categories of benchmark points. The explicit parameter choices for these scenarios are displayed in Tab.~\ref{Tab:Coupling_Scenarios}, and they approximately correspond to the average of the values reported in Tab.~7 of Ref.~\cite{King:2014uha}.

\begin{table}[th!]
\centering
 \begin{tabular}{c|c|c|c}
  & \textcolor{red}{\bf red} & \textcolor{purple}{\bf purple} & \textcolor{blue}{\bf blue} \\
 \hline \hline
  $f_{ee}$ & $10^{-16}$ & $10^{-15}$ & $10^{-1}$ \\
 \hline
   $f_{e\mu}$ & $10^{-2}$ & $10^{-3}$ & $10^{-4}$ \\ 
 \hline
   $f_{e\tau}$ & $10^{-19}$ & $10^{-2}$ & $10^{-2}$ \\ 
 \hline
   $f_{\mu\mu}$ & $10^{-4}$ & $10^{-3}$ & $10^{-3}$ \\ 
 \hline
   $f_{\mu\tau}$ & $10^{-5}$ & $10^{-4}$ & $10^{-4}$ \\ 
\hline \hline
   $f^*_{ee}\,f_{e\mu}$ & $10^{-18}$ & $10^{-18}$ & $\mathbf{10^{-5}}$ \\
 \hline
   $f^*_{e\mu}\,f_{\mu\mu}$ & $\mathbf{10^{-6}}$ & $\mathbf{10^{-6}}$ & $10^{-7}$ \\
 \hline
   $f^*_{e\tau}\,f_{\mu\tau}$ & $10^{-24}$ & $\mathbf{10^{-6}}$ & $10^{-6}$ \\
 \end{tabular}
 \caption{ \label{Tab:Coupling_Scenarios}\emph{Upper part}: Couplings for the three scenarios discussed in the text. \emph{Lower part}: Combinations of couplings entering the $\mu$-$e$ conversion amplitude. Bold figures indicate the dominant contributions.}
 \label{Tab:Coupling_Scenarios}
\end{table}

\begin{figure}[th!]
\centering
 \includegraphics[width=10cm]{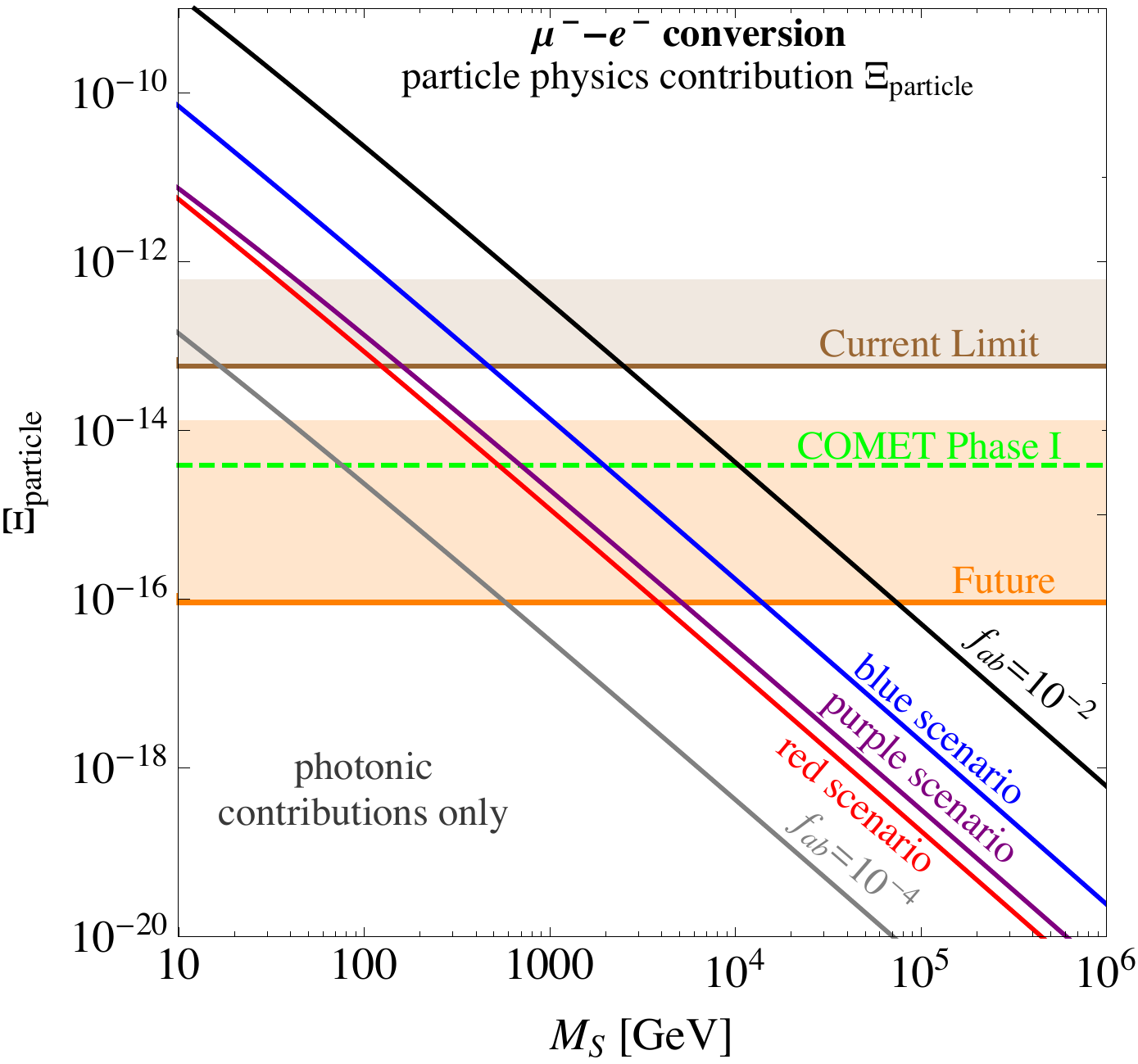}
 \caption{\label{fig:Xi_photonic}Bounds on the particle physics contribution $\Xi_{\rm particle}$ arising from the photonic (long-range) contributions only.}
\end{figure}

We are now ready to present our results for $\mu^-$-$e^-$ conversion when only taking the photonic (long-range) contributions into account. Fig.~\ref{fig:Xi_photonic} summarises all the information we have collected so far, and it also illustrates how strongly the doubly charged scalar mass can be constrained. We have displayed the particle physics parts of the amplitude as functions of the doubly charged scalar mass $M_S$, i.e., the photonic/long-range contribution $\Xi_{\rm particle}$ from Eq.~\eqref{eq:Xi_simpl}. The next step is to compare the predictions to the experimental bounds. As already indicated, we have collected several current (SINDRUM~II~\cite{Dohmen:1993mp,Bertl:2006up,Honecker:1996zf}) and future (DeeMe~\cite{Aoki:2010zz}, COMET~\cite{COMET}, Mu2e~\cite{Kutschke:2011ux}, PRISM/PRIME~\cite{Barlow:2011zza}) limits on the branching ratio of $\mu^-$-$e^-$ conversion. However, due to both nuclear physics uncertainties and experiments on different isotopes potentially pushing one and the same particle physics observable, we have decided to display a range of bounds in Fig.~\ref{fig:Xi_photonic}. Thereby, the nominally best limits are represented by the bold horizontal lines, and the variation among the different isotopes and/or experiments is indicated by the lightly coloured rectangles which absorb all uncertainties as long as the particle physics part of the amplitude can be extracted. Moreover, we have included the sensitivity expected to be reached in Phase~I of COMET. The corresponding bound of $\Xi^{\rm Al}_{\rm particle}=3.87\cdot 10^{-15}$ on the particle physics observable is represented by the dashed green line and stems from the single event sensitivity of ${\rm BR}(\mu^- {\rm Al} \to e^- {\rm Al})=3.1\cdot 10^{-15}$ reported in Ref.~\cite{COMET2014}. Note that we have \emph{not} indicated the variation with nuclear physics uncertainties, because we have not found any reliable up-to-date information. It is however evident how to include information on this point, so that it will be easy to update our plot once this information is available.

Looking at the numbers, it is evident that we can in fact obtain very strong bounds on the doubly charged scalar mass from not having observed $\mu$-$e$ conversion. In Tab.~\ref{tab:photonic}, we have displayed both the current limits and the future sensitivities as well as the sensitivity that will be reached within COMET's Phase~I. The ranges displayed in Tab.~\ref{tab:photonic} are obtained by taking both the most optimistic (i.e., the bold horizontal lines in Fig.~\ref{fig:Xi_photonic}) and the most pessimistic (i.e., the upper edges of the lightly coloured rectangles in Fig.~\ref{fig:Xi_photonic}) bounds at face value. This accounts for the possible variations among the different experiments. However, we would like to stress once more that further variations due to nuclear physics uncertainties may well be possible. While these are not expected to dramatically change our results, they may be able to at least change the last few digits in the figures quoted in Tab.~\ref{tab:photonic}. Nevertheless, it is evident that even the most pessimistic limits are in fact quite impressive, revealing that, for doubly charged scalars, $\mu$-$e$ conversion may be able to lead to bounds stronger than those obtained by colliders~\cite{Geib:2015tvt}.

\begin{table}[t]
\centering
\hspace{-0.2cm}
\begin{tabular}{c||l|l|l}
& current limit [GeV]& future sensitivity [GeV] & COMET~I (Al-27) [GeV] \\
\hline \hline
black curve & $M_S \textgreater 708.6 - 2390.2$ &  $M_S \textgreater 5500.0 - 70369.3$ & $M_S \textgreater 10401.9$\\
\hline
blue curve & $M_S \textgreater 131.9 - 447.1$ &  $M_S \textgreater 1031.5 - 13271.3$ &  $M_S \textgreater 1954.1$\\
\hline
purple curve & $M_S \textgreater 42.5 - 152.3$ &  $M_S \textgreater 360.7 - 4885.2$ & $M_S \textgreater 694.5$ \\
\hline
red curve & $M_S \textgreater 33.9 - 118.1$ &  $M_S \textgreater 276.3 - 3656.1$ & $M_S \textgreater 528.0$ \\
\hline
gray curve & $M_S \textgreater 4.1 - 15.9$ &  $M_S \textgreater 38.7 - 548.7$ & $M_S \textgreater 75.7$ \\
\end{tabular}
\caption{\label{tab:photonic}Lower limits on the mass $M_S$ resulting from $\mu$-$e$ conversion, displaying the range from the most pessimistic to the most optimistic values. Figures are deliberately shown with a too good precision, in order to ease the comparison with Tab.~\ref{tab:non-photonic}.}
\end{table}

The question to answer is why the bounds from $\mu$-$e$ conversion seem to be significantly stronger than those for $\mu \to e\,\gamma$ obtained in Ref.~\cite{King:2014uha}. This is particularly surprising when disregarding the short-range contributions, as we do, since then at first sight $\mu$-$e$ conversion looks just like a $\mu \to e\,\gamma$ diagram attached to a nucleus, cf.\ Fig.~\ref{fig:Diagrams_mu_e_conversion}. However, the result can be understood by carefully comparing the amplitudes for both processes. The branching ratio of $\mu \to e+\gamma$ depends on an amplitude of the form:
\begin{equation}
 \mathcal{A}\propto \big|f^*_{ee}\,f_{e\mu}+f^*_{e\mu}\,f_{\mu\mu}+f^*_{e\tau}\,f_{\tau\mu} \big|\cdot C,
 \label{eq:meg-amp}
\end{equation}
where $C$ is a flavour-independent constant incorporating all non-Yukawa couplings. As explained, the benchmark points in Ref.~\cite{King:2014uha} had been chosen such that all experimental bounds are fulfilled. In particular for the purple and blue points, cancellations appear in Eq.~\eqref{eq:meg-amp}, which allow to evade the (quite strong) bound from $\mu \to e\,\gamma$. On the other hand, glancing at Eq.~\eqref{eq:Xi_simpl}, the amplitude for $\mu$-$e$ conversion is of the form:
\begin{equation}
 \mathcal{A}\propto \big| C_e\,f^*_{ee}\,f_{e\mu}+C_\mu\,f^*_{e\mu}\,f_{\mu\mu}+C_\tau\,f^*_{e\tau}\,f_{\tau\mu}\big|,
 \label{eq:mecon-amp}
\end{equation}
where now the ``constant'' $C$ from Eq.~\eqref{eq:meg-amp} has gained a flavour dependence, $C\to C_{e, \mu, \tau}$. Thus, one cannot simply extract this factor from the amplitude in Eq.~\eqref{eq:mecon-amp} and, in particular, the cancellations at work to evade the $\mu \to e\,\gamma$ bound will not work for $\mu$-$e$ conversion anymore. Instead, comparatively large values of the Yukawa couplings are strongly constrained by the experimental limits. This is perfectly consistent with the figures quoted in the lower part of Tab.~\ref{Tab:Coupling_Scenarios}, where the sizes of the combinations $(f^*_{ee}\,f_{e\mu}, f^*_{e\mu}\,f_{\mu\mu}, f^*_{e\tau}\,f_{\tau\mu})$ appearing in Eq.~\eqref{eq:mecon-amp} are estimated for the three scenarios. The largest such combination appears for the blue scenario, $|f^*_{ee}\,f_{e\mu}| \sim 10^{-5}$, while the red and purple scenarios instead seem to yield a very similar size. Indeed this tendency is perfectly visible in both Fig.~\ref{fig:Xi_photonic} and Tab.~\ref{tab:photonic}, where the bounds on the blue scenario indeed turn out to be stronger than those on the red and purple scenarios, which are quite similar.\\

Summing up, we have shown that already the photonic (long-range) contributions to $\mu$-$e$ conversion lead to comparatively strong lower bounds on the scalar mass $M_S$.

\section{\label{sec:mue_minus-non-photonic}Short-range (non-photonic) contributions}

The next step is to include the non-photonic (short-range) contributions to $\mu$-$e$ conversion.

\subsection{\label{sec:mue_minus-non-photonic_form}Computing the form factors}

The non-photonic contributions to the $\mu$-$e$ conversion amplitude are commonly subsumed into four fermion interactions, i.e., we are considering point-like (short-range) operators coupling one $\mu$ and one $e$ to two quarks. It is a priori not clear whether these contributions could modify the $\mu$-$e$ conversion rate significantly. Quite generally, including these terms spoils the factorisation of the branching ratio into nuclear physics and particle physics parts, such that Eq.~\eqref{eq:mu-e_BR_long} is not applicable anymore. In general, the effect on the particle physics amplitude will be to now turn into a combined amplitude incorporating both photonic (long-range) and non-photonic (short-range) contributions, the latter being dependent on $Z$ and $N$:
\begin{equation*}
 \Xi_\text{particle} \to \Xi_\text{combined}(Z,N) = \Xi_\text{photonic} + \Xi_\text{non-photonic}(Z,N).
\end{equation*}
However, as we will see, in our case the short-range contributions turn out to be completely \emph{subdominant}. Thus, although Eq.~\eqref{eq:mu-e_BR_long} is in general \emph{not} correct, applying it would introduce only a very small error, and we can thus approximate $\Xi_\text{particle} \simeq \Xi_\text{photonic}$ to a very good precision. We will in the following illustrate how to explicitly compute the short-range contributions to $\mu$-$e$ conversion.

\begin{figure}[t]
  \begin{minipage}[c]{14cm}
    \begin{subfigure}[c]{6.6cm}
    \includegraphics[width=6.6cm]{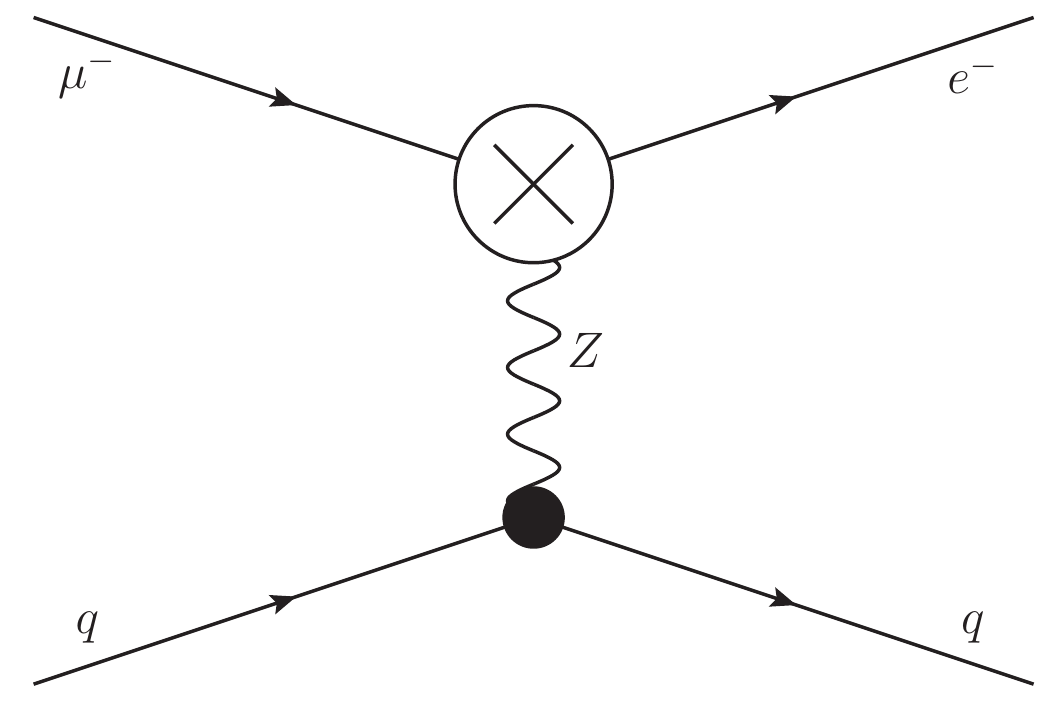}
    \end{subfigure}
   $\longrightarrow \qquad$
   \begin{subfigure}[c]{6.6cm}
    \includegraphics[width=6.6cm]{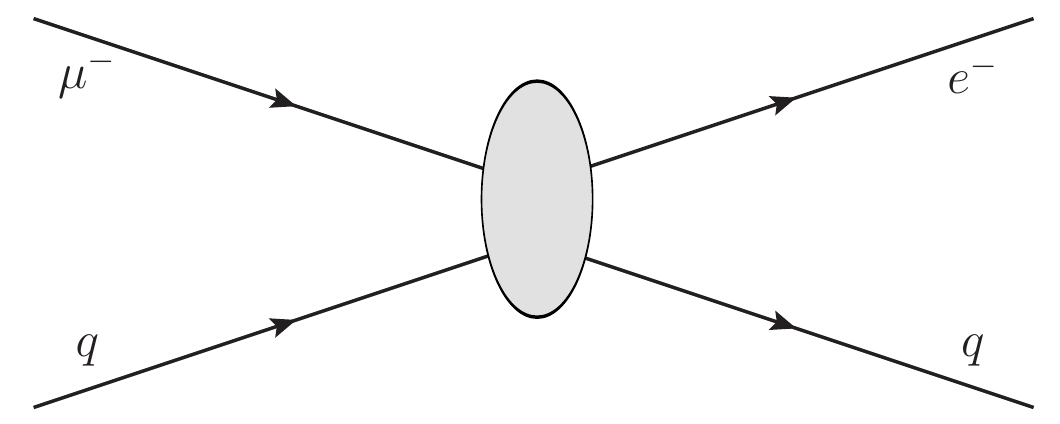}
     \end{subfigure}  
  \end{minipage}
    \caption{Integrating out the $Z$-boson results into a short-range contribution.}
    \label{fig:Eff_Vertex_Z}
\end{figure}

Considering effective operators up to dimension-six, a general interaction of an electron and a muon with two quarks is described by~\cite{Kuno:1999jp}:
\begin{eqnarray}
 \mathcal{L}_{\text{non-photonic}}&=&-\frac{G_F}{\sqrt{2}}\,\sum_{q=u,d,s,\cdots} \Bigg[\Big(g_{LS(q)}\overline{e_L}\,\mu_R+g_{RS(q)}\overline{e_R}\,\mu_L\Big)\overline{q}\,q+\Big(g_{LP(q)}\overline{e_L}\,\mu_R+g_{RP(q)}\overline{e_R}\,\mu_L\Big)\overline{q}\,\gamma_5\,q\nonumber\\
 &&+\Big(g_{LV(q)}\overline{e_L}\,\gamma^\nu\,\mu_L+g_{RV(q)}\overline{e_R}\,\gamma^\nu\,\mu_R\Big)\overline{q}\,\gamma_\nu\,q+\Big(g_{LA(q)}\overline{e_L}\,\gamma^\nu\,\mu_L+g_{RA(q)}\overline{e_R}\,\gamma^\nu\,\mu_R\Big)\overline{q}\,\gamma_\nu\,\gamma_5\,q\nonumber\\
 &&+\frac{1}{2}\Big(g_{LT(q)}\overline{e_L}\,\sigma^{\nu \rho}\,\mu_R+g_{RT(q)}\overline{e_R}\,\sigma^{\nu \rho}\mu_L\Big)\overline{q}\,\sigma_{\nu \rho}\,q+\text{h.c.} \Bigg]\,.
\label{eq:nonPhot_1}
\end{eqnarray}
The effective four fermion couplings given above originate from integrating out all particles which could possibly be exchanged between two quarks and two charged leptons. In our setup, the dominant non-photonic contribution arises from the $Z$-boson exchange between two quarks in the nucleus and the particle physics loop, depicted in Diagrams~I to~IV in Figs.~\ref{fig:Diagrams_mu_e_conversion}. The terms involving neutrinos in the loops are again GIM-suppressed~\cite{Glashow:1970gm}, which is the case for both categories, penguin diagrams (Diagrams~V to~VIII) and box-diagrams (Diagrams~IX and~X). The diagrams based on Higgs exchange are suppressed even further, a back-of-the-envelope estimate resulting in a suppression of $\mathcal{O}(10^{-3})$ compared to the other short-range contributions, which are already suppressed themselves. We will thus completely disregard the diagrams based on Higgs-exchange. Note that, in order to consistently obtain the form factors $g_{XK(q)}$ ($X=R,L$ and $K=S,P,V,A,T$), we match the relevant set of diagrams to the four fermion operators using a generic $\mu$-$e$-$Z$ interaction $\Gamma_\nu$, see Fig.~\ref{fig:Eff_Vertex_Z}. 

The Feynman rules tell us:
\begin{equation}
 i\mathcal{M}=\overline{u}_e (p_e)\,\Gamma_\nu\,\,u_\mu(p_\mu)\, \cdot \frac{-i}{q'^2-M_Z^2}\,\big(g^{\nu \rho}-\frac{q'^\nu q'^\rho}{M_Z^2}\big)\cdot \overline{q}\,\frac{i g}{4 \cos \theta_W}\,\gamma_\rho \,\Big[1+k_q\,\sin^2 \theta_W +s_q\,\gamma_5 \Big]q\,,
\end{equation}
for the ``full theory'' diagram on the left. Here, the coefficients $k_q$ and $s_q$ depend on the quark being up- or down-type: $k_{d,s,b}=4/3$, $s_{d,s,b}=1$, $k_{u,c,t}=-8/3$, and $s_{u,c,t}=-1$. By contracting the bosonic propagator, i.e. taking the limit $M^2_Z \gg q'^2$, the matrix element takes the form:
\begin{equation}
\begin{split}
 i\mathcal{M}&=\overline{u}_e (p_e)\,\Gamma_\nu\,\,u_\mu(p_\mu)\,\frac{i}{M_Z^2}\,g^{\nu \rho}\,\overline{q}\,\frac{i g}{4 \cos \theta_W}\,\gamma_\rho \,\Big[1+k_q\,\sin^2 \theta_W +s_q\,\gamma_5 \Big]q\\
 &=-\frac{g}{4 M^2_Z\,\cos \theta_W}\, \overline{u}_e (p_e)\,\Gamma_\nu\,u_\mu(p_\mu)\, \Big[\big(1+k_q\,\sin^2 \theta_W \big)\underbrace{\tikz[baseline]{\node [fill=white!20,anchor=base,draw=dunkelgruen] (e1){$\overline{q}\,\gamma^\nu \,q$};}}_{\textcolor{dunkelgruen}{\mathclap{\text{vector coupling}}}}+s_q\,\underbrace{\tikz[baseline]{\node [fill=white!20,anchor=base,draw=dunkelblau] (e2){$\overline{q}\,\gamma^\nu\,\gamma_5\,q$};}}_{\textcolor{dunkelblau}{\mathclap{\substack{\text{axial vector}\\ \text{coupling}}}}} \Big]\,.
 \end{split}
 \label{eq:nonPhot_2}
\end{equation}
Apparently, only the vector and axial vector structures are realised. Since we consider coherent $\mu$-$e$ conversion, however, only the vector coupling will ultimately contribute to the branching ratio. Taking into account gauge invariance, the most general form for the generic coupling $\Gamma_\nu$ can be written as~\cite{Patel:2015tea}:
\begin{eqnarray}
 \Gamma_\nu&=&\gamma_\nu\,P_L\,A_L\,(q'^2)+\frac{i \sigma_{\nu \rho}\,q'^\rho}{m_\mu+m_e}\,P_L\,B_L(q'^2)+2\frac{q'_\nu}{m_\mu+m_e}\,P_LC_L(q'^2)+\gamma_\nu\,P_R\,A_R\,(q'^2)\nonumber\\
 &&+\frac{i \sigma_{\nu \rho}\,q'^\rho}{m_\mu+m_e}\,P_R\,B_R(q'^2)+2\frac{q'_\nu}{m_\mu+m_e}\,P_RC_L(q'^2)\,.
\end{eqnarray}

However, as mentioned earlier, we only take into account effective operators with mass dimension up to six. Since the combined mass dimension of four spin-$1/2$ fields and the momentum $q'$ already exceeds dimension six, we can consistently drop such terms. Moreover, the doubly charged scalar solely couples to right-handed leptons. Since we assume the electron to be massless, all form factors $g_{LK}$ vanish identically. Thus, the dominant contribution to the short-range part of coherent $\mu$-$e$ conversion emerges from just one single term. After rewriting the couplings in Eq.~\eqref{eq:nonPhot_2} such that they match those in Eq.~\eqref{eq:nonPhot_1}, the relevant effective Lagrangian is given by:
\begin{equation}
 \mathcal{L}_{\text{non-photonic}}=-\frac{G_F}{\sqrt{2}}\,\underbrace{\frac{2\big(1+k_q\sin^2 \theta_W \big)\cos \theta_W}{g}\,A_R(q'^2)}_{=g_{RV(q)}}\,\ \overline{e_R}\,\gamma_\nu\,\mu_R\,\ \overline{q}\,\gamma^\nu\,q\,.
 \label{eq:nonPhot_3}
\end{equation}
However, this Lagrangian still operates at quark level, while what we are interested in is the analogous vertex coupling the muon and the electron to nucleons. Converting the Lagrangian in Eq.~\eqref{eq:nonPhot_3} to nucleon level, the new coupling constants $g^{(0)}_{XK}$ and $g^{(1)}_{XK}$ can be re-expressed in terms of the nucleon form factors $G_K^{(q,p)}$ and $G_K^{(q,n)}$, see Ref.~\cite{Kuno:1999jp} for details:
\begin{equation}
\begin{split}
 g^{(0)}_{XK}&=\frac{1}{2}\sum_{q=u,d,s} g_{XK(q)}\Big(G_K^{(q,p)}+G_K^{(q,n)}\Big)\,,\\
 g^{(1)}_{XK}&=\frac{1}{2}\sum_{q=u,d,s} g_{XK(q)}\Big(G_K^{(q,p)}-G_K^{(q,n)}\Big)\,.
 \end{split}
 \label{eq:nonPhoto_couplings}
\end{equation}
Taking the limit of isospin invariance, we can relate the proton and neutron form factors~\cite{Bernabeu:1993ta}: $G_K^{(u,p)}=G_K^{(d,n)}$, $G_K^{(u,n)}=G_K^{(d,p)}$, and $G_K^{(s,p)}=G_K^{(s,n)}$. Furthermore, it is $G_V^{(u,p)}=2$, $G_V^{(u,n)}=1$, and $G_V^{(s,p)}=0$ for the vector current. Again employing the non-relativistic approximation for the muon wave function, the branching ratio of coherent $\mu$-$e$ conversion takes the general form~\cite{Kuno:1999jp}:
\begin{eqnarray}
 \mathrm{BR}(\mu^- N \to e^- N)&=&\frac{|\vec{p}_e| E_e m^3_\mu G^2_F \alpha^3 Z^4_{\text{eff}} F^2_p}{8\pi^2 Z\,\Gamma_{\rm capt}} \Bigg[\Big|  \big(Z+N\big)\big(g^{(0)}_{LS}+g^{(0)}_{LV}\big)+\big(Z-N\big)\big(g^{(1)}_{LS}+g^{(1)}_{LV}\big)\Big|^2\nonumber\\
 &&+\Big|\big(Z+N\big)\big(g^{(0)}_{RS}+g^{(0)}_{RV}\big)+\big(Z-N\big)\big(g^{(1)}_{RS}+g^{(1)}_{RV}\big)\Big|^2\Bigg],
\end{eqnarray}
under the assumptions of equal proton and neutron densities as well as a quasi-constant muon wave function within the nucleus. Here, $G_F$ is Fermi's constant and $\alpha=e^2/(4\pi)=g^2 \sin^2 \theta_W/(4\pi)$. All other quantities are defined as in Eq.~\eqref{eq:mu-e_BR_long}.

Within our framework there are neither scalar contributions, i.e.\ $g^{(0,1)}_{LS}=g^{(0,1)}_{RS}=0$, nor contributions that include left-handed electrons, i.e.\ $g^{(0,1)}_{LV}=0$. Moreover, we take the electron to be massless, which leads to $E_e=|\vec{p}_e| \simeq m_\mu$. In combination with Eqs.~\eqref{eq:nonPhot_3} and~\eqref{eq:nonPhoto_couplings}, the branching ratio hence simplifies to:
\begin{equation}
\begin{split}
& \mathrm{BR}(\mu^- N \to e^- N)=\\
& \frac{8 \alpha^5 m_\mu Z^4_{\text{eff}} Z F^2_p}{\Gamma_{\rm capt}}\,\underbrace{\frac{m^4_\mu \cos^2 \theta_W}{128\,\pi \alpha Z^2 M^4_W  \sin^2 \theta_W}\,\Big|\Big(3\big(Z+N\big)-4\,Z \sin^2 \theta_W \Big) A_R(-m^2_\mu)\big)\Big|^2}_{\equiv \Xi^2_{\text{non-photonic}}},
 \end{split}
 \label{eq:BR_nonphontic}
\end{equation}
where we have used $G_F=\frac{\alpha \pi}{\sqrt{2}M^2_W \sin^2 \theta_W}$. Here, we have rewritten the non-photonic branching ratio such that we can extract a $\Xi_\text{non-photonic}$ in analogy to the photonic contributions. However, in contrast to the photonic part $\Xi_{\rm photonic}$, one cannot factorise the particle and nuclear physics contributions, in the sense that $\Xi_\text{non-photonic}$ depends on the nuclear characteristics $(Z, N)$: $\Xi_\text{non-photonic} = \Xi_\text{non-photonic}(Z, N)$. While this looks like as if it made the distinction between particle physics and nuclear physics parts impossible, it will turn out that the dependence on $(Z, N)$ is in reality so weak that it can be dropped without changing the results. This is again a reflection of the short-range contribution being subdominant by far.

In order to determine the form factor $A_R(q'^2)$, we proceed in a way similar as we did for the photonic form factors, meaning that we consider the process $\mu \to e\,Z$ for an off-shell gauge boson. From Diagram~I in Fig.~\ref{fig:Diagram_I}, we obtain the matrix element:
\begin{eqnarray}
 i\mathcal{M}_{\rm I} &=&-8\,f^*_{ea} f_{a\mu}\,g' \sin \theta_W\,Z^\nu(q')\,\\
 && \overline{u}_e(p_e)\, \int \frac{\diff^d k}{(2\pi)^d}\,\frac{P_L\,\slashed{k}\big(2p_\mu-2k+q'\big)_\nu}{[k^2-m^2_a][(p_\mu-k+q')^2-M^2_S][(p_\mu-k)^2-M^2_S]}\,u_\mu (p_\mu)\,,\nonumber
\end{eqnarray}
where we have dropped the '$+i\epsilon$' terms for brevity. For Diagram~II, see Fig.~\ref{fig:Diagram_II}, the matrix element is given by:
\begin{eqnarray}
 i\mathcal{M}_{\rm II} &=& -f^*_{ea} f_{a\mu}\,\frac{g}{\cos \theta_W}\,Z^\nu(q')\,\\
 && \overline{u}_e(p_e)\, \int \frac{\diff^d k}{(2\pi)^d}\,\frac{P_L\big(\slashed{k}+\slashed{q}'+m_a\big)\gamma_\nu\big(1-4\sin^2 \theta_W +\gamma_5\big)\big(\slashed{k}+m_a\big)P_R}{[k^2-m^2_a][(p_\mu-k)^2-M^2_S][(k+q')^2-m^2_a]}\,u_\mu (p_\mu)\,.\nonumber
\end{eqnarray}
From Fig.~\ref{fig:Diagram_III}, we extract:
\begin{equation}
 i\mathcal{M}_{\rm III}=-f^*_{ea} f_{a\mu}\,\frac{g}{\cos \theta_W}\,Z^\nu(q')\,\overline{u}_e(p_e)\,\int \frac{\diff^d k}{(2\pi)^d}\,\frac{\gamma_\nu\big(-1+4 \sin^2 \theta_W +\gamma_5\big)\slashed{p}_\mu\,\slashed{k}\,P_R}{p^2_\mu\,[k^2-m^2_a][(p_\mu-k)^2-M^2_S]}\,u_\mu (p_\mu)\,.
\end{equation}
And, finally, from Fig.~\ref{fig:Diagram_IV}:
\begin{equation}
 i\mathcal{M}_{\rm IV}=-f^*_{ea} f_{a\mu}\,\frac{g}{\cos \theta_W}\,Z^\nu(q')\,\overline{u}_e(p_e)\,\int \frac{\diff^d k}{(2\pi)^d}\,\frac{P_L\,\slashed{k}\big(\slashed{p}_e+m_\mu\big)\gamma_\nu\big(-1+4 \sin^2 \theta_W +\gamma_5\big)}{-m^2_\mu\,[k^2-m^2_a][(p_e-k)^2-M^2_S]}\,u_\mu (p_\mu)\,.
\end{equation}
Again using \texttt{Package-X}, we compute each diagram's contribution to $A_R$, and combine them using $g'=g \tan \theta_W$. Due to the absence of a tree-level 3-point vertex connecting muon, electron and Z-boson, the form factor $A_R$ has to be UV-finite. Similarly to the photonic case, the UV divergences occurring in the individual diagrams~I~-~IV cancel each other, thus leaving $A_R$ finite. As before, we can simplify the form factor by exploiting the mass hierarchy $M_S \gg m_{e, \mu, \tau}$ to obtain:
\begin{equation}
\begin{split}
 A_R (-m^2_\mu)&=\frac{-ig}{24\pi^2\,\cos \theta_W\,M^2_S\,m_\mu}\,f^*_{ea} f_{a\mu}\,\Bigg(m_\mu\,m^2_a\big(-3+8\sin^2 \theta_W\big)+2\sqrt{4 m^2_a+m^2_\mu}\Big(2m^2_\mu \sin^2 \theta_W \\
 &+ m^2_a\big(3-4\sin^2 \theta_W\big)\Big)\arctanh\Big[\frac{m_\mu}{\sqrt{4 m^2_a+m^2_\mu}}\Big]+\Big(3m^2_a m_\mu+2 m^3_\mu \sin^2 \theta_W \Big)\ln \Big[\frac{m^2_a}{M^2_S}\Big]\Bigg)\,,
 \end{split}
\end{equation}
where the sum over $a=e,\mu,\tau$ is implied.\\
So, the somewhat artificial (because in reality not dominating) non-photonic particle physics factor $\Xi_{\text{non-photonic}}$ can be deduced to be:
\begin{eqnarray}
 && \Xi^2_{\text{non-photonic}}=\frac{m^2_\mu \, \big|3\big(Z+N\big)-4\,Z  \sin^2 \theta_W \big|^2}{18432\, \pi^4  Z^2 M^4_W\,M^4_S \sin^4 \theta_W}\,\Bigg|\sum_{a=e,\mu,\tau}  f^*_{ea} f_{a\mu}\,\Bigg(m_\mu\,m^2_a\big(-3+8\sin^2 \theta_W\big)\nonumber\\
 && +2\sqrt{4 m^2_a+m^2_\mu} \Big(2m^2_\mu \sin^2 \theta_W+ m^2_a\big(3-4\sin^2 \theta_W\big)\Big)\arctanh\Big[\frac{m_\mu}{\sqrt{4 m^2_a+m^2_\mu}}\Big]\nonumber\\
 && +\Big(3m^2_a m_\mu+2 m^3_\mu \sin^2 \theta_W \Big)\ln \Big[\frac{m^2_a}{M^2_S}\Big]\Bigg)\Bigg|^2,
\end{eqnarray}
at leading order.

\subsection{The Total Branching Ratio}

In general, both the photonic and non-photonic processes contribute to $\mu$-$e$ conversion. Kinematics dictate that $q'^2 = -m^2_\mu$, which in combination with the non-relativistic approximation of the muon wave function implies that the photonic (long-range) contribution can effectively be treated as an addition $\Delta g^{(0,1)}_{RV}$ to the vectorial coupling constants $g^{(0,1)}_{RV}$, see Eq.~(141) of Ref.~\cite{Kuno:1999jp}. We thus obtain:
\begin{equation}
 g^{(0,1)}_{RV} \to  g^{(0,1)}_{RV}+\Delta g^{(0,1)}_{RV}\,, \quad \text{where} \quad \Delta g^{(0,1)}_{RV}=\frac{4\sqrt{2}\,\alpha \,\pi}{G_F\,m^2_\mu}\Big(F_2(-m^2_\mu)-F_1(-m^2_\mu)\Big),
 \label{eq:total-form-factors}
\end{equation}
with the form factors $F_1$ and $F_2$ explicitly given in Eqs.~\eqref{eq:FFexpl_1} and~\eqref{eq:FFexpl_2}, respectively. We can now understand why the non-photonic (short-range) contributions are subdominant: while both $|F_2(-m^2_\mu)-F_1(-m^2_\mu)|$ and $|A_R (-m^2_\mu)|$ are of $\mathcal{O}(m_a^2/M_S^2)$, we can see from Eq.~\eqref{eq:total-form-factors} that the photonic (long-range) contributions are considerably \emph{less suppressed}, receiving a relative enhancement factor that should naively be of the order of $\alpha/(G_F\,m^2_\mu) \sim M^2_W/m^2_\mu \sim 10^5$.

Replacing the purely non-photonic couplings in favour of the ones given above in Eq.~\eqref{eq:total-form-factors}, we can derive the general branching ratio in analogy to the derivation of Eq.~\eqref{eq:BR_nonphontic}. The combined branching ratio, incorporating both photonic (long-range) and non-photonic (short-range) contributions, takes the form:
\begin{eqnarray}
 &&\mathrm{BR}(\mu^- N \to e^- N)=\frac{8 \alpha^5 m_\mu Z^4_{\text{eff}} Z F^2_p}{\Gamma_{\rm capt}}\,\Xi^2_{\text{combined}}(Z,N)\,, \label{eq:BR-total}\\
 &&\text{with}\quad \Xi^2_{\text{combined}} =\frac{m^4_\mu}{8192\,\pi^4\,Z^2\, M^4_W \sin^4 \theta_W}\,\Bigg|\sum_{a=e,\mu,\tau}  f^*_{ea} f_{a\mu}\,\Bigg(-\frac{2\big(3\big(Z+N\big)-4\,Z \sin^2 \theta_W\big)}{3\,m_\mu\,M^2_S}\Big(m_\mu\,m^2_a\nonumber\\
 &&\big(-3+8\sin^2 \theta_W\big)+2\sqrt{4 m^2_a+m^2_\mu}\Big(2m^2_\mu \sin^2 \theta_W+ m^2_a\big(3-4\sin^2 \theta_W\big)\Big)\arctanh\Big[\frac{m_\mu}{\sqrt{4 m^2_a+m^2_\mu}}\Big]\nonumber\\
 &&+\Big(3m^2_a m_\mu+2 m^3_\mu \sin^2 \theta_W \Big)\ln \Big[\frac{m^2_a}{M^2_S}\Big]\Big) +\frac{16}{3}\,\frac{M^2_W\,Z \sin^2 \theta_W}{m^3_\mu\,M^2_S}\Big(4m^2_a\,m_\mu-m^3_\mu\nonumber\\
 &&+2\big(-2m^2_a+m^2_\mu\big)\sqrt{4m^2_a+m^2_\mu}\,\arctanh \Big[\frac{m_\mu}{\sqrt{4 m^2_a+m^2_\mu}}\Big]+m^3_\mu\,\ln \Big[\frac{m^2_a}{M^2_S}\Big]\Big)\Bigg)\Bigg|^2,\nonumber
\end{eqnarray}
at leading order in the small ratios $m^2_a/M^2_S$.

As already pointed out and as is now clearly visible from Eq.~\eqref{eq:BR-total}, $\Xi_{\text{combined}}$ is not pure particle physics quantity, in the sense that it also depends on the nuclear characteristics $Z$ and $N$. However, we can nevertheless use it to compare the impact of a certain bound on the new physics parameters, as long as we take into account the variation with $Z$ and $N$. Thus, when plotting $\Xi_{\text{combined}}$ as a function of the scalar mass $M_S$, one would not only obtain a simple line but a band, the width arising from varying $Z$ and $N$. However, as we will see, numerically this variation is very mild, since it only affects the subdominant contribution to the decay -- in a logarithmic plot, the width of the band would not even be visible. Thus, in practice, we can disregard the variation with $Z$ and $N$ whenever presenting a bound just for illustrative purposes.

\begin{figure}[h!]
\centering
 \includegraphics[width=10cm]{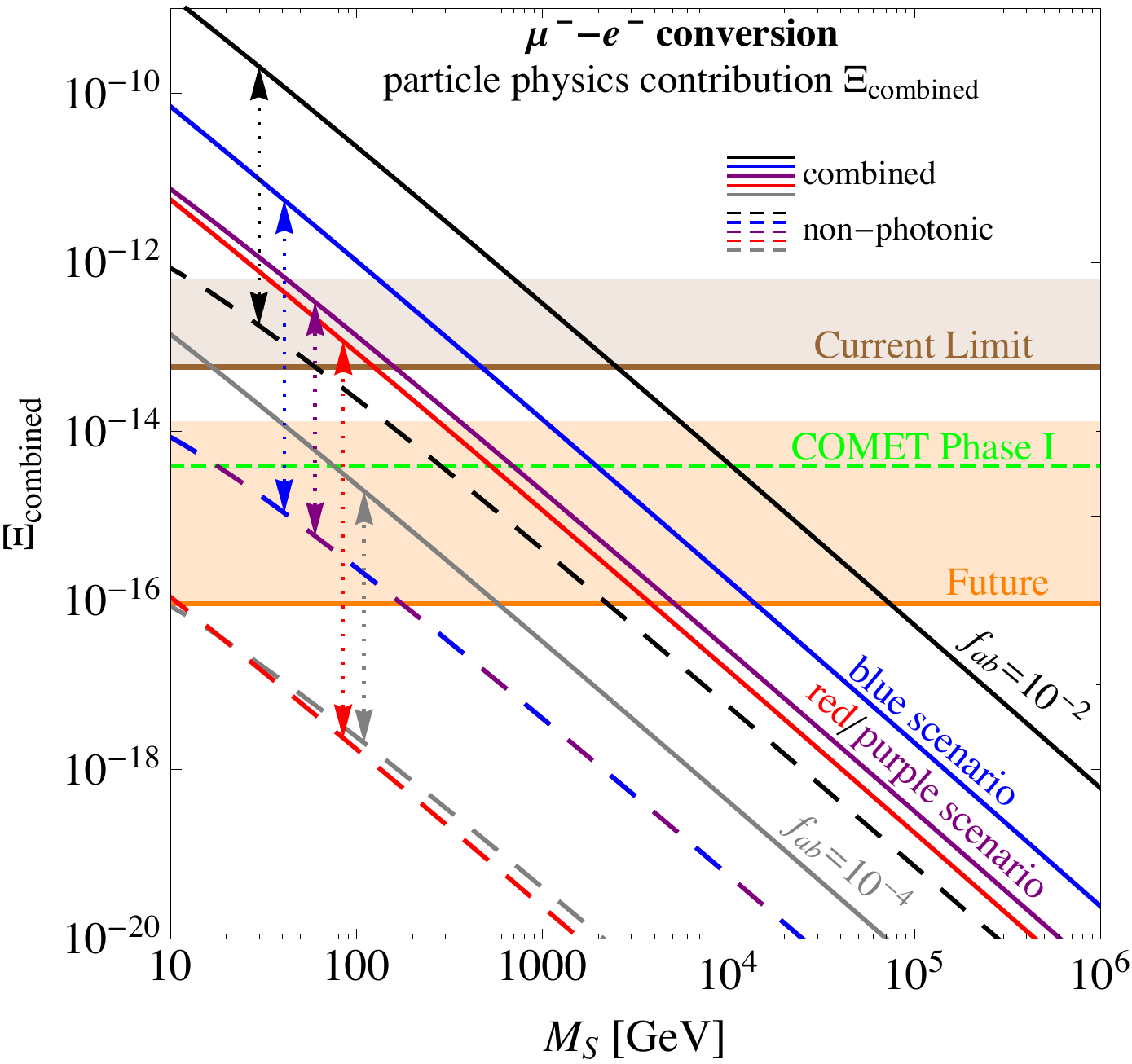}
 \caption{\label{fig:Xi_nonphotonic}Bounds on the full particle physics contribution $\Xi_{\rm combined}$.}
\end{figure}

We are now ready to present our final results for $\mu^-$-$e^-$ conversion, which are displayed in Fig.~\ref{fig:Xi_nonphotonic}. In contrast to Fig.~\ref{fig:Xi_photonic}, we now present both the total contribution [$\Xi_{\rm combined}$, cf.\ Eq.~\eqref{eq:BR-total}] and the non-photonic/short-range contribution [$\Xi_\text{non-photonic}$, cf.\ Eq.~\eqref{eq:BR_nonphontic}]. Note that the latter quantity is in fact not physical, as explained, in the sense that in reality it does not occur in isolation, i.e.~without the long-range contributions. However, artificially separating them makes it evident that the short-range contributions are indeed very subdominant, by several orders of magnitude for each of the benchmark scenarios displayed. Thus, it is an excellent approximation to take $\Xi_{\rm combined} \simeq \Xi_{\rm photonic}$ and to completely disregard the short-range part, effectively going back to our intermediate result from Eqs.~\eqref{eq:Xi_simpl} and~\eqref{eq:mu-e_BR_long}. Furthermore, as explained above, the lines representing the non-photonic contributions for the different scenarios are in fact bands with finite widths, due to their dependence on the isotope under consideration. However, the widths are so small that they would hardly be visible in the logarithmic plot presented in Fig.~\ref{fig:Xi_nonphotonic}.

Furthermore, we extract the bounds on the scalar mass $M_S$ obtained from the combination of photonic and non-photonic contributions in analogy to Sec.~\ref{sec:mue_minus-nuclear}. The resulting ranges of lower limits displayed in Tab.~\ref{tab:non-photonic} differ from the values for the purely photonic contributions only at the per mille level, cf.\ Tab.~\ref{tab:photonic}. While this confirms that we can render the non-photonic contributions negligible, however, it is also visible that -- depending on the combinations of couplings -- the naive estimate of the effect of the non-photonic contributions may underestimate them by several orders of magnitude. Thus, it is in fact not a priori clear that the short-range diagrams are always negligible, contrary to what had been claimed earlier in Ref.~\cite{Raidal:1997hq}.

\begin{table}[t]
\centering
\hspace{-0.2cm}
\begin{tabular}{c||l|l|l}
& current limit [GeV] & future sensitivity [GeV] & COMET I (Al-27) [GeV]\\
\hline \hline
black curve & $M_S \textgreater 708.1 - 2388.6$ &  $M_S \textgreater 5497.1 - 70326.3$ & $M_S \textgreater 10396.1$ \\
\hline
blue curve & $M_S \textgreater 131.9 - 447.1$ &  $M_S \textgreater 1031.4 - 13269.4$ & $M_S \textgreater 1953.9$ \\
\hline
purple curve & $M_S \textgreater 42.5 - 152.2$&  $M_S \textgreater 360.6 - 4880.6$ & $M_S \textgreater 693.9$ \\
\hline
red curve & $M_S \textgreater 33.9 - 118.1$ &  $M_S \textgreater 276.3 - 3656.1$ & $M_S \textgreater 528.0$ \\
\hline
gray curve & $M_S \textgreater 4.1 - 15.9$ &  $M_S \textgreater 38.7 - 548.4$ & $M_S \textgreater 75.7$ \\
\end{tabular}
\caption{\label{tab:non-photonic}Lower limits on the mass $M_S$ resulting from the total branching ratio for $\mu$-$e$ conversion, displaying the range from the most pessimistic to the most optimistic values. Figures are deliberately shown with a too-good precision, in order to ease the comparison with Tab.~\ref{tab:photonic}. Indeed, the figures are nearly identical to those obtained when only taking into account the photonic contribution, just as to be expected from Fig.~\ref{fig:Xi_nonphotonic}.}
\end{table}

Although we can neglect the non-photonic contributions due to their smallness, there are two interesting observations related to them, which we want to briefly discuss. First, we cannot distinguish the blue from the purple non-photonic contributions, while they differ by about an order of magnitude in the photonic case. This can again be understood by having a close look at the amplitudes for both processes. The amplitude that enters the non-photonic $\Xi_\text{non-photonic}$ takes the form:
\begin{equation*}
 \mathcal{A}\propto \big|f^*_{ee}\,f_{e\mu}\,D(m_e)+f^*_{e\mu}\,f_{\mu\mu}\,D(m_\mu)+f^*_{e\tau}\,f_{\tau\mu}\,D(m_\tau)\big|\,,
\end{equation*}
where the function $D(m_a)$, which is proportional to the form factor $A_R$ for a fixed $m_a$, strongly varies with $m_a$. The dominant term (without including the couplings $f^*_{e a}f_{a\mu}$) stems from the $\tau$ propagating in the loop, i.e.\ $D(m_\tau)$. It exceeds the $\mu$ and $e$ contributions by about three to four orders of magnitude. Furthermore, neither the combination $f^*_{ee}f_{e\mu}$ nor $f^*_{e\mu}f_{\mu\mu}$, see Tab.~\ref{Tab:Coupling_Scenarios}, can bypass this difference in the blue and purple scenarios. Thus, the equality of the non-photonic contribution of blue and purple scenarios is traced back to the identical combination of $f^*_{e\tau}f_{\tau\mu}$ in both scenarios.

The second observation is that -- in contrast to the photonic case where the red scenario consistently attains values more than an order of magnitude higher -- the red and gray scenarios are comparable in the non-photonic case. Following the argument given above, the gray scenario should dominate, due to $f^*_{e\tau}f_{\tau\mu}=10^{-8}$ (gray) in comparison to $f^*_{e\tau}f_{\tau\mu}=10^{-24}$ (red), which seems to contradict the observations from the plot. However, for the red scenario, $f^*_{e\tau}f_{\tau\mu}$ is smaller than the combinations $f^*_{ee}f_{e\mu}$ and $f^*_{e\mu}f_{\mu\mu}$ by at least six orders of magnitude, see Tab.~\ref{Tab:Coupling_Scenarios}. It hence overcompensates the dominance of $D(m_\tau)$ such that $f^*_{e\mu}\,f_{\mu\mu}\,D(m_\mu)$ is the relevant contribution in the red scenario. The latter yields the same order of magnitude results as the $f^*_{e\tau}\,f_{\tau\mu}\,D(m_\tau)$ contribution of the gray scenario.\\

Summing up, we have presented a detailed computation of $\mu$-$e$ conversion mediated by a doubly charged $SU(2)$ singlet scalar coupling to pairs of right-handed charged leptons. The formulae obtained are general, however, for illustration the numerical results focus on the scenarios obtained in Ref.~\cite{King:2014uha}. In all cases, the current/future lower bounds on the doubly charged scalar mass $M_S$ resulting from the non-observation of $\mu$-$e$ conversion turn out to be very strong, which illustrates the value of new measurements of $\mu$-$e$ conversion.

\section{\label{sec:conc}Conclusions and Outlook}

In this paper, we have presented the first detailed computation of $\mu$-$e$ conversion, i.e., a reaction turning a muon bound to a nucleus into an electron, for the case of the process being mediated by a doubly charged singlet scalar particle. After having identified the decisive Feynman diagrams, we have computed the resulting amplitude for the conversion and we have mapped it to the known most general amplitude for the process. We have taken into account both the long-range and short-range contributions, the latter of which are however subdominant and can be neglected in practice. Our results are fully general and hold for any doubly charged singlet scalar coupling to pairs of right-handed charged leptons, thereby closing a big gap in our current knowledge on $\mu$-$e$ conversion. Even for doubly charged scalars which are no singlets under $SU(2)$, such as the doubly charged component of a Higgs triplet field, most of the computation presented practically stays the same -- a generalisation of our results is both possible and doable with moderate effort.

In addition, we have investigated how strongly the parameters related to the doubly charged scalar can be constrained by future experimental limits on the process, which are expected to dramatically improve within the coming years. For illustrative purposes, we have also included an explicit example of a model that generates a valid light neutrino mass at 2-loop level and which contains our general setting as a subset. As we have seen, despite intrinsic nuclear physics uncertainties, the limits to be expected strongly constrain the mass of the doubly charged scalar, so much so that future indirect limits from $\mu$-$e$ conversion are even likely to be more stringent than the direct limits which will be obtained by the LHC. Thus, realistically, experiments on lepton flavour violation can serve as a valuable addition to collider studies in the hunt for new physics beyond the SM.

\section*{Acknowledgements}

We would like to thank S.~F.~King, J.~M.~No, L.~Panizzi, and K.~Zuber for useful discussions and comments. AM acknowledges partial support by the European Union Grant No.~FP7 ITN-INVISIBLES (Marie Curie Actions, Grant No.~PITN-GA-2011-289442) and by the Micron Technology Foundation, Inc.~.\\
\\
\paragraph{Note added:} As this paper was being finalised a related paper~\cite{Chakrabortty:2015zpm} appeared, which focuses more on the muon $(g-2)$ computation, but also treats low-energy LFV, in particular $\mu$-$e$ conversion, and collider phenomenology. Our paper however gives much more details on the computation of $\mu$-$e$ conversion, and in particular it allows to reproduce our results and to adopt them to similar cases. Thus, Ref.~\cite{Chakrabortty:2015zpm} and the present paper complement each other.

\appendix

\section{\label{app:FeynmanRules}Appendix: Feynman Rules}

\begin{figure}[h]
  \begin{minipage}[c]{14cm}
    \begin{subfigure}[c]{7cm}
    \includegraphics[width=7cm]{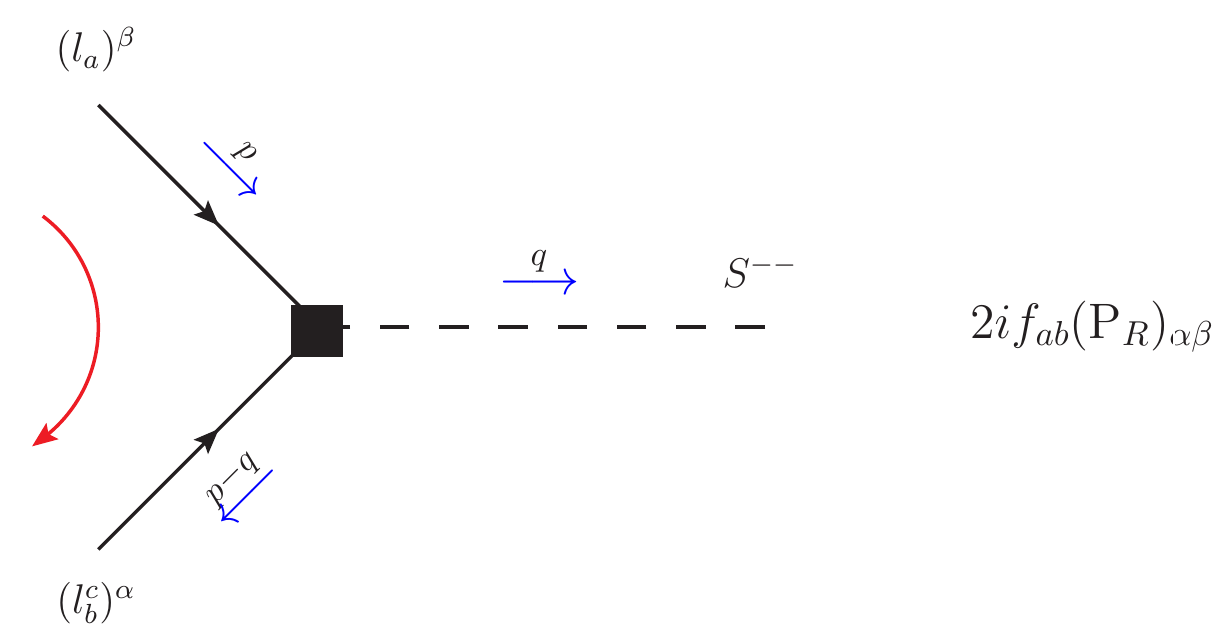}
    \caption{Outgoing $S^{--}$}
    \label{fig:FeynRule1a}
    \end{subfigure}
    \hspace{1cm}
   \begin{subfigure}[c]{7cm}
    \includegraphics[width=7cm]{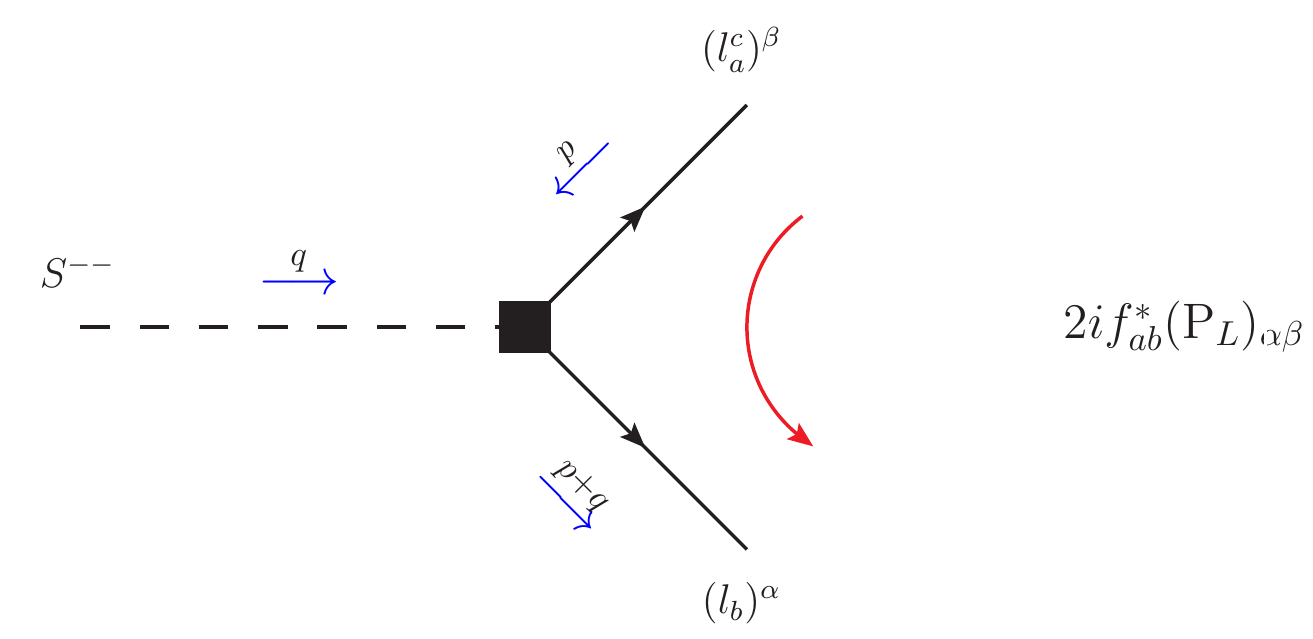}
    \caption{Incoming $S^{--}$}
    \label{fig:FeynRule1b}
     \end{subfigure}  
  \end{minipage}
    \caption{Two-lepton and $S^{--}$ interactions with $f^{(*)}_{ab}=f^{(*)}_{ba}$.}
    \label{fig:FeynRule1}
  \end{figure}

\begin{figure}[h]
  \begin{minipage}[c]{14cm}
     \begin{subfigure}[c]{7cm}
    \includegraphics[width=7cm]{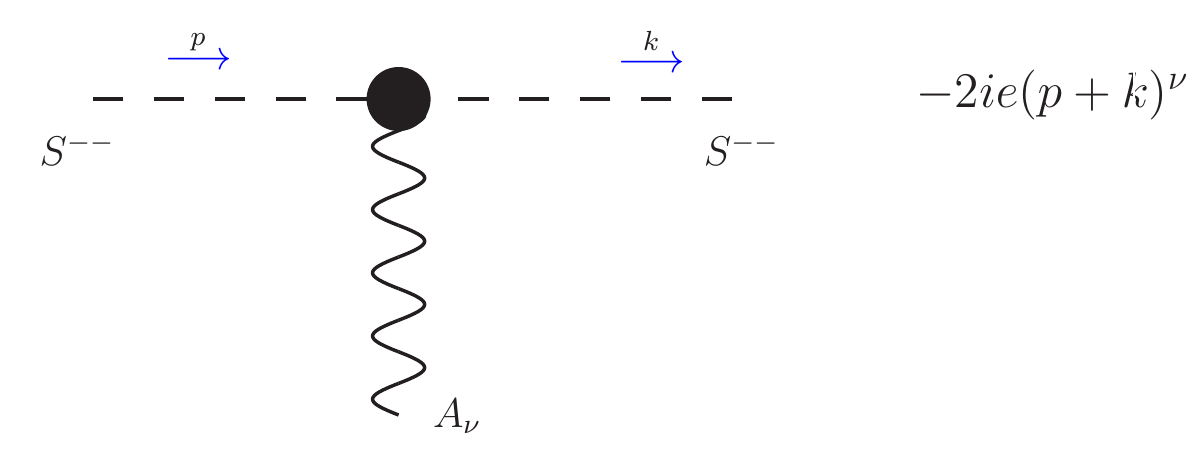}
    \caption{3-point vertex: two doubly charged scalars and photon}
    \label{fig:FeynRule2b}
     \end{subfigure}
     \hspace{1cm}
    \begin{subfigure}[c]{7cm}
    \includegraphics[width=7cm]{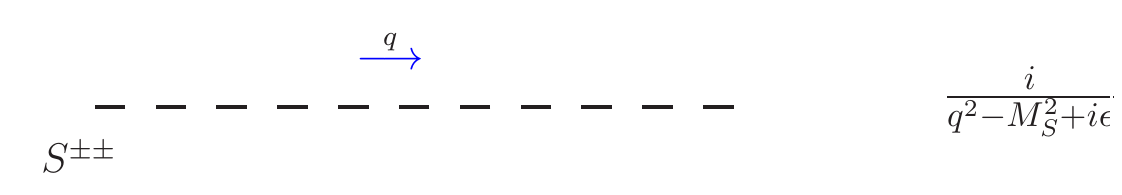}
    \caption{$S^{--}$ propagator}
    \label{fig:FeynRule2a}
    \end{subfigure}  
    \begin{subfigure}[c]{7cm}
    \includegraphics[width=7cm]{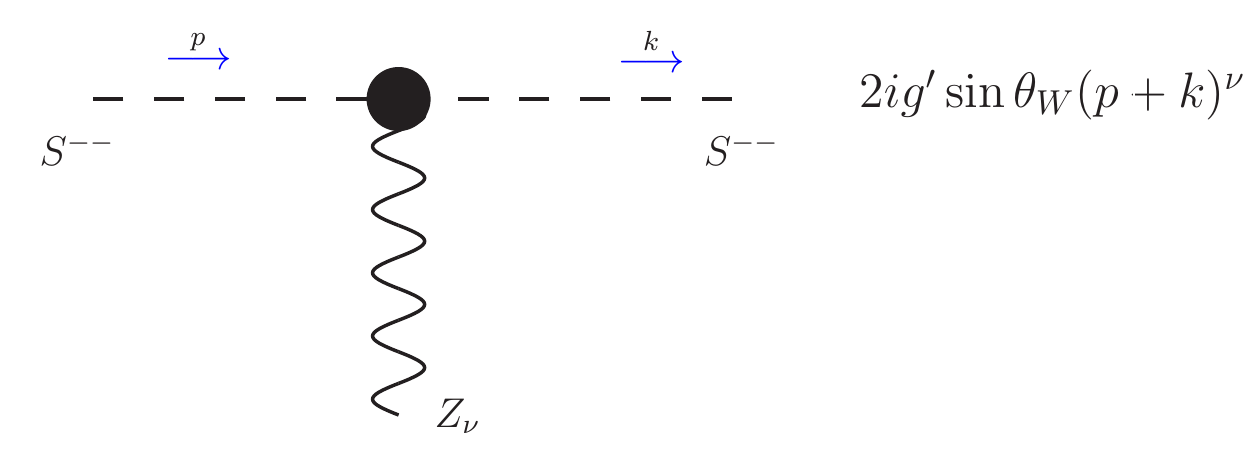}
    \caption{3-point vertex: two doubly charged scalars and $Z$-boson}
    \label{fig:FeynRule2c}
     \end{subfigure}
  \end{minipage}
    \caption{$S^{--}$ and its interaction with neutral gauge bosons.}
    \label{fig:FeynRule2}
  \end{figure}

\begin{figure}[h]
  \begin{minipage}[c]{14cm}
    \begin{subfigure}[c]{7cm}
    \includegraphics[width=7cm]{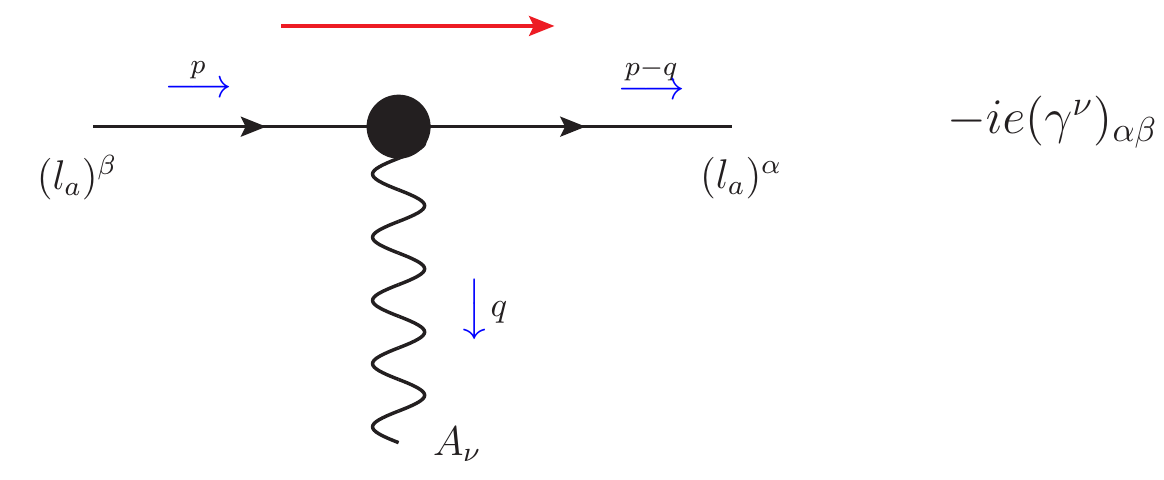}
    \caption{Usual orientation of 'fermion flow'}
    \label{fig:FeynRule3a}
    \end{subfigure}
    \hspace{1cm}
   \begin{subfigure}[c]{7cm}
    \includegraphics[width=7cm]{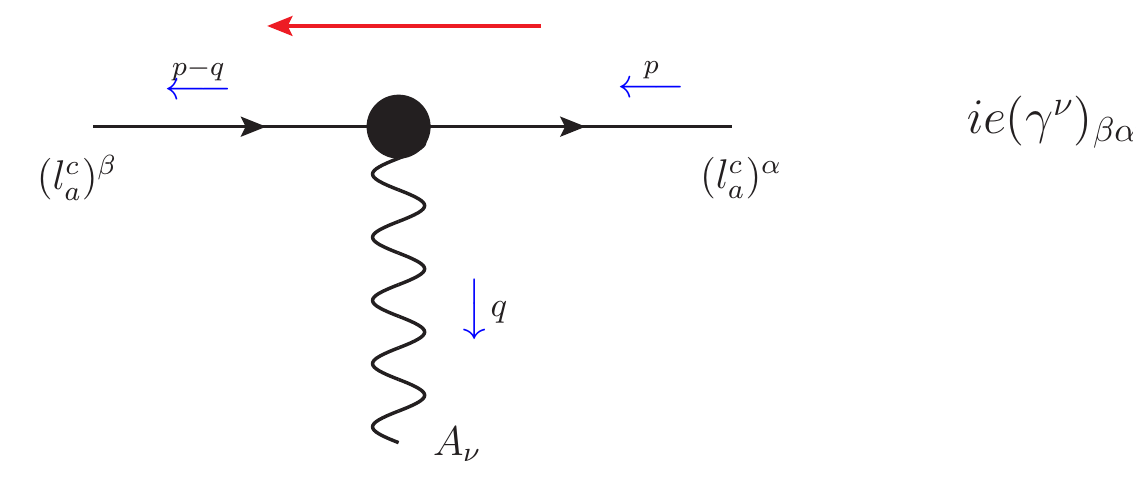}
    \caption{Reversed 'fermion flow'}
    \label{fig:FeynRule3b}
     \end{subfigure}  
  \end{minipage}
    \caption{Electromagnetic vertex.}
    \label{fig:FeynRule3}
  \end{figure}

\begin{figure}[h!]
  \begin{minipage}[c]{14cm}
    \begin{subfigure}[c]{7cm}
    \includegraphics[width=7cm]{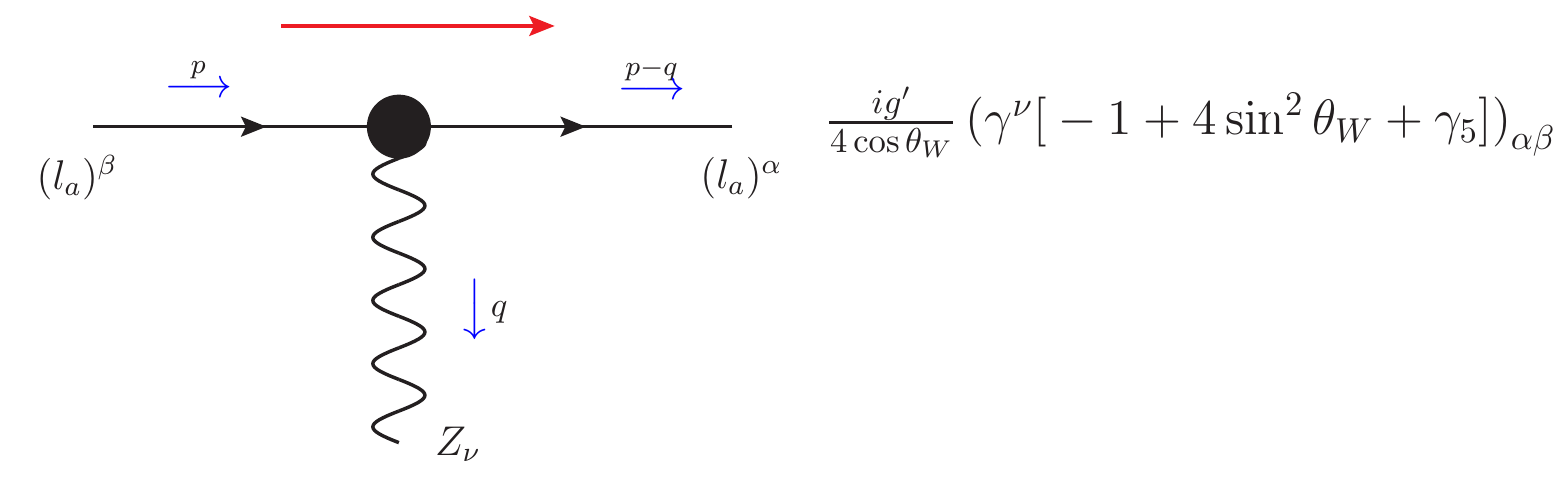}
    \caption{Usual orientation of 'fermion flow'}
    \label{fig:FeynRule3c}
    \end{subfigure}
    \hspace{1cm}
   \begin{subfigure}[c]{7cm}
    \includegraphics[width=7cm]{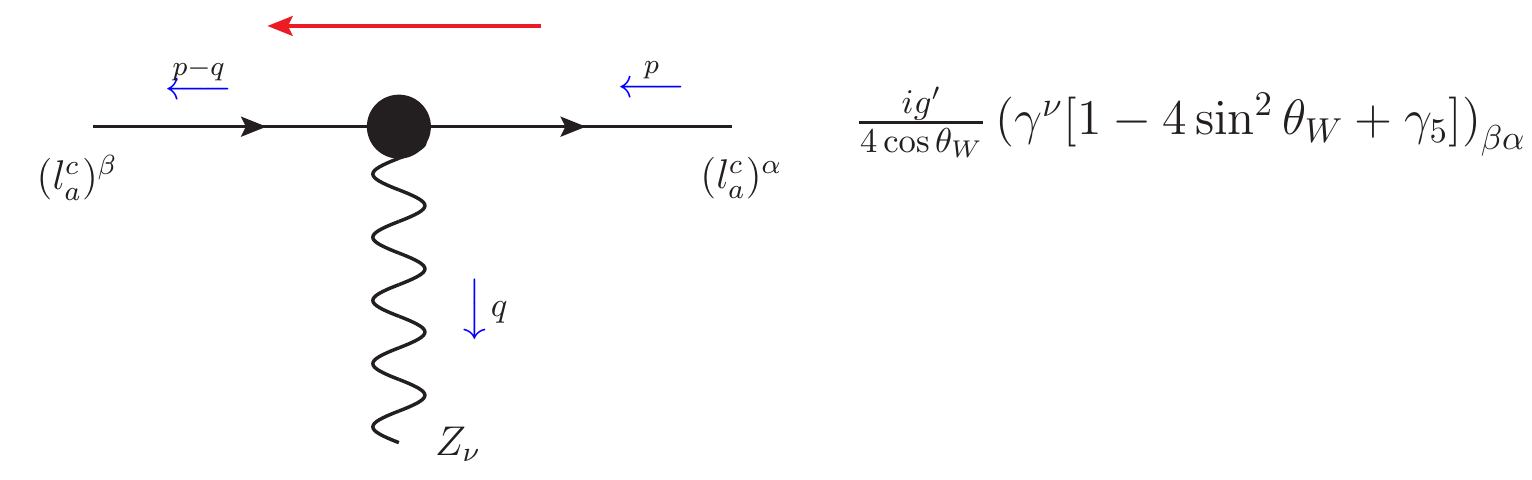}
    \caption{Reversed 'fermion flow'}
    \label{fig:FeynRule3c}
     \end{subfigure}  
  \end{minipage}
    \caption{$Z$-boson vertex.}
    \label{fig:FeynRule3bc}
  \end{figure}

\begin{figure}[h!]
  \begin{minipage}[c]{14cm}
    \begin{subfigure}[c]{7cm}
    \includegraphics[width=7cm]{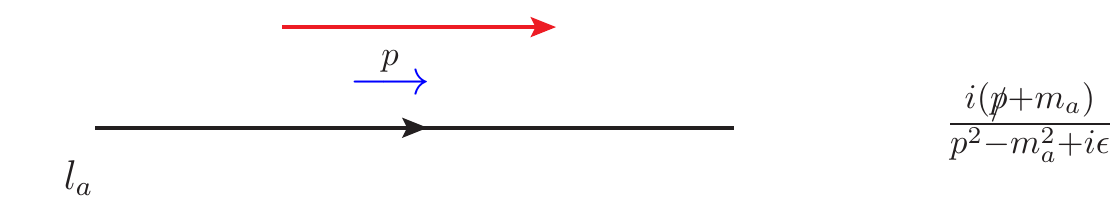}
    \caption{Usual orientation of 'fermion flow' for a propagating particle}
    \label{fig:FeynRule4a}
    \end{subfigure}
   \hspace{1cm}
   \begin{subfigure}[c]{7cm}
    \includegraphics[width=7cm]{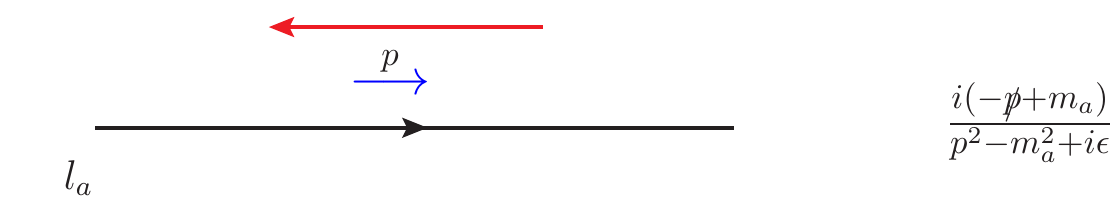}
    \caption{Reversed 'fermion flow' for a particle}
    \label{fig:FeynRule4b}
     \end{subfigure}  
     \begin{subfigure}[c]{7cm}
    \includegraphics[width=7cm]{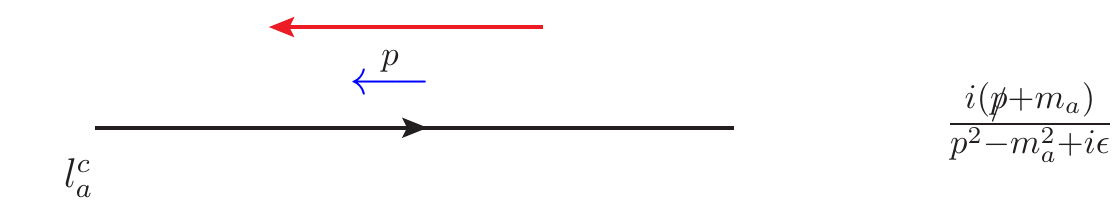}
    \caption{Reversed 'fermion flow' for an antiparticle }
    \label{fig:FeynRule4c}
     \end{subfigure}  
  \end{minipage}
    \caption{(Anti-) leptonic propagator and its alteration with the 'fermion flow'.}
    \label{fig:FeynRule4}
  \end{figure}

In order to obtain the decisive matrix elements in Sections \ref{sec:mue_minus-matching} and \ref{sec:mue_minus-non-photonic_form}, we make use of the Feynman rules given in Figs.~\ref{fig:FeynRule1} to \ref{fig:FeynRule4}. Here, $P_{L,R}$ are the left-/right-handed projectors, the indices $\alpha,\,\beta$ are Dirac spinor indices, and $a,\,b=e,\,\mu,\,\tau$ denote the lepton flavour. The doubly charged scalar's interactions are described by means of the covariant derivative $D_\mu=\partial_\mu +i g' Y B_\mu$. The hypercharge is given by $Y=Q-I_3$ ($=\pm 2$ for $S^{\pm \pm}$), such that the covariant derivative takes the form $D_\mu=\partial_\mu\pm 2i e A_\mu\mp 2i g' \sin \theta_W Z_\mu$. Note that, since there are lepton number violating (LNV) vertices in our theory, we naturally encounter vertices with clashing arrows. For a consistent treatment using the Feynman rule language, we choose a fixed orientation of the ``fermion flow'' for each diagram, i.e.\ the order in which each fermionic chain is written down, and adjust the Feynman rules~\cite{Denner:1992vza}. For example, when reversing the ``fermion flow'' from Fig.~\ref{fig:FeynRule3a} to the opposite direction as displayed in Fig.~\ref{fig:FeynRule3b}, we instead work with the anti-field $l^c_a=C\,\overline{l_a}^T$ and alter the Feynman rules accordingly. In Figs.~\ref{fig:FeynRule1} to \ref{fig:FeynRule4}, the red arrow indicates the orientation of the ``fermion flow'', i.e., of lepton number.

\clearpage

\section{\label{app:Passarino}Appendix: The scalar three-point function}

\begin{figure}[h!]
\centering
 \includegraphics[width=0.35\textwidth]{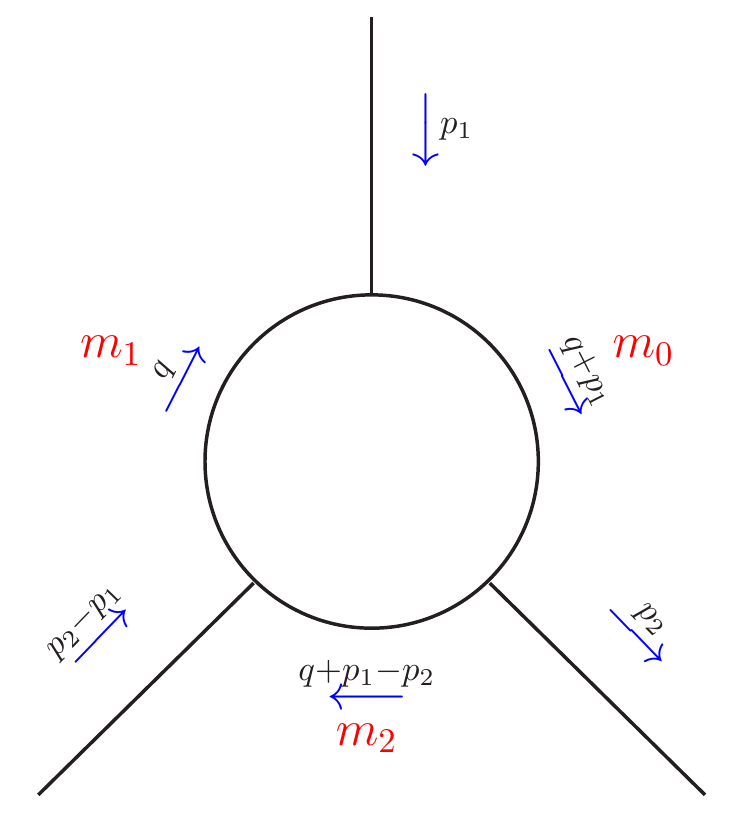}
 \caption{Kinematic set-up corresponding to the scalar three-point function Eq.~\eqref{eq:C_0}.}
 \label{fig:C0_Notation}
\end{figure}

The kinematical configuration corresponding to the scalar three-point function given in Eq.~\eqref{eq:C_0} is displayed in Fig.~\ref{fig:C0_Notation}. 


\clearpage
\bibliographystyle{./apsrev}
\bibliography{Nu_LHC}

\begin{thebibliography}{10}
\expandafter\ifx\csname bibnamefont\endcsname\relax
  \def\bibnamefont#1{#1}\fi
\expandafter\ifx\csname bibfnamefont\endcsname\relax
  \def\bibfnamefont#1{#1}\fi
\expandafter\ifx\csname url\endcsname\relax
  \def\url#1{\texttt{#1}}\fi
\expandafter\ifx\csname urlprefix\endcsname\relax\def\urlprefix{URL }\fi
\providecommand{\bibinfo}[2]{#2}
\providecommand{\eprint}[2][]{\url{#2}}

\bibitem{Gorringe:2015cma}
\bibinfo{author}{\bibfnamefont{T.~P.} \bibnamefont{Gorringe}} \bibnamefont{and}
  \bibinfo{author}{\bibfnamefont{D.~W.} \bibnamefont{Hertzog}},
  \bibinfo{journal}{Prog. Part. Nucl. Phys.} \textbf{\bibinfo{volume}{84}},
  \bibinfo{pages}{73} (\bibinfo{year}{2015}), \eprint{1506.01465}.

\bibitem{Barbieri:1987fn}
\bibinfo{author}{\bibfnamefont{R.}~\bibnamefont{Barbieri}} \bibnamefont{and}
  \bibinfo{author}{\bibfnamefont{G.~F.} \bibnamefont{Giudice}},
  \bibinfo{journal}{Nucl. Phys.} \textbf{\bibinfo{volume}{B306}},
  \bibinfo{pages}{63} (\bibinfo{year}{1988}).

\bibitem{Peccei:1977hh}
\bibinfo{author}{\bibfnamefont{R.~D.} \bibnamefont{Peccei}} \bibnamefont{and}
  \bibinfo{author}{\bibfnamefont{H.~R.} \bibnamefont{Quinn}},
  \bibinfo{journal}{Phys. Rev. Lett.} \textbf{\bibinfo{volume}{38}},
  \bibinfo{pages}{1440} (\bibinfo{year}{1977}).

\bibitem{Bertone:2004pz}
\bibinfo{author}{\bibfnamefont{G.}~\bibnamefont{Bertone}},
  \bibinfo{author}{\bibfnamefont{D.}~\bibnamefont{Hooper}}, \bibnamefont{and}
  \bibinfo{author}{\bibfnamefont{J.}~\bibnamefont{Silk}},
  \bibinfo{journal}{Phys. Rept.} \textbf{\bibinfo{volume}{405}},
  \bibinfo{pages}{279} (\bibinfo{year}{2005}), \eprint{hep-ph/0404175}.

\bibitem{Mohapatra:2005wg}
\bibinfo{author}{\bibfnamefont{R.~N.} \bibnamefont{Mohapatra}} \emph{et~al.},
  \bibinfo{journal}{Rept. Prog. Phys.} \textbf{\bibinfo{volume}{70}},
  \bibinfo{pages}{1757} (\bibinfo{year}{2007}), \eprint{hep-ph/0510213}.

\bibitem{Fukuda:1998mi}
\bibinfo{author}{\bibfnamefont{Y.}~\bibnamefont{Fukuda}} \emph{et~al.}
  (\bibinfo{collaboration}{Super-Kamiokande}), \bibinfo{journal}{Phys. Rev.
  Lett.} \textbf{\bibinfo{volume}{81}}, \bibinfo{pages}{1562}
  (\bibinfo{year}{1998}), \eprint{hep-ex/9807003}.

\bibitem{Ahmad:2002jz}
\bibinfo{author}{\bibfnamefont{Q.~R.} \bibnamefont{Ahmad}} \emph{et~al.}
  (\bibinfo{collaboration}{SNO}), \bibinfo{journal}{Phys. Rev. Lett.}
  \textbf{\bibinfo{volume}{89}}, \bibinfo{pages}{011301}
  (\bibinfo{year}{2002}), \eprint{nucl-ex/0204008}.

\bibitem{Araki:2004mb}
\bibinfo{author}{\bibfnamefont{T.}~\bibnamefont{Araki}} \emph{et~al.}
  (\bibinfo{collaboration}{KamLAND}), \bibinfo{journal}{Phys. Rev. Lett.}
  \textbf{\bibinfo{volume}{94}}, \bibinfo{pages}{081801}
  (\bibinfo{year}{2005}), \eprint{hep-ex/0406035}.

\bibitem{Michael:2006rx}
\bibinfo{author}{\bibfnamefont{D.~G.} \bibnamefont{Michael}} \emph{et~al.}
  (\bibinfo{collaboration}{MINOS}), \bibinfo{journal}{Phys. Rev. Lett.}
  \textbf{\bibinfo{volume}{97}}, \bibinfo{pages}{191801}
  (\bibinfo{year}{2006}), \eprint{hep-ex/0607088}.

\bibitem{An:2012eh}
\bibinfo{author}{\bibfnamefont{F.~P.} \bibnamefont{An}} \emph{et~al.}
  (\bibinfo{collaboration}{Daya Bay}), \bibinfo{journal}{Phys. Rev. Lett.}
  \textbf{\bibinfo{volume}{108}}, \bibinfo{pages}{171803}
  (\bibinfo{year}{2012}), \eprint{1203.1669}.

\bibitem{Ahn:2012nd}
\bibinfo{author}{\bibfnamefont{J.~K.} \bibnamefont{Ahn}} \emph{et~al.}
  (\bibinfo{collaboration}{RENO}), \bibinfo{journal}{Phys. Rev. Lett.}
  \textbf{\bibinfo{volume}{108}}, \bibinfo{pages}{191802}
  (\bibinfo{year}{2012}), \eprint{1204.0626}.

\bibitem{Abe:2011sj}
\bibinfo{author}{\bibfnamefont{K.}~\bibnamefont{Abe}} \emph{et~al.}
  (\bibinfo{collaboration}{T2K}), \bibinfo{journal}{Phys. Rev. Lett.}
  \textbf{\bibinfo{volume}{107}}, \bibinfo{pages}{041801}
  (\bibinfo{year}{2011}), \eprint{1106.2822}.

\bibitem{Abe:2011fz}
\bibinfo{author}{\bibfnamefont{Y.}~\bibnamefont{Abe}} \emph{et~al.}
  (\bibinfo{collaboration}{Double Chooz}), \bibinfo{journal}{Phys. Rev. Lett.}
  \textbf{\bibinfo{volume}{108}}, \bibinfo{pages}{131801}
  (\bibinfo{year}{2012}), \eprint{1112.6353}.

\bibitem{Adam:2013mnn}
\bibinfo{author}{\bibfnamefont{J.}~\bibnamefont{Adam}} \emph{et~al.}
  (\bibinfo{collaboration}{MEG}), \bibinfo{journal}{Phys. Rev. Lett.}
  \textbf{\bibinfo{volume}{110}}, \bibinfo{pages}{201801}
  (\bibinfo{year}{2013}), \eprint{1303.0754}.

\bibitem{Aubert:2009ag}
\bibinfo{author}{\bibfnamefont{B.}~\bibnamefont{Aubert}} \emph{et~al.}
  (\bibinfo{collaboration}{BaBar}), \bibinfo{journal}{Phys. Rev. Lett.}
  \textbf{\bibinfo{volume}{104}}, \bibinfo{pages}{021802}
  (\bibinfo{year}{2010}), \eprint{0908.2381}.

\bibitem{ChengLi}
\bibinfo{author}{\bibfnamefont{T.~P.} \bibnamefont{Cheng}} \bibnamefont{and}
  \bibinfo{author}{\bibfnamefont{L.~F.} \bibnamefont{Li}},
  \emph{\bibinfo{title}{{Gauge Theory Of Elementary Particle Physics}}}
  (\bibinfo{publisher}{Oxford Science Publications}, \bibinfo{address}{Oxford,
  UK}, \bibinfo{year}{1984}).

\bibitem{Glashow:1970gm}
\bibinfo{author}{\bibfnamefont{S.~L.} \bibnamefont{Glashow}},
  \bibinfo{author}{\bibfnamefont{J.}~\bibnamefont{Iliopoulos}},
  \bibnamefont{and} \bibinfo{author}{\bibfnamefont{L.}~\bibnamefont{Maiani}},
  \bibinfo{journal}{Phys. Rev.} \textbf{\bibinfo{volume}{D2}},
  \bibinfo{pages}{1285} (\bibinfo{year}{1970}).

\bibitem{Cheng:1980tp}
\bibinfo{author}{\bibfnamefont{T.~P.} \bibnamefont{Cheng}} \bibnamefont{and}
  \bibinfo{author}{\bibfnamefont{L.-F.} \bibnamefont{Li}},
  \bibinfo{journal}{Phys. Rev. Lett.} \textbf{\bibinfo{volume}{45}},
  \bibinfo{pages}{1908} (\bibinfo{year}{1980}).

\bibitem{Cheng:1980qt}
\bibinfo{author}{\bibfnamefont{T.~P.} \bibnamefont{Cheng}} \bibnamefont{and}
  \bibinfo{author}{\bibfnamefont{L.-F.} \bibnamefont{Li}},
  \bibinfo{journal}{Phys. Rev.} \textbf{\bibinfo{volume}{D22}},
  \bibinfo{pages}{2860} (\bibinfo{year}{1980}).

\bibitem{Blum:2007he}
\bibinfo{author}{\bibfnamefont{A.}~\bibnamefont{Blum}} \bibnamefont{and}
  \bibinfo{author}{\bibfnamefont{A.}~\bibnamefont{Merle}},
  \bibinfo{journal}{Phys. Rev.} \textbf{\bibinfo{volume}{D77}},
  \bibinfo{pages}{076005} (\bibinfo{year}{2008}), \eprint{0709.3294}.

\bibitem{Bellgardt:1987du}
\bibinfo{author}{\bibfnamefont{U.}~\bibnamefont{Bellgardt}} \emph{et~al.}
  (\bibinfo{collaboration}{SINDRUM}), \bibinfo{journal}{Nucl. Phys.}
  \textbf{\bibinfo{volume}{B299}}, \bibinfo{pages}{1} (\bibinfo{year}{1988}).

\bibitem{Hayasaka:2010np}
\bibinfo{author}{\bibfnamefont{K.}~\bibnamefont{Hayasaka}} \emph{et~al.},
  \bibinfo{journal}{Phys. Lett.} \textbf{\bibinfo{volume}{B687}},
  \bibinfo{pages}{139} (\bibinfo{year}{2010}), \eprint{1001.3221}.

\bibitem{Raidal:2008jk}
\bibinfo{author}{\bibfnamefont{M.}~\bibnamefont{Raidal}} \emph{et~al.},
  \bibinfo{journal}{Eur. Phys. J.} \textbf{\bibinfo{volume}{C57}},
  \bibinfo{pages}{13} (\bibinfo{year}{2008}), \eprint{0801.1826}.

\bibitem{Weinberg:1959zz}
\bibinfo{author}{\bibfnamefont{S.}~\bibnamefont{Weinberg}} \bibnamefont{and}
  \bibinfo{author}{\bibfnamefont{G.}~\bibnamefont{Feinberg}},
  \bibinfo{journal}{Phys. Rev. Lett.} \textbf{\bibinfo{volume}{3}},
  \bibinfo{pages}{111} (\bibinfo{year}{1959}).

\bibitem{Marciano:1977cj}
\bibinfo{author}{\bibfnamefont{W.~J.} \bibnamefont{Marciano}} \bibnamefont{and}
  \bibinfo{author}{\bibfnamefont{A.~I.} \bibnamefont{Sanda}},
  \bibinfo{journal}{Phys. Rev. Lett.} \textbf{\bibinfo{volume}{38}},
  \bibinfo{pages}{1512} (\bibinfo{year}{1977}).

\bibitem{Dinh:2012bp}
\bibinfo{author}{\bibfnamefont{D.~N.} \bibnamefont{Dinh}},
  \bibinfo{author}{\bibfnamefont{A.}~\bibnamefont{Ibarra}},
  \bibinfo{author}{\bibfnamefont{E.}~\bibnamefont{Molinaro}}, \bibnamefont{and}
  \bibinfo{author}{\bibfnamefont{S.~T.} \bibnamefont{Petcov}},
  \bibinfo{journal}{JHEP} \textbf{\bibinfo{volume}{08}}, \bibinfo{pages}{125}
  (\bibinfo{year}{2012}), \bibinfo{note}{[Erratum: JHEP09,023(2013)]},
  \eprint{1205.4671}.

\bibitem{Bernabeu:1993ta}
\bibinfo{author}{\bibfnamefont{J.}~\bibnamefont{Bernabeu}},
  \bibinfo{author}{\bibfnamefont{E.}~\bibnamefont{Nardi}}, \bibnamefont{and}
  \bibinfo{author}{\bibfnamefont{D.}~\bibnamefont{Tommasini}},
  \bibinfo{journal}{Nucl. Phys.} \textbf{\bibinfo{volume}{B409}},
  \bibinfo{pages}{69} (\bibinfo{year}{1993}), \eprint{hep-ph/9306251}.

\bibitem{Crivellin:2014cta}
\bibinfo{author}{\bibfnamefont{A.}~\bibnamefont{Crivellin}},
  \bibinfo{author}{\bibfnamefont{M.}~\bibnamefont{Hoferichter}},
  \bibnamefont{and} \bibinfo{author}{\bibfnamefont{M.}~\bibnamefont{Procura}},
  \bibinfo{journal}{Phys. Rev.} \textbf{\bibinfo{volume}{D89}},
  \bibinfo{pages}{093024} (\bibinfo{year}{2014}), \eprint{1404.7134}.

\bibitem{Frank:2000sn}
\bibinfo{author}{\bibfnamefont{M.}~\bibnamefont{Frank}}, \bibinfo{journal}{Eur.
  Phys. J.} \textbf{\bibinfo{volume}{C17}}, \bibinfo{pages}{501}
  (\bibinfo{year}{2000}).

\bibitem{Arganda:2007jw}
\bibinfo{author}{\bibfnamefont{E.}~\bibnamefont{Arganda}},
  \bibinfo{author}{\bibfnamefont{M.~J.} \bibnamefont{Herrero}},
  \bibnamefont{and} \bibinfo{author}{\bibfnamefont{A.~M.}
  \bibnamefont{Teixeira}}, \bibinfo{journal}{JHEP}
  \textbf{\bibinfo{volume}{10}}, \bibinfo{pages}{104} (\bibinfo{year}{2007}),
  \eprint{0707.2955}.

\bibitem{Zhang:2013jva}
\bibinfo{author}{\bibfnamefont{H.-B.} \bibnamefont{Zhang}},
  \bibinfo{author}{\bibfnamefont{T.-F.} \bibnamefont{Feng}},
  \bibinfo{author}{\bibfnamefont{G.-H.} \bibnamefont{Luo}},
  \bibinfo{author}{\bibfnamefont{Z.-F.} \bibnamefont{Ge}}, \bibnamefont{and}
  \bibinfo{author}{\bibfnamefont{S.-M.} \bibnamefont{Zhao}},
  \bibinfo{journal}{JHEP} \textbf{\bibinfo{volume}{07}}, \bibinfo{pages}{069}
  (\bibinfo{year}{2013}), \bibinfo{note}{[Erratum: JHEP10,173(2013)]},
  \eprint{1305.4352}.

\bibitem{Raidal:1997hq}
\bibinfo{author}{\bibfnamefont{M.}~\bibnamefont{Raidal}} \bibnamefont{and}
  \bibinfo{author}{\bibfnamefont{A.}~\bibnamefont{Santamaria}},
  \bibinfo{journal}{Phys. Lett.} \textbf{\bibinfo{volume}{B421}},
  \bibinfo{pages}{250} (\bibinfo{year}{1998}), \eprint{hep-ph/9710389}.

\bibitem{King:2014uha}
\bibinfo{author}{\bibfnamefont{S.~F.} \bibnamefont{King}},
  \bibinfo{author}{\bibfnamefont{A.}~\bibnamefont{Merle}}, \bibnamefont{and}
  \bibinfo{author}{\bibfnamefont{L.}~\bibnamefont{Panizzi}},
  \bibinfo{journal}{JHEP} \textbf{\bibinfo{volume}{1411}}, \bibinfo{pages}{124}
  (\bibinfo{year}{2014}), \eprint{1406.4137}.

\bibitem{Geib:2015tvt}
\bibinfo{author}{\bibfnamefont{T.}~\bibnamefont{Geib}},
  \bibinfo{author}{\bibfnamefont{S.~F.} \bibnamefont{King}},
  \bibinfo{author}{\bibfnamefont{A.}~\bibnamefont{Merle}},
  \bibinfo{author}{\bibfnamefont{J.~M.} \bibnamefont{No}}, \bibnamefont{and}
  \bibinfo{author}{\bibfnamefont{L.}~\bibnamefont{Panizzi}},
  \bibinfo{journal}{Phys. Rev.}
  \textbf{\bibinfo{volume}{D93}}(\bibinfo{number}{7}), \bibinfo{pages}{073007}
  (\bibinfo{year}{2016}), \eprint{1512.04391}.

\bibitem{Geib:2016atx}
\bibinfo{author}{\bibfnamefont{T.}~\bibnamefont{Geib}},
  \bibinfo{author}{\bibfnamefont{A.}~\bibnamefont{Merle}}, \bibnamefont{and}
  \bibinfo{author}{\bibfnamefont{K.}~\bibnamefont{Zuber}}
  (\bibinfo{year}{2016}), \eprint{1609.09088}.

\bibitem{Kuno:1999jp}
\bibinfo{author}{\bibfnamefont{Y.}~\bibnamefont{Kuno}} \bibnamefont{and}
  \bibinfo{author}{\bibfnamefont{Y.}~\bibnamefont{Okada}},
  \bibinfo{journal}{Rev. Mod. Phys.} \textbf{\bibinfo{volume}{73}},
  \bibinfo{pages}{151} (\bibinfo{year}{2001}), \eprint{hep-ph/9909265}.

\bibitem{Kitano:2002mt}
\bibinfo{author}{\bibfnamefont{R.}~\bibnamefont{Kitano}},
  \bibinfo{author}{\bibfnamefont{M.}~\bibnamefont{Koike}}, \bibnamefont{and}
  \bibinfo{author}{\bibfnamefont{Y.}~\bibnamefont{Okada}},
  \bibinfo{journal}{Phys. Rev.} \textbf{\bibinfo{volume}{D66}},
  \bibinfo{pages}{096002} (\bibinfo{year}{2002}), \eprint{hep-ph/0203110}.

\bibitem{Lavoura:2003xp}
\bibinfo{author}{\bibfnamefont{L.}~\bibnamefont{Lavoura}},
  \bibinfo{journal}{Eur. Phys. J.} \textbf{\bibinfo{volume}{C29}},
  \bibinfo{pages}{191} (\bibinfo{year}{2003}), \eprint{hep-ph/0302221}.

\bibitem{Patel:2015tea}
\bibinfo{author}{\bibfnamefont{H.~H.} \bibnamefont{Patel}},
  \bibinfo{journal}{Comput. Phys. Commun.} \textbf{\bibinfo{volume}{197}},
  \bibinfo{pages}{276} (\bibinfo{year}{2015}), \eprint{1503.01469}.

\bibitem{Berestetsky:1982aq}
\bibinfo{author}{\bibfnamefont{V.}~\bibnamefont{Berestetsky}},
  \bibinfo{author}{\bibfnamefont{E.}~\bibnamefont{Lifshitz}}, \bibnamefont{and}
  \bibinfo{author}{\bibfnamefont{L.}~\bibnamefont{Pitaevsky}},
  \emph{\bibinfo{title}{{Quantum Electrodynamics}}}
  (\bibinfo{publisher}{Pergamon Press}, \bibinfo{address}{Oxford, UK},
  \bibinfo{year}{1982}).

\bibitem{Chiang:1993xz}
\bibinfo{author}{\bibfnamefont{H.~C.} \bibnamefont{Chiang}},
  \bibinfo{author}{\bibfnamefont{E.}~\bibnamefont{Oset}},
  \bibinfo{author}{\bibfnamefont{T.~S.} \bibnamefont{Kosmas}},
  \bibinfo{author}{\bibfnamefont{A.}~\bibnamefont{Faessler}}, \bibnamefont{and}
  \bibinfo{author}{\bibfnamefont{J.~D.} \bibnamefont{Vergados}},
  \bibinfo{journal}{Nucl. Phys.} \textbf{\bibinfo{volume}{A559}},
  \bibinfo{pages}{526} (\bibinfo{year}{1993}).

\bibitem{Shanker:1979ap}
\bibinfo{author}{\bibfnamefont{O.~U.} \bibnamefont{Shanker}},
  \bibinfo{journal}{Phys. Rev.} \textbf{\bibinfo{volume}{D20}},
  \bibinfo{pages}{1608} (\bibinfo{year}{1979}).

\bibitem{Alonso:2012ji}
\bibinfo{author}{\bibfnamefont{R.}~\bibnamefont{Alonso}},
  \bibinfo{author}{\bibfnamefont{M.}~\bibnamefont{Dhen}},
  \bibinfo{author}{\bibfnamefont{M.~B.} \bibnamefont{Gavela}},
  \bibnamefont{and} \bibinfo{author}{\bibfnamefont{T.}~\bibnamefont{Hambye}},
  \bibinfo{journal}{JHEP} \textbf{\bibinfo{volume}{01}}, \bibinfo{pages}{118}
  (\bibinfo{year}{2013}), \eprint{1209.2679}.

\bibitem{Passarino:1978jh}
\bibinfo{author}{\bibfnamefont{G.}~\bibnamefont{Passarino}} \bibnamefont{and}
  \bibinfo{author}{\bibfnamefont{M.~J.~G.} \bibnamefont{Veltman}},
  \bibinfo{journal}{Nucl. Phys.} \textbf{\bibinfo{volume}{B160}},
  \bibinfo{pages}{151} (\bibinfo{year}{1979}).

\bibitem{'tHooft:1978xw}
\bibinfo{author}{\bibfnamefont{G.}~\bibnamefont{'t~Hooft}} \bibnamefont{and}
  \bibinfo{author}{\bibfnamefont{M.~J.~G.} \bibnamefont{Veltman}},
  \bibinfo{journal}{Nucl. Phys.} \textbf{\bibinfo{volume}{B153}},
  \bibinfo{pages}{365} (\bibinfo{year}{1979}).

\bibitem{Bardin:1999ak}
\bibinfo{author}{\bibfnamefont{D.~Y.} \bibnamefont{Bardin}} \bibnamefont{and}
  \bibinfo{author}{\bibfnamefont{G.}~\bibnamefont{Passarino}},
  \emph{\bibinfo{title}{{The standard model in the making: Precision study of
  the electroweak interactions}}} (\bibinfo{publisher}{Clarendon Press},
  \bibinfo{address}{Oxford, UK}, \bibinfo{year}{1999}).

\bibitem{DeJager:1987qc}
\bibinfo{author}{\bibfnamefont{H.}~\bibnamefont{De~Vries}},
  \bibinfo{author}{\bibfnamefont{C.~W.} \bibnamefont{De~Jager}},
  \bibnamefont{and} \bibinfo{author}{\bibfnamefont{C.}~\bibnamefont{De~Vries}},
  \bibinfo{journal}{Atom. Data Nucl. Data Tabl.} \textbf{\bibinfo{volume}{36}},
  \bibinfo{pages}{495} (\bibinfo{year}{1987}).

\bibitem{Fricke:1995zz}
\bibinfo{author}{\bibfnamefont{G.}~\bibnamefont{Fricke}},
  \bibinfo{author}{\bibfnamefont{C.}~\bibnamefont{Bernhardt}},
  \bibinfo{author}{\bibfnamefont{K.}~\bibnamefont{Heilig}},
  \bibinfo{author}{\bibfnamefont{L.~A.} \bibnamefont{Schaller}},
  \bibinfo{author}{\bibfnamefont{L.}~\bibnamefont{Schellenberg}},
  \bibinfo{author}{\bibfnamefont{E.~B.} \bibnamefont{Shera}}, \bibnamefont{and}
  \bibinfo{author}{\bibfnamefont{C.~W.} \bibnamefont{de~Jager}},
  \bibinfo{journal}{Atom. Data Nucl. Data Tabl.} \textbf{\bibinfo{volume}{60}},
  \bibinfo{pages}{177} (\bibinfo{year}{1995}).

\bibitem{NuclearData}
\bibinfo{author}{\bibfnamefont{A.}~\bibnamefont{Brody}},
  \bibinfo{author}{\bibfnamefont{D.}~\bibnamefont{Day}},
  \bibinfo{author}{\bibfnamefont{B.}~\bibnamefont{Lewis}}, \bibnamefont{and}
  \bibinfo{author}{\bibfnamefont{S.}~\bibnamefont{Washington}}
  \bibinfo{note}{{\emph{The Nuclear Charge Density Archive},
  \url{http://faculty.virginia.edu/ncd/index.html}}}.

\bibitem{Gonzalez:2013rea}
\bibinfo{author}{\bibfnamefont{M.}~\bibnamefont{Gonzalez}},
  \bibinfo{author}{\bibfnamefont{T.}~\bibnamefont{Gutsche}},
  \bibinfo{author}{\bibfnamefont{J.~C.} \bibnamefont{Helo}},
  \bibinfo{author}{\bibfnamefont{S.}~\bibnamefont{Kovalenko}},
  \bibinfo{author}{\bibfnamefont{V.~E.} \bibnamefont{Lyubovitskij}},
  \bibnamefont{and} \bibinfo{author}{\bibfnamefont{I.}~\bibnamefont{Schmidt}},
  \bibinfo{journal}{Phys. Rev.}
  \textbf{\bibinfo{volume}{D87}}(\bibinfo{number}{9}), \bibinfo{pages}{096020}
  (\bibinfo{year}{2013}), \eprint{1303.0596}.

\bibitem{Faessler:2005hx}
\bibinfo{author}{\bibfnamefont{A.}~\bibnamefont{Faessler}},
  \bibinfo{author}{\bibfnamefont{T.}~\bibnamefont{Gutsche}},
  \bibinfo{author}{\bibfnamefont{S.}~\bibnamefont{Kovalenko}},
  \bibinfo{author}{\bibfnamefont{V.~E.} \bibnamefont{Lyubovitskij}},
  \bibnamefont{and} \bibinfo{author}{\bibfnamefont{I.}~\bibnamefont{Schmidt}},
  \bibinfo{journal}{Phys. Rev.} \textbf{\bibinfo{volume}{D72}},
  \bibinfo{pages}{075006} (\bibinfo{year}{2005}), \eprint{hep-ph/0507033}.

\bibitem{Faessler:2004ea}
\bibinfo{author}{\bibfnamefont{A.}~\bibnamefont{Faessler}},
  \bibinfo{author}{\bibfnamefont{T.}~\bibnamefont{Gutsche}},
  \bibinfo{author}{\bibfnamefont{S.}~\bibnamefont{Kovalenko}},
  \bibinfo{author}{\bibfnamefont{V.~E.} \bibnamefont{Lyubovitskij}},
  \bibinfo{author}{\bibfnamefont{I.}~\bibnamefont{Schmidt}}, \bibnamefont{and}
  \bibinfo{author}{\bibfnamefont{F.}~\bibnamefont{Simkovic}},
  \bibinfo{journal}{Phys. Rev.} \textbf{\bibinfo{volume}{D70}},
  \bibinfo{pages}{055008} (\bibinfo{year}{2004}), \eprint{hep-ph/0405164}.

\bibitem{Faessler:2004jt}
\bibinfo{author}{\bibfnamefont{A.}~\bibnamefont{Faessler}},
  \bibinfo{author}{\bibfnamefont{T.}~\bibnamefont{Gutsche}},
  \bibinfo{author}{\bibfnamefont{S.}~\bibnamefont{Kovalenko}},
  \bibinfo{author}{\bibfnamefont{V.~E.} \bibnamefont{Lyubovitskij}},
  \bibinfo{author}{\bibfnamefont{I.}~\bibnamefont{Schmidt}}, \bibnamefont{and}
  \bibinfo{author}{\bibfnamefont{F.}~\bibnamefont{Simkovic}},
  \bibinfo{journal}{Phys. Lett.} \textbf{\bibinfo{volume}{B590}},
  \bibinfo{pages}{57} (\bibinfo{year}{2004}), \eprint{hep-ph/0403033}.

\bibitem{Kosmas:1997eca}
\bibinfo{author}{\bibfnamefont{T.~S.} \bibnamefont{Kosmas}},
  \bibinfo{author}{\bibfnamefont{A.}~\bibnamefont{Faessler}},
  \bibinfo{author}{\bibfnamefont{F.}~\bibnamefont{Simkovic}}, \bibnamefont{and}
  \bibinfo{author}{\bibfnamefont{J.~D.} \bibnamefont{Vergados}},
  \bibinfo{journal}{Phys. Rev.} \textbf{\bibinfo{volume}{C56}},
  \bibinfo{pages}{526} (\bibinfo{year}{1997}), \eprint{nucl-th/9704021}.

\bibitem{Faessler:2012ku}
\bibinfo{author}{\bibfnamefont{A.}~\bibnamefont{Faessler}},
  \bibinfo{author}{\bibfnamefont{V.}~\bibnamefont{Rodin}}, \bibnamefont{and}
  \bibinfo{author}{\bibfnamefont{F.}~\bibnamefont{Simkovic}},
  \bibinfo{journal}{J. Phys.} \textbf{\bibinfo{volume}{G39}},
  \bibinfo{pages}{124006} (\bibinfo{year}{2012}), \eprint{1206.0464}.

\bibitem{Barea:2015kwa}
\bibinfo{author}{\bibfnamefont{J.}~\bibnamefont{Barea}},
  \bibinfo{author}{\bibfnamefont{J.}~\bibnamefont{Kotila}}, \bibnamefont{and}
  \bibinfo{author}{\bibfnamefont{F.}~\bibnamefont{Iachello}},
  \bibinfo{journal}{Phys. Rev.}
  \textbf{\bibinfo{volume}{C91}}(\bibinfo{number}{3}), \bibinfo{pages}{034304}
  (\bibinfo{year}{2015}), \eprint{1506.08530}.

\bibitem{Hyvarinen:2015bda}
\bibinfo{author}{\bibfnamefont{J.}~\bibnamefont{Hyv{\"a}rinen}}
  \bibnamefont{and} \bibinfo{author}{\bibfnamefont{J.}~\bibnamefont{Suhonen}},
  \bibinfo{journal}{Phys. Rev.}
  \textbf{\bibinfo{volume}{C91}}(\bibinfo{number}{2}), \bibinfo{pages}{024613}
  (\bibinfo{year}{2015}).

\bibitem{Engel:2015wha}
\bibinfo{author}{\bibfnamefont{J.}~\bibnamefont{Engel}}, \bibinfo{journal}{J.
  Phys.} \textbf{\bibinfo{volume}{G42}}(\bibinfo{number}{3}),
  \bibinfo{pages}{034017} (\bibinfo{year}{2015}).

\bibitem{Simkovic:2013qiy}
\bibinfo{author}{\bibfnamefont{F.}~\bibnamefont{Simkovic}},
  \bibinfo{author}{\bibfnamefont{V.}~\bibnamefont{Rodin}},
  \bibinfo{author}{\bibfnamefont{A.}~\bibnamefont{Faessler}}, \bibnamefont{and}
  \bibinfo{author}{\bibfnamefont{P.}~\bibnamefont{Vogel}},
  \bibinfo{journal}{Phys. Rev.}
  \textbf{\bibinfo{volume}{C87}}(\bibinfo{number}{4}), \bibinfo{pages}{045501}
  (\bibinfo{year}{2013}), \eprint{1302.1509}.

\bibitem{Barea:2013bz}
\bibinfo{author}{\bibfnamefont{J.}~\bibnamefont{Barea}},
  \bibinfo{author}{\bibfnamefont{J.}~\bibnamefont{Kotila}}, \bibnamefont{and}
  \bibinfo{author}{\bibfnamefont{F.}~\bibnamefont{Iachello}},
  \bibinfo{journal}{Phys. Rev.}
  \textbf{\bibinfo{volume}{C87}}(\bibinfo{number}{1}), \bibinfo{pages}{014315}
  (\bibinfo{year}{2013}), \eprint{1301.4203}.

\bibitem{Suhonen:2012ii}
\bibinfo{author}{\bibfnamefont{J.}~\bibnamefont{Suhonen}} \bibnamefont{and}
  \bibinfo{author}{\bibfnamefont{O.}~\bibnamefont{Civitarese}},
  \bibinfo{journal}{J. Phys.} \textbf{\bibinfo{volume}{G39}},
  \bibinfo{pages}{124005} (\bibinfo{year}{2012}).

\bibitem{Suhonen:2012zzc}
\bibinfo{author}{\bibfnamefont{J.}~\bibnamefont{Suhonen}} \bibnamefont{and}
  \bibinfo{author}{\bibfnamefont{O.}~\bibnamefont{Civitarese}},
  \bibinfo{journal}{J. Phys.} \textbf{\bibinfo{volume}{G39}},
  \bibinfo{pages}{085105} (\bibinfo{year}{2012}).

\bibitem{Dohmen:1993mp}
\bibinfo{author}{\bibfnamefont{C.}~\bibnamefont{Dohmen}} \emph{et~al.}
  (\bibinfo{collaboration}{SINDRUM II}), \bibinfo{journal}{Phys. Lett.}
  \textbf{\bibinfo{volume}{B317}}, \bibinfo{pages}{631} (\bibinfo{year}{1993}).

\bibitem{Bertl:2006up}
\bibinfo{author}{\bibfnamefont{W.~H.} \bibnamefont{Bertl}} \emph{et~al.}
  (\bibinfo{collaboration}{SINDRUM II}), \bibinfo{journal}{Eur. Phys. J.}
  \textbf{\bibinfo{volume}{C47}}, \bibinfo{pages}{337} (\bibinfo{year}{2006}).

\bibitem{Honecker:1996zf}
\bibinfo{author}{\bibfnamefont{W.}~\bibnamefont{Honecker}} \emph{et~al.}
  (\bibinfo{collaboration}{SINDRUM II}), \bibinfo{journal}{Phys. Rev. Lett.}
  \textbf{\bibinfo{volume}{76}}, \bibinfo{pages}{200} (\bibinfo{year}{1996}).

\bibitem{Aoki:2010zz}
\bibinfo{author}{\bibfnamefont{M.}~\bibnamefont{Aoki}}
  (\bibinfo{collaboration}{DeeMe}), \bibinfo{journal}{PoS}
  \textbf{\bibinfo{volume}{ICHEP2010}}, \bibinfo{pages}{279}
  (\bibinfo{year}{2010}).

\bibitem{COMET}
\bibinfo{author}{\bibfnamefont{Y.~G.} \bibnamefont{Ciu}} \emph{et~al.},
  \emph{\bibinfo{title}{{Conceptual Design Report for Experimental Search for
  Lepton Flavor Violating mu-e Conversion at Sensitivity of $10^{16}$ with a
  Slow-Extracted Bunched Proton Beam (COMET) J-PARC P21}}}
  (\bibinfo{year}{2009}),
  \bibinfo{note}{\url{http://comet.phys.sci.osaka-u.ac.jp:8080/comet/internal/publications/comet-cdr-v1.0.pdf/view}.}

\bibitem{Kutschke:2011ux}
\bibinfo{author}{\bibfnamefont{R.~K.} \bibnamefont{Kutschke}}
  (\bibinfo{year}{2011}), \eprint{1112.0242}.

\bibitem{Barlow:2011zza}
\bibinfo{author}{\bibfnamefont{R.~J.} \bibnamefont{Barlow}},
  \bibinfo{journal}{Nucl. Phys. Proc. Suppl.} \textbf{\bibinfo{volume}{218}},
  \bibinfo{pages}{44} (\bibinfo{year}{2011}).

\bibitem{COMET2014}
\bibinfo{author}{\bibfnamefont{R.}~\bibnamefont{Akhmetshin}} \emph{et~al.}
  (\bibinfo{collaboration}{COMET}), \emph{\bibinfo{title}{{COMET International
  Review Document (TDR-2014)}}} (\bibinfo{year}{2014}),
  \bibinfo{note}{\url{http://comet.kek.jp/Documents_files/IPNS-Review-2014.pdf}.}

\bibitem{Chakrabortty:2015zpm}
\bibinfo{author}{\bibfnamefont{J.}~\bibnamefont{Chakrabortty}},
  \bibinfo{author}{\bibfnamefont{P.}~\bibnamefont{Ghosh}},
  \bibinfo{author}{\bibfnamefont{S.}~\bibnamefont{Mondal}}, \bibnamefont{and}
  \bibinfo{author}{\bibfnamefont{T.}~\bibnamefont{Srivastava}}
  (\bibinfo{year}{2015}), \eprint{1512.03581}.

\bibitem{Denner:1992vza}
\bibinfo{author}{\bibfnamefont{A.}~\bibnamefont{Denner}},
  \bibinfo{author}{\bibfnamefont{H.}~\bibnamefont{Eck}},
  \bibinfo{author}{\bibfnamefont{O.}~\bibnamefont{Hahn}}, \bibnamefont{and}
  \bibinfo{author}{\bibfnamefont{J.}~\bibnamefont{Kublbeck}},
  \bibinfo{journal}{Nucl. Phys.} \textbf{\bibinfo{volume}{B387}},
  \bibinfo{pages}{467} (\bibinfo{year}{1992}).

\end{thebibliography}

\end{document}